\documentclass[12pt,numberedappendix]{emulateapj}

\slugcomment{Submitted to PASP}

\shorttitle{Instrumentation on Large Optical Telescopes}
\shortauthors{S. R. Kulkarni}

\begin{document}

\title{Instruments on Large  Optical Telescopes -- A Case Study}

\author{S. R. Kulkarni}
\affil{Caltech Optical Observatories 249-17\\
California Institute of Technology, Pasadena, CA 91108}

\begin{abstract}

In the distant past, telescopes were known, first and foremost, for
the sizes of their apertures. However, the astronomical output of
a telescope is determined by both the size of the aperture as well
as the capabilities of the attached instruments.  Advances in
technology (not merely those related to astronomical detectors) are
now enabling astronomers to build extremely powerful instruments
to the extent that instruments have now achieved  importance
comparable or even exceeding the usual importance accorded to the
apertures of the telescopes. However, the cost of successive
generations of instruments has risen at a rate noticeably above that of
the rate of inflation. Indeed, the cost of instruments, when spread
over their prime lifetime, can be a significant expense for
observatories.  Here, given the vast sums of money now being expended
on optical telescopes and their instrumentation, I argue that
astronomers must undertake ``cost-benefit'' analysis for future
planning.  I use the scientific output of the first two decades of
the W.~M.~Keck Observatory  as a laboratory for this purpose. I
find, in the absence of upgrades, that the time to reach peak paper
production for an instrument  is about  six years. The prime lifetime
of instruments (sans upgrades), as measured by citations returns,
is about a decade. Well thought out and timely upgrades  increase
and sometimes even double the useful lifetime. Thus, upgrades  are
highly cost effective.  I investigate how well  instrument builders
are rewarded (via citations by users of their instruments). I find
acknowledgements ranging from almost 100\% to as low as 60\%. Next,
given the increasing cost of operating optical telescopes, the
management of existing observatories continue to seek  new partnerships.
This naturally raises the question ``What is the cost of a single
night of telescope time". I provide a rational basis to compute
this quantity.  I then end the paper with some thoughts on the
future of large ground-based optical telescopes, bearing in mind
the explosion of synoptic precision photometric, astrometric and
imaging surveys across the electromagnetic spectrum, the increasing
cost of instrumentation and the rise of mega instruments.

\end{abstract}

%\keywords{instruments: efficiency}

\date{\today}

\section{Background \&\ Motivation}
\label{sec:Motivation}
 
Historically, ground-based optical telescopes have been the primary
experimental method by which astronomers investigated the heavens.
The serendipitous discovery of cosmic radio emission and later
cosmic X-ray sources led to a flood of exploration in other
electromagnetic bands.  Today it is routine for an active astronomer
to call upon  data from radio (decameter to the sub-millimeter),
thermal infrared (mid infrared, MIR), near infrared (NIR),  space
ultra-violet (UV) and high energy (X-ray, $\gamma$-ray) bands to
study and draw conclusions about celestial objects.

Space-based astronomy offers exquisite performance in several ways.
For certain bands (e.g.\ UV, X-ray, THz and others) either the poor
transmission through the atmosphere or a high atmospheric background
leave us with no choice but to go to space.  For other bands (e.g.\
MIR; see Appendix~\S\ref{sec:NIR} for definitions of IR bands)
ground-based observations suffer from high but (barely) acceptable
background noise.  Next, atmospheric turbulence degrades the
wave-front leading to poor  image quality and a corresponding
decrease in precision and accuracy of photometry and astrometry.
Adaptive optics (AO) offers some solace but with limitations (e.g.\
narrow field, requirement for guide stars).  Independent  of this
discussion, it is hard to beat space-based instruments when one
desires ultra-fine measurements in photometry (e.g.\ color-magnitude
of globular clusters, extra-solar planet transits, astero-seismology,
CMB observations) or wide field astrometry (e.g.\ {\it Gaia}).

Separately, there is now a substantial investment in non-electromagnetic
astronomical facilities:  neutrinos, gravitational waves and cosmic
rays (and primarily pursued by physicists). These very large
investments are a testimony to the fecundity of astronomy.

Despite investments in flagship space-based  electromagnetic missions
and flagship non-electromagnetic facilities, the fact remains that
ground-based optical and infrared (0.3--2\,$\mu$m; hereafter,
optical-IR or OIR) telescopes continue to play a {\it leading} role
in the overall development of astronomy.  In the optical band, the
atmosphere is relatively quiet and the absorption is low. At the
same time, in the optical band, celestial sources exhibit a moderate
number  of spectral lines  from which astronomers can infer distance
(via redshift), masses (via velocities), temperatures (via line
width or line ratios) and the abundances of a number of elements.

\subsection{Maturity of Optical Telescope \& Observatory Technology}
\label{sec:MaturityTelescopes}

We are now in the fifth century since a patent application for a
``spyglass" (the forerunner of telescope) was made  by H.\ Lippershey
of Zeeland (a province of the Netherlands). On hearing of the
invention, G. Galileo who was then working in Venice, put together
a small telescope. With the double advantage of being ``first on
the block" and possessing deep physical insight Galileo  went on
to make revolutionary advances in astronomy, physics and theology.
It is not a surprise that later generation astronomers aspire to
at least have the same external advantage as Galileo himself had
(namely, first access to a revolutionary observational facility).

The early refractors gave way to reflectors. Over the course of
time  there have been improvements in every aspect related to the
engineering of telescopes: mirror coatings, materials (e.g.\ low
expansion glass such as Zerodur); opto-mechanical solutions which
abandon rigidity for knowledge and control  (thin mirrors with
active optics; e.g.\ European Southern Observatory's  New Technology
Telescope); and large  monolithic mirrors with nearly unity $f$-ratios
(made possible by honey-comb light weighting and spin casting; e.g.\
the Large Binocular 8.4-m mirrors).  In my view, in my lifetime,  the
greatest advance  in telescope engineering is finely segmented
telescopes (e.g.\ the Keck 10-m telescope). This approach has opened
up an elegant path for the realization of larger telescopes at lower
cost (on a per unit area basis).

Thanks to all the advances discussed above  the cost of  large
telescopes (per unit area) is decreasing.  As a result
the global astronomical community now enjoys a dozen large aperture
(8-m and 10-m) telescopes. Even bigger telescopes are now being planned
or are under construction.

So far the discussion has been about telescopes  which are ultimately
based on a glass-based parabolic mirror to collect the light.  There
have been attempts at alternative approaches.  Liquid mirrors could
offer an inexpensive way to realize large apertures (e.g.\ the Large
Zenith Telescope\footnote{http://www.astro.ubc.ca/LMT/} based on
liquid mercury). Another approach is spherical
reflectors fixed to the ground (cf.\ the Arecibo radio telescope). 
However, to date there is no liquid
mirror telescope in routine operation and there are only two
operational spherical mirror telescopes   (see
\S\ref{sec:EraOfLargeTelescopes}).

The ``delivered image quality" (DIQ) of  a telescope, even if
perfectly engineered, is limited by ``seeing" which has several
components: high-altitude seeing, ground-layer seeing and dome
seeing.  Astronomers have become painfully aware of these issues.
As a result, nowadays, astronomers undertake extensive studies of
telescope sites before finalizing the site selection (e.g.\
\citealt{ser+09}).  Thermal and seeing (turbulence) control is
another explicit engineering consideration in the design of modern
observatories (e.g.\ \citealt{rsc+91,bbb+14}).  Domes are designed
keeping in mind prevailing winds (and with computer controlled
louvers to prevent buildup of turbulence within the dome) and cooled
to temperatures anticipated for the coming night (e.g.\ \citealt{bbt+12}).
Thinner mirrors, cooling lines and carefully engineered  heat
dissipation by instruments are key inputs for good thermal control
of  the telescope.  As a result, the DIQ of telescopes has consistently
increased with each generation.  It is fair to say that a modern
well-designed telescope can be expected to routinely perform at a
level  limited by overall site seeing.

I end this section by a parenthetical remark, namely that the technology
for fabricating small and moderate size telescopes is now quite
mature. The primary advance (and gains) lie in  reduction of unit
costs.\footnote{e.g.\ a fully robotic 70-cm telescope from Planewave
Instruments of Los Angeles costs \$200K (not including the burdensome
California sales taxes).} This trend combined with continued
improvements in detector technology (particularly the possibility of
low or nearly
zero read noise) opens  up the possibility of realizing a large
aperture via a number of small diameter telescopes (``Large Aperture
via Small Telescopes" or LAST; this can be compared to ``Large Number of
Small Diameter dishes" or LNSD architecture in radio astronomy).  Separately,
it may not be surprising that within this decade astronomers will
have farms of 1-m telescopes, each dedicated for a specific target
or a specific cause.

\subsection{The Rising Cost of Instrumentation}

While the telescope gathers light it is the instrument that delivers
the science.  The costs of instruments were minor for the first
generation of modern telescopes (e.g.\ the Lick 36-inch refractor
or the Mt.\ Wilson 60-inch reflector).  Imaging was provided by a
simple camera with a photographic plate. The imaging was, at best,
seeing limited and thus the optics were simple (the plates could
also be curved, if needed, thus further simplifying the optics).
The focus was on single object spectroscopy and this simplified the
design of the spectrographs.  In both cases, the observer was
responsible for the most delicate part of the observation -- the
guiding.

Advances in technology have made it possible to build instruments
which can fill a significant fraction of the available focal plane.
As a result, modern spectrographs have the ability to return spectra
of multiple objects (large {\it reach}).  A new development is
``mega" instruments which are instruments with extra-ordinarily
large reach (Appendix~\ref{sec:MegaInstruments}).  These instruments
have already had a big impact and are poised to fundamentally change the
landscape of optical telescopes.  While in the past, say about three
decades ago, one talked of the aperture of telescopes, today
astronomers talk of the capabilities of the mega instruments just
as much as (and sometimes even more than) the apertures of telescopes.

However, it appears to be the case that the cost of instruments has
risen faster than the nominal and the real GDP.  In addition, rapid changes
in technology are accelerating obsolescence. This combination is
deadly in that the instrumentation ``line" (the annual cost for
instrumentation, averaged over say a decade) can become  financially
draining.

Next, in the not-so-distant past, astronomers were not accustomed
to the word ``pipeline" or ``user ready data products". It was
expected that the data reduction was undertaken by each astronomer
using their own tools or within a framework supplied by the Observatory
(e.g.\ IRAF). This worked reasonably well since most astronomers
were quite specialized and typically wedded to a single facility
or a narrow suite of instruments.

In view of the large sums expended for flagship projects funding
agencies like to see {\it maximal} and {\it timely} exploitation
of data.  The expectation of great returns, in turn, mandates
sophisticated algorithms for optimal extraction. Next, instruments
with large reach produce such large amounts of data that the
traditional ``hand" data reduction is not practical.  These two
drivers have led to the growth of high quality data reduction
pipelines (DRP).   DRPs with such high expectations are not cheap.
After all each DRP has to contend with data taken under different
observing conditions and account for instrumental idiosyncrasies
whilst still delivering optimal returns.   Finally, the increased
cost of astronomical facilities has naturally led to the development
of archives so as to maximize the returns from the mission or
facility.  Unfortunately, archives, if they are to be useful at all
(which means those which produce high value product on request) do
not come cheap, also.

\subsection{ The Thesis \&\ the Motivation}

The fundamental thesis of this paper is, given the maturity of
telescope technology,  that the output of an Observatory following
the commissioning of the telescope is determined primarily by its
instrumentation.  Given the discussion in the previous section  the
term ``instrumentation" includes quality DRPs and powerful archives.

Large optical telescopes are expensive. The capitalization cost is
in excess of \$150M (for a single telescope). A full suite of high
quality instrumentation could easily run up to \$50M (or more). The
operating cost including new instrumentation and upgrades start at
\$15M (and up).  Clearly, observatories hosting large optical
telescope must be regarded as ``large" science. As such optical
astronomers must undertake ``cost-benefit" analysis and come to
grips with ``opportunity cost"\footnote{The formal definition is
``the loss of potential gain from other alternatives when one
alternative is chosen".  This important concept was developed by
the French economist Fr\'ed\'eric Bastiat and the classic reference is
his 1850 essay entitled ``What is Seen and What is Unseen".} of their
decisions.

One could argue that, since astronomical research is far removed
from ordinary life,  the very concept of cost-benefit analysis is
meaningless. I do not agree with this sentiment for two reasons.
First, when hundreds of millions of dollars are being spent,  funding
agencies necessarily demand a greater level of scrutiny and
justification.  Next, to me it is a self-evident truth that research
is simply another human activity and as such subject to the same
set of issues as one faces in ordinary life.

Here, I use the scientific output of the W.~M.~Keck Observatory
(WMKO) -- one of the two observatories that I am familiar with --
as a laboratory for the ``business" of large OIR telescope
observatories.    The first goal of this paper is to measure the
impact of instrumentation. Next, the increased cost of operating
large optical facilities is motivating the operators of Observatories
to seek partnerships (and inversely those lacking access to seek
partnership on existing telescopes).  This development leads to the
second goal: the construction of a framework in which the  {\it
value} for each night can be computed and accepted by a {\it rational}
market.

\subsection{The Organization of the Paper}

The paper is organized as follows.  In \S\ref{sec:MeasuringProgress}
I argue that the annual flux of citations is a good measure of the
productivity of an observatory. This is followed by a brief history
of WMKO (\S\ref{sec:BriefHistory}).  In \S\ref{sec:TheInstruments}
I summarize the principal instruments that have been or continue
to be employed at the W.~M.~Keck Observatory (or simply the Observatory) followed by the Adaptive Optics
facilities (\S\ref{sec:AdaptiveOptics}).  The primary input for
this report are the papers which have resulted from data based on
the Keck Observatory. In \S\ref{sec:PrimaryDataMethodology} I
summarize the methodologies used and metrics employed in this paper.
The analysis and basic inferences can be found in \S\ref{sec:Analysis}
and \S\ref{sec:Inferences}.  In \S\ref{sec:Archives} I summarize a
recent development, the Keck Observatory Archive. This archive
enables further exploitation of Keck data and in the process is
augmenting the productivity of the Observatory.  In
\S\ref{sec:CostAndValue} I propose that the value of one  night of
telescope time should be tied to the productivity of the Observatory.
I end the paper first by summarizing the rapidly evolving landscape
for optical/IR astronomy (\S\ref{sec:FutureLandscape}) followed by
my views of the future of large optical telescopes
(\S\ref{sec:LargeOpticalTelescopes}) and that of the W.\ M.\ Keck
Observatory (\S\ref{sec:FutureWMKO}).

\section{Measuring Progress} 
\label{sec:MeasuringProgress}

The cost of an astronomical instrument or facility is easy to define.
For telescopes it is the money spent to  design and fabricate the
telescope through the commissioning of the first light instruments.
This sum is usually referred to as  the ``capital cost''.  For
facilities one must also include the annual
operation or ``ops" cost.  Ops cost must include expenses
for  infra-structure
improvements, instrument upgrades, and developing and
maintaining archives.  The benefits are much harder to quantify and
some may even argue that benefits cannot even be agreed upon by a
group of astronomers (with disparate interests).

However, the situation is not entirely hopeless. There exists a
rich literature of astronomers defining and measuring progress. A
good review of astronomical ``bibilometrica'' (or ``scientometrica")
is provided by \citet{Abt2005}.  I found myself entirely in agreement
with the opening paragraph of Abt's paper: ``Astronomers insist
upon seeing quantitative evidence in scientific papers or they will
not believe the results claimed. However, when discussing policies
or making decisions about funding, instrumentation, promotions,
etc., they depend mostly upon impressions, feelings and intuition.
But measures of productivity, success and importance can be
quantitative, and quantitative measures should replace impressions.''

In this paper I will be using two metrics to measure  progress.
Most research consists of making gradual progress. Thus an active
area of astronomy (almost by definition) will have a  flux of papers,
and necessarily this flux will be associated with a flux of citations.
In most cases,  activity  can be reasonably expected to measure
progress. We thus use the {\it citation flux} as a measure of routine
progress.

Next, Abt ({\it ibid}) demonstrates that the top cited papers are
almost always agreed to be landmark papers by eminent astronomers
and inversely those considered to be  landmark papers are  also
heavily cited.  Abt arrives at this conclusion by using the Centennial
Issue\footnote{\url{http://www.amazon.com/American-Astronomical-Society-Centennial-Astrophysical/dp/0226001857}}
of the Astrophysical Journal as the input sample. He cleverly builds
the control sample (papers which, in the Astrophysical Journal,
merely precede highly cited papers).  As a simple check, I went
through my list of papers and composed a list of what I thought were my top ten
papers. I compared this list to a list of ten papers with the highest
citations.  I found an excellent concordance between the two lists.
Thus, as a second measure of progress, I will be using {\it the
collection of the most cited papers}.

Returning to the subject of ``bibliometrics'' I refer the reader
to a series  of papers by V. Trimble and associates and by  H.\ Abt
(e.g.\ \citealt{tzb05,tz06,tc08,Abt2012}). These authors use citation
rates and investigate  the productivity and impact of telescopes
of various apertures, of different vintages, sorted by wavelength
and so on and so forth.

Before proceeding further I would like to acknowledge that the
statistics of citation are, in part, dependent on fashion and
certainly influenced by the number of people who work in a given
field (which is directly correlated with funding). In
astronomy, currently, the two most popular and fashionable fields
are cosmology and extra-solar planets.
\cite{pk12} define  a new index ``Total Research Impact" or {\it
tori} which takes into account (1) field-dependent citation rates
(popular versus less popular fields), (2) the number of co-authors
(papers with many co-authors are likely to be cited more often than
single author papers) and (3) shot noise (some papers become very
popular for reasons that are not clear even after the fact, cf.\
{\it Gangnam Style}
phenomenon\footnote{\url{http://en.wikipedia.org/wiki/Gangnam\_Style}.
\citet{backovic16} provides analytical models for equivalent phenomena
in astro-particle physics, CMB and particle physics.}).  As noted
above, funding directly determines the number of researchers working
in a field. In turn, funding has several drivers including particularly
the choice of missions or facilities.
Here, I will stick to the two
measures, both based on citations, but add the caution that, for
all the reasons mentioned above, it may not be appropriate to compare
the citation returns from, say,  ground-based optical facilities
to, say, those resulting from ground-based radio facilities or
space-based facilities.

\section{The W.~M.~Keck Observatory: A Brief History}
\label{sec:BriefHistory}

%CFHT 1979
%NTT 1989

The history of optical/IR astronomy has been, for a long time,
driven by ever increasing apertures. Larger collecting areas allow
for spectroscopy of faint objects-- an almost unique contribution
of ground-based optical astronomy.  However, as noted in
\S\ref{sec:MaturityTelescopes}, getting the best DIQ starts off
with cold sites (critical for operations in K-band and longer wavelengths)
with excellent and stable seeing and preferably with
little variation in night time temperature.
Thanks to the pioneering astronomer Gerard
Kuiper and the continued efforts of astronomers at the University
of Hawaii (UH), in particular John T.\ Jeffries, Mauna Kea was found
to be a high quality site for astronomical observations.  

The UH
88-inch telescope, commissioned in 1970,
 was the first research telescope atop Mauna Kea.
The year 1979 saw the commissioning of
NASA's (National Aeronautics \&\ Space Administration) Infrared 3-m
Telescope Facility (IRTF),  the Canada-France-Hawaii (CFH) 3.6-m
telescope (hereafter, CHFT) and the United Kingdom Infrared 3.6-m
telescope (UKIRT).  In particular, CFHT was a highly visible
international project. The great success of this telescope demonstrated
the value of locating a modern large telescope at a site with superb
seeing. It was only natural that Mauna Kea was chosen as the site
for the next large telescope coming from the West Coast of the US
-- the Keck 10-m telescope(s).

\subsection{The Keck 10-m Telescopes}

Breaking the tradition of monolithic primary mirror, the large
aperture of the 10-m Keck telescope was realized by 36 hexagonal
segments.  This approach was pioneered by Jerry Nelson and Terry
Mast of the Lawrence Berkeley Laboratory (LBL), University of
California at Berkeley (UCB).  The Keck project began with a grant,
in 1985, of \$70M from the W.~M.~Keck Foundation to California
Institute of Technology (Caltech) in support of the construction
of the first Keck telescope. The University of California (UC) and
Caltech formed a non-profit entity, the California Association for
Research in Astronomy (CARA), and jointly led the Keck project. As
a part of this agreement, UC signed up to pay for operations of the
Observatory for the first twenty five years.\footnote{The specific 
financial arrangement
ends by March 2018, after which both UC and Caltech will bear
equal financial responsibility.}  Following ground-breaking in 1986, first light
on Keck~I (with all segments) was obtained on 14 April 1992.  The
first light instruments were three workhorses: NIRC, LRIS and HIRES
(described below in \S\ref{sec:TheInstruments}).  The construction
costs of these instruments were included as a part of the construction
cost of Keck~I.  The run-out cost\footnote{Throughout this paper,
costs are ``then-year" costs, unless otherwise stated.} through
first light for Keck~I was \$94.3M.

In 1992, the Keck Foundation donated a second tranche, to the tune
of \$74.5M, to Caltech for the construction of the Keck~II telescope.
The construction was completed in early 1996 and routine observations
began in October of 1996.  The runout cost\footnote{All the cost
numbers reported here, including the extended commissioning costs,
were obtained by the author from  Gerald (``Jerry") Smith, the Project Manager for
the Keck Telescopes.} for Keck~II was \$77.8M.     In return for
hosting the telescopes on the Mauna Kea Science Reserve, the
University of Hawaii receives 10\% of Keck~I and 15\% of Keck~II
time.

Separately,  what eventually became the Keck Interferometer emerged
as a major recommendation from the TOPS (Toward Other Planetary
Systems) study commissioned by NASA.  In 1996 NASA joined CARA as
a  partner and did so by contributing \$30M as capital contribution
for a sixth share and a proportional fraction of the  ops cost.  
Soon thereafter, in response to the recommendations
of TOPS  and other advisory committees, NASA embarked on a program
to implement the Keck Interferometer project. NASA selected JPL to
implement the interferometer jointly with WMKO.

%NY Times article 
%http://www.nytimes.com/1996/05/09/us/world-s-biggest-telescope-has-finally-met-its-match-a-twin.html

The incurred (capital) cost for the two Keck telescopes was \$172M
(or \$187.6M, if post-construction commissioning costs are included).
Usually the average of these two numbers is often quoted in the
media\footnote{\label{ft:COSTKECK}\url{http://www.nytimes.com/1996/05/09/us/world-s-biggest-telescope-has-finally-met-its-match-a-twin.html}\\
New York Times, May 9, 1996. Money left over from the construction
of Keck~II, including interest earned, was applied towards the
development and construction of the first AO system
(\S\ref{sec:AdaptiveOptics}).}. This low cost is a testament to
both the ingenuity of the designers of the telescope as well as
vivid demonstration of the segmented architecture in breaking the
cost scaling law for monolithic telescopes \citep{sdg+03}.

The Keck telescopes had a major impact
\citep{Crabtree08,Kim2011}\footnote{Crabtree issues an annual update
to his 2008 paper.} because not only the telescopes represent a
huge jump in collecting area (relative to the earlier generation
of large telescopes with usable  effective diameters of about 5-m)
but were also able to produce superb images limited only by the
exquisite seeing at Mauna Kea. Next, at first light, astronomers had
access  to a suite of powerful instruments.

\subsection{The Era of Large Telescopes}
\label{sec:EraOfLargeTelescopes}

%ESO-VLT (May 1998, March 1999, January 2000, September 2000)
%Subaru: Jan 1999
%Gemini: 1999 (N), 2000 (S)
%Magellan: Baade (2000), Clay (2002)

The next group of large telescopes, the 8.2-m European Southern
Observatory (ESO) Very Large Telescope (VLT; at Paranal, Chile;
1998--2002),
the Subaru 8.2-m telescope atop Mauna Kea (1999), the 6.5-m Magellan
telescopes (at Las Campanas, Chile; 2000-2002) and the two Gemini 8.2-m
telescopes (one located on Cerron Pach\'on, Chile and the other on
Mauna Kea;1999--2000) came into operation starting mid 1998 through
2002.

A different approach was taken by astronomers at the University of
Texas and the Pennsylvania State University: the realization of
large aperture but with a fixed spherical primary (cf.\ Arecibo).
The Hobby-Eberly telescope (HET; McDonald Observatory, Texas) was
the first such telescope. It used fixed segmented hexagonal segments
for the primary.  The telescope was nominally commissioned in 1996,
but keeping the segments phased was problematic. Fixes were designed
\citep{bwf+03} and implemented by 2004 \citep{bab+04}.  The lessons
learnt were applied to the South African Large Telescope (Sutherland,
South Africa; commissioned 2005). Both these telescopes achieve
large apertures (effective aperture size of about 9-m) at low cost
(but with observations limited to regions near to the zenith and
also, relative to conventional telescopes,  a limited field-of-view (FOV)\footnote{
The FOV of the HET at the time of first light was circle of diameter 4 arc minutes. A major
upgrade was undertaken for the HETDEX project (see Appendix~\S\ref{sec:MegaInstruments})
and the FOV increased to 22 arc minutes}).

\begin{deluxetable}{rrrrr}
\tabletypesize{\scriptsize}
\tablecaption{Facility Instruments at the W. M. Keck Observatory}
\tablewidth{0pt}
\tablehead{
\colhead{Inst.} & Upgrade& \colhead{Period} &
\colhead{Refs} & \colhead{Cost}\\
& & & & \$M
}
\startdata
NIRC & - & 1989-1994 & [1a,1b] &  1.9\\
LRIS & \checkmark & 1988-1994 & [2a] &  4.3\\
. & LRIS-Blue & 1995-2000 & [2b,2c] &  4.3\\
. & LRIS-ADC & 2003-2007 & [2d] &  0.9\\
. & LRIS-Red & 2007-2010 & [2e] &  1.6\\
HIRES & \checkmark & 1988-1994 & [3] &  4.2\\
. & HIRES-3-CCD & 2002-2004 & - &  1.5\\
ESI & - & 1996-2000 & [4a,4b] &  4.0\\
NIRC2 & - & 1994-2000 & - &  6.0\\
NIRSPEC & - & 1994-2000 & [6] &  4.4\\
DEIMOS & - & 1993-2002 & [7] & 11.0\\
OSIRIS & \checkmark & 2000-2005 & [8a,8b] &  5.6\\
. & H2RG & 2014-2015 & - &  1.1\\
MOSFIRE & - & 2005-2012 & [9] & 14.6 
\enddata
\tablecomments{
From left to right.: the name of the instrument, the upgrades
($\checkmark$, if one was undertaken;``-", otherwise).  any), the
period of construction, the reference to the project and the run-out
cost (marked to first light or thereabout; in ``then'' dollars).
The references are as follows:
[1a] Matthews \&\ Soifer (1994a)\nocite{ms94a}. 
[1b] Matthews \&\ Soifer (1994b)\nocite{ms94b}.
[2a] \cite{occ+95}. 
[2b] \cite{mcb+98}. [2c] \cite{ssp+04}. 
[2d] \cite{pmc+08}.
[2e] \cite{rck+10}.
[3] \cite{vab+94}. 
[4a]  \cite{sbe+02}. [4b] \cite{smb+00}.
[6] \cite{mbb+98}. 
[7] \cite{fpk+03}.
[8a] \cite{lbk+06a}. [8b]\cite{lbk+06b}.
[9] \cite{mse+12}.
}
\label{tab:TheInstruments}
\end{deluxetable}

\section{The Instruments}
\label{sec:TheInstruments}

There are (or have been) nine ``facility'' (major) instruments at
the Keck Observatory (see  Table~\ref{tab:TheInstruments} for summary
and \S\ref{sec:NIRC}--\ref{sec:MOSFIRE} for details).  There were
three other major instruments: the Long Wavelength Infrared
Camera\footnote{which was built but never commissioned}, the
Long-Wavelength Spectrometer and the Keck Interferometer. The latter
two are no longer operational.  In addition, WMKO hosted a few
``visitor" instruments. Further details or mention of these two
instruments and the visitor instruments can be found in
\S\ref{sec:OtherInstruments}.

Adaptive optics (both with natural guide star, NGS, and laser guide
star, LGS) is not an instrument but is integral to the performance
of some instruments (NIRC2, OSIRIS; see below). The performance of
such instruments is almost entirely dependent on the improvement
in image quality provided by AO. As such I have included a detailed
discussion of AO (\S\ref{sec:AdaptiveOptics}).

\subsection{Near-Infrared Camera (NIRC)}
\label{sec:NIRC}

NIRC was the first instrument to be commissioned at the W. M. Keck
Observatory.  The instrument was located in the forward Cassegrain
module of the Keck~I telescope which was fed by a gold-coated $f$/25
chopping secondary mirror.  The Principal Investigators (PIs) of the
project were Keith Matthews and B. Thomas Soifer of Caltech.

The preliminary study for NIRC began in 1987 in response to a call
for first light instruments for the Keck~I telescope.    Construction
for NIRC was initiated in 1989 and completed by the end of 1992.
The primary detector for NIRC was a Santa Barbara Research Corporation
(SBRC) ALADDIN ({\it Astronomical Large Area Detector Development
on InSb}) $256\times 256$ pixel array.  First light was obtained in March
of 1993 on the Keck~I telescope (Matthews \&\ Soifer 1994a,
1994b)\nocite{ms94a}\nocite{ms94b}.

%The total cost through first light was \$1.8M (per KYM).  
%An additional 0.1M was spent at WMKO (per Margarita).
%Final cost is the sum of the two 

Thanks to a careful optical design, NIRC achieved low background
levels which allowed for sensitive imaging and  grism (low resolution)
spectroscopy in the wavelength range of 1--5\,$\mu$m.  In 1995, an
image expander module was added and this allowed for high resolution
imaging via speckle imaging\footnote{this mode was listed as ``NIRCs''
in the scheduling logs.} \citep{mgw+96}.  The same mode was used
later on for aperture masking experiments \citep{tmd+00}.  The
instrument was decommissioned following the run of 30 January 2010.
NIRC can now be found in the lobby area of the WMKO head quarters
in Waimea (Kamuela), Hawaii.

\subsection{Low Resolution Imaging Spectrometer (LRIS)}
\label{sec:LRIS}

As with NIRC, the study for LRIS began in 1987.  LRIS, following the
venerable Double Beam Spectrograph (DBSP; \citealt{og82})\footnote{
This workhorse spectrograph,  built by J.\ Beverley Oke and James
E.\ Gunn, is still in operation at the 200-inch Palomar telescope. It
has undergone more than six detector upgrades over its lifetime.}
had one arm optimized for blue bands and the other for red  bands.
LRIS, as implied by its name, also had an imaging mode. Unlike the
previous generation of (long-) slit spectrographs, LRIS was designed to
routinely undertake multi-object spectroscopy.  The 
PIs were J. Beverly Oke and Judith G.  Cohen, both
of Caltech.

Construction of LRIS  was completed in 1992 and installed at the
Cassegrain focus of the Keck~I telescope. First science light was
achieved in the summer of 1993 (see \citealt{occ+95}).  Owing to
financial reasons only the red arm was populated for first light.
Following first light some repairs were undertaken between 1994 and
1996.

%The total cost was estimated to be \$3.6M (FY1992 base; J. Cohen, pers.
%comm.).
%There was additional work done at Keck (turret problem) and M carries
%these costs on her books. The total cost is \$4.3M

The blue arm of LRIS was populated as a part of the ``LRIS-Blue''
(LRIS-B) upgrade project. This project was led by James K. McCarthy
and Charles C. Steidel, both at Caltech, and lasted from 1995 through
2000. The addition of the blue channel thus doubled the data (with
the existing channel providing the red spectrum or red image).  In
2002 the  original Tektronix (SITe) 2K$\times$2K 24-micron 
pixel array detector was replaced by a blue-optimized Charge Coupled Device (CCD)
mosaic of two
EEV 2K$\times$4K pixel array CCDs with 15\,$\mu$m pitch. The new CCD
mosaic not only offered a better match to the spectral resolution
but also increased   the nominal spectral coverage by 25\%.  The
primary references for the LRIS-B project  are \citealt{mcb+98}
(the design) and \citealt{ssp+04} (the performance).

%The total cost of the LRIS-B project including the CCD that was 
%installed in 2002 was \$3.6M (C. Steidel, pers. comm.) 
%Adding in WMKO costs (per M) the total rises to \$4.3M

The availability of red sensitive CCDs (deep-depletion CCDs) made
it attractive to replace the original Tektronix chip by a mosaic
of two 2K$\times$4K pixel fully depleted, high resistivity  CCDs for the
red arm.  In addition, the electronics were upgraded and a new focus
mechanism installed. This project was led by Constance M. Rockosi
of the University of California at Santa Cruz (UCSC).  The initial
CCD was found to be unreliable and a replacement was installed by
end of 2010. The official reference for this upgrade is
\citet{rck+10}.

%The estimated cost for this upgrade was \$1.3M.
%When WMKO costs are included the total rises to \$1.6M

The ``Atmospheric Dispersion Corrector" (ADC)  project was headed
by Joseph S. Miller and   A. ``Drew'' Phillips, both from UCSC.
The project was initiated in 2003 and the ADC was commissioned in
2007 \citep{pmc+08}.  The ADC increases the flexibility of the
multi-object spectrograph mode (the slit mask can be designed without
paying attention to parallactic angle) and also makes possible
increased target throughput for single object spectroscopy.

%ADC subsystem: a total cost of about \$0.6M (Cowley). 
%Margarita is carrying \$0.9M on her books.
%Hilton thinks that additional labor expenses at UCSC have not
%bee included and that the true cost may exceed \$1M
%I am awaiting better numbers from M. 

\subsection{High Resolution Spectrograph (HIRES)}
\label{sec:HIRES}

As with the previous two instruments HIRES was selected  following
a call for first-light instruments for the Keck~I telescope (although
the conceptual idea and early design started in 1983).  The project
was led by Steven S. Vogt of UCSC. It took five years (1988--1993)
to design and build the instrument.   First light was achieved on
July 16, 1993.  Further details on the instrument can be found in
\citet{vab+94}.  HIRES is mounted on one of the Nasmyth ports of
the Keck~I telescope. Consequently,  as the telescope moves in the
sky (tracking the source), the sky image rotates with respect to
the detector. The image motion then limits the integration time.
The ``de-rotator"  project was led by David R. Tytler of University
of California at San Diego (UCSD; during the period 1997--1999).

%The cost of the project was estimated to be
%\$4.1M (Vogt, pers. comm.) and did not include an image rotator
%I do not know the cost of the image rotator.

HIRES was originally built for high resolution spectroscopy of stars
and quasar absorption line studies. The optical design is versatile
to accommodate operation in the entire band 0.3--1.2\,$\mu$m.  Over
time it has been extensively used for extra-solar planet searches
via precision radial velocity (RV) studies. To this end an insertable
Iodine cell and an exposure meter were added.

In 2004, Vogt led a project to replace the original engineering grade
2K$\times$2K pixel Tektronix CCD with a mosaic of three science grade
CCDs (2K$\times$4K pixel MIT Lincoln Lab).  The smaller pixel size
(15\,$\mu$m) of the new detectors was better suited to the HIRES
camera.  Furthermore, the three CCDs are each optimized for the
wavebands of the dispersed spectrum (more precisely, two are blue
sensitive and one is red sensitive). The upgrade contributed to
both an increase in the spectral coverage by a factor of three and
also improved the precision in RV from 3\,m\,s$^{-1}$ to 1\,m\,s$^{-1}$
\citep{bwm+06}.  To my knowledge there is no official reference
which summarizes the technical details of the upgrade.

%The cost of the upgrade project was \$1.5M
%(S. Vogt, pers. comm.).
%M and concluded that the image rotator costs (\$0.27M) are
%included in the above "upgrade" costs. 

HIRES is noteworthy for two reasons. First, early on,  a pipeline
to reduce the data was available (MAKEE) -- a novelty (at least for
the California community) in those days.  The pipeline allowed for
rapid exploitation of HIRES data. This became particularly important
following the upgrade of HIRES.  Second, starting 2004 the data
from HIRES were archived at the newly formed Keck Observatory Archive
(KOA).  The success of the HIRES archive project led NASA to mandate
that KOA begin a phased approach to ingesting data from all other
Keck instruments (see \S\ref{sec:Archives}).

\subsection{Echellette Spectrograph \&\ Imager (ESI)}
\label{sec:ESI}

ESI is a medium-resolution spectrograph with imaging capability
\citep{smb+00}. The instrument has an echellete grating and two
prisms for cross-dispersion.  In the low dispersion mode, 50 to
300\,km\,s$^{-1}$, the dispersion is provided by  prisms.  This
mode has high throughput but owing to the large number of sky lines
(in the red region of the spectrum) this mode is only popular with
astronomers interested in the study of blue objects. In the echellete
mode, the two prisms cross-disperse the beam diffracted by the
echellete grating. The spectral resolution is moderate, about
50\,km\,s$^{-1}$ over the entire range 0.39--1\,$\mu$m.  The moderate
spectral resolution is well suited to kinematics, abundance studies
of faint stars (especially giant stars in the Local Group) and faint
galaxies and absorption line studies of quasars.

The project was led by J. Miller of UCSC.  The instrument was
officially commissioned towards the end of 1999 \citep{sbe+02}.  In
early 2010 an Integral Field Unit (IFU) capability was commissioned.

%The estimated cost was a little under \$4M (Michael Bolte, pers. comm.).
%Note: On the Keck books the cost is \$3.1M. I can only assume that
%UCO must have spent additional funds to account for the discrepancy
%between Bolte's estimate and WMKO books.

\subsection{Near-Infrared Echelle Spectrograph (NIRSPEC)}
\label{sec:NIRSPEC}

NIRSPEC is a cross-dispersed echelle spectrograph that operates in
the 0.95--5\,$\mu$m band.  The instrument has two spectral modes:
high spectral resolution mode with a resolution of about 25,000 and
a low spectral resolution mode with a resolution of 2,300. An SBRC
SBRC ALADDIN-3  $1024\times 1024$ pixel array (27\,$\mu$m pitch) served
as the detector for the spectroscopic channel while a Rockwell
$256\times 256$ pixel PICNIC  array (see Appendix~\ref{sec:NIR})
served as the detector to view the slit (``SCAM").  An Inmos T805
transputer was used for data acquisition and processing.

NIRSPEC can be mounted at either of the two Nasmyth ports of the Keck~II
telescope.  It can be used in a stand-alone mode (seeing-limited)
or behind the Keck~II AO system which is mounted on the ``right"
Nasmyth port (\S\ref{sec:AdaptiveOptics}). This latter mode is
referred as ``NIRSPAO''.  The NIRSPEC project was led by Ian S.
McLean of the University of California at Los Angeles (UCLA).  The
primary reference paper for the instrumentation is \citet{mbb+98}.

The NIRSPEC project ran from October 1994 through September 1999.
First light was achieved on April 23, 1999.  A refurbishment of
some gears and motors was also performed in 2000. In the same year
the NIRSPAO mode  was implemented. This necessitated fore optics
for zooming the input image and a corresponding smaller pupil stop
in the filter wheel. While the main strength of NIRSPEC is spectroscopy
some astronomers have used SCAM for purely imaging purposes.

As we go to press there are major plans to upgrade NIRSPEC.  The
ALADDIN-3 detector will be replaced by an H2RG (with 18\,$\mu$m
pixels). The expected increase in sensitivity is a factor of {\it
six} (photon limited case)! For SCAM the PICNIC detector will be
replaced by an H1RG (but with a long wavelength cutoff of 5\,$\mu$m).
The transputers (which were already recognized to be obsolescent
at the time of commissioning) will be replaced with current digital
gateware and computer hardware.  There are also plans to enable a
precision radial velocity mode, replete with an NIR laser comb.

%The final run out cost for NIRSPEC was
%about \$4.2M, on an original estimate of \$3.9M.
%Including WMKO costs it is \$4.4M  

\subsection{Deep Imaging Multi-Object Spectrograph (DEIMOS)}
\label{sec:DEIMOS}

DEIMOS is a multi-object optical spectrograph optimized for studying
large scale structure of the Universe (via spectroscopy of galaxies).
It is mounted at the ``left" Nasmyth focus of the Keck~II telescope.
The spectrograph employs an array of eight red-sensitive CCDs.
Sufficient spectral resolution in the red band allows for minimization
of bright terrestrial OH lines. The effective slit length on the
sky is 17 arc minutes (a second barrel, if built, will add an equal
length slit in an adjacent field). The key feature of DEIMOS was
the wide-angle camera, which offered both a long slit length and a
wide spectral coverage.  The project was led by Sandra M. Faber of
UCSC and the official reference is \citet{fpk+03}.  The project ran
from 1993 to 2002.  First light was achieved in Spring of 2002.
DEIMOS was unique (140 galaxy spectra at a time) at the time  it
was built and was only matched by IMACS which was commissioned on
Magellan in 2004 \citep{dbh+11}.
DEEP\footnote{\url{http://deep.ucolick.org//}} (Deep Extragalactic
Evolutionary Probe) was a major survey undertaken at WMKO (PIs: M.
Davis of UCB and S. Faber of UCSC)  and the primary motivation for
DEIMOS.  Other notable studies with DEIMOS include ``galactic
archaeology" studies (multiplexed spectroscopy of stars in the
Galactic disk, in the near and distant halo, in satellite dwarf
galaxies and in M31).

%The run out cost of DEIMOS was about \$11M.
%On WMKO's books the cost is \$9.4M. However, there was an MRI that
%went to UCSC and we believe that the difference between the one
%I list (obtained from Sandy, I think) and the WMKO cost is the MRI

\subsection{Near Infra-Red Camera 2 (NIRC2)}
\label{sec:NIRC2}

The Near Infra-Red Camera 2 (NIRC2) was designed to be the primary
imager for the Observatory's Adaptive Optics system (both Laser
Guide Star and Natural Guide Star; \S\ref{sec:AdaptiveOptics}).
The instrument is located behind the AO bench at the right Nasmyth
focus of the Keck~II telescope. Three pixel scales allow for
diffraction limited imagery in  $z$ through M bands. The detector
is a 1024$\times$1024 pixel  ALADDIN-3 array. The filter wheel
accommodates a large number of filters over the spectral range
0.93--5.3\,$\mu$m.  Two prisms allow for low and medium-resolution
slit spectroscopy.  A choice of pupil masks (including non-redundant
pupil masks) and coronagraphic stops  (including an L-band vortex
coronagraph, installed in 2015) allow for low background and high
contrast imaging and spectroscopy.  The principal investigators
were K.  Matthews and B. T. Soifer.

With the view of undertaking decade-long astrometry, careful attention
was paid to keep NIRC very stable.  Construction for NIRC2 began
in 1994 and concluded in 2000. First light was achieved in the
summer of 2001.   Since there is no paper detailing the design and
performance of the instrument the reader is directed to the instrument
homepage\footnote{\url{http://www2.keck.hawaii.edu/inst/nirc2/}}
for further details.

%The run-out cost marked to first light was \$4.8M (per KYM). 
%Extended NIRC2 phase (running to 2001) was \$1.2M. M will investigate
%what was this all about

\subsection{OH-Suppressing Infrared Imaging Spectrograph (OSIRIS)}
\label{sec:OSIRIS}

OSIRIS is an IFU spectrograph operating in the NIR band. It was
designed to take advantage of diffraction limited images made
possible by the Observatory's Adaptive Optics system
(\S\ref{sec:AdaptiveOptics}).  The principal investigator  (PI) of
the project was James Larkin (UCLA) and the co-PI was Alfred Krabbe
(UCB). A lenslet array feeds a rectangular patch (1000 spaxels) of
the sky into a moderate spectral resolution ($R\sim 3800$) spectrograph
which can operate from the $z$ band through K band.  The 1000-spaxel
format is suitable for imaging compact objects (0.3 arc seconds to
3 arc seconds in the short axis).  With the advent of a second LGS
system on Keck~I (see \S\ref{sec:AdaptiveOptics}) OSIRIS was moved
to Keck~I in late 2012.

The design study for OSIRIS was undertaken in 1999.  First light
was achieved during 2005.  The primary reference for OSIRIS is
supposed to be Larkin {\it et al.} (2006a)\nocite{lbk+06a}.  However,
I have also included the reference Larkin {\it et al.}
(2006b)\nocite{lbk+06b} since it  appears to have garnered more
citations than the officially favored instrument reference.

%The total cost including early design and through construction was about 
%\$4.8M (this is the UCLA cost). 
%But when WMKO costs are included (per Margarita on 24-Apr-2014) the 
%total is \$5.6M 

Shortly after OSIRIS was commissioned it became clear that the
throughput of the instrument was lower than expected. It was traced
to a grating which was not manufactured to specifications. Finally
in 2013, a new grating was installed. As a result OSIRIS achieved
the sensitivity that was expected from the initial design \citep{mwl+14}.
In early 2016 the spectrograph detector (a Hawaii-2) was replaced
with a Hawaii-2RG.  An ongoing project is to replace the current
imaging detector (H1) to an H2RG (the FOV remains unchanged at 20
arc seconds but the finer pitch will lead to 10\,mas pixels).

%* Note to myself: include Mieda paper for "Builder" section.

\subsection{Multi-Object Spectrograph for Infra-Red Exploration (MOSFIRE)}
\label{sec:MOSFIRE}

MOSFIRE, a multi-object near-IR (0.97--2.1\,$\mu$m) spectrograph
and imager, is the latest addition to the stable of facility
instruments \citep{mse+12}.  The instrument is notable for its
``on-the-fly'' configurable slit mask. The   user can obtain moderate
resolution ($\lambda/\delta\lambda\approx 3600$) slit spectra of
46 objects spread over a field-of-view (FOV) of  6\,arc minutes by
6\,arc minutes.  Cryogenic cooling of the slit mask, a low-noise
2K$\times$2K pixel Hawaii-2RG detector and the large collecting area of
the Keck telescope makes MOSFIRE perhaps the most sensitive NIR
multi-object spectrograph at the present time.  The instrument can be 
mounted at the Cassegrain focus of the Keck~I telescope. The principal
investigators are I. S. McLean of UCLA and C. C. Steidel of Caltech.
The project\footnote{ The first attempt for a multi-slit IR
spectrograph was KIRMOS.  Following the preliminary design phase
(2002--2005) the estimated cost of the rather ambitious instrument
was deemed to be too high to warrant construction. KIRMOS was then
abandoned.} began in 2005 and the instrument completed by April
2011.  However, just prior to shipping the instrument from Caltech
to Hawaii, it was discovered that the WMKO rotator bearing assigned
for MOSFIRE was defective.  A new bearing had to be manufactured.
The long delay and unanticipated manufacturing increased the cost
of the project.  First light was achieved in early April 2012.

%The total cost of the project was \$14.6M (S. Adkins, pers. comm.).
%[This includes the $1.5M due to delay caused by defective rotator bearing.
%The cost reported here included \$2.6M spend in the KIRMOS phase.
%There are differing views of the value of KIRMOS work for MOSFIRE.
%Per someone (Sean Adkins?, not sure) the usable fraction of was 50\% 
%Steidel & McLean state (upon inquiry) "The *only* results we used from KIRMOS were the development
%of the concept for mounting lenses using bonded flexuresÑ otherwise we began with the KONSAG list of requirements
%for a near-IR multi-object spectrometer developed post-KIRMOS
%As a result of this interchange my valuation for MOSFIRE no longer
%carries the KIRMOS spent money -- SRK

\subsection{Other Instruments}
\label{sec:OtherInstruments}

The same forward Cassegrain module that housed NIRC had the ability
to also  accommodate both NIR/MIR IR instrument.  The facility
Long Wavelength Spectrometer (LWS; \citealt{cj04})  was on the
Keck~I telescope for a total of 363 nights. The primary detector
was a $128\times 128$ pixel Boeing Si:As moderate flux array (with
75\,$\mu$m pitch). The wavelength range for the detector was
3.5--25\,$\mu$m. LWS had both imaging and long slit spectroscopic
modes\footnote{\url{https://www2.keck.hawaii.edu/inst/lws/}}.  The
Long Wavelength Infrared Camera
(LWIRC)\footnote{\url{https://www2.keck.hawaii.edu/inst/lwirc/lwirc2.html}}
was an imaging camera in the $10\,\mu$m band. It too was based on
$128\times 128$ pixel Si:As doctor array and was a part of the
NIRC/LWS suite. However, LWIRC did not proceed to
commissioning.

The Keck Interferometer used both telescopes and was entirely funded
by NASA \citep{cwa+13}. Originally it was envisaged to include a
collection of smaller telescopes (``outriggers'' or ``side-Kecks'')
for year-round precision astrometry and occasional Keck~I--Keck~II
interferometry (visibility and nulling) to characterize the
distribution of zodiacal dust in a sample of nearby Sun-like stars.
The first phase of the project was the development of the standard
visibility mode (``V$^2$''; commissioned in 2001) followed by the
``Nuller'' mode.   Phase referencing methodology (first demonstrated
at the Palomar Testbed Interferometer; \citealt{cwh+99}) was successfully undertaken
with the Keck~I--Keck~II interferometer (the ``ASTRA'' project;
\citealt{wwa+14}).  

%Rough number of Keck-Int papers < 160 (simple ADS abstract search)

%Phasing Camera System, or PCS \citep{cnm+94}.

To complete the census of the allocated nights I  note  ``guest" or
Principal Investigator (PI)
instruments\footnote{\url{http://www2.keck.hawaii.edu/realpublic/observing/\\public\_instrument\_info/vis/index.html}}:
MAPS, STEPS, MIRLIN and OSCIR. These together obtained a
total of about four months.  Finally, about 5\% of the nights appear
to have been used for engineering, commissioning new instruments
and other purposes.

\begin{deluxetable}{rrrr}
\tabletypesize{\scriptsize}
\tablecaption{Adaptive Optics at the W. M. Keck Observatory}
\tablewidth{0pt}
\tablehead{
\colhead{System} & \colhead{Tel} & \colhead{Year} & \colhead{Cost}\\
                 &               &                & \colhead{(\$M)}
}
\startdata
NGS  & II & 1999 &  4.0
\cr
LGS  & II & 2004 &  7.5
\cr
WF-Upgrade  & II & 2007 &  2.2
\cr
Center-Launch  & II & 2014 &  2.6
\cr
TOPTICA-Laser  & II & 2015 &  4.0
\cr
NGS  & I & 2002 &  3.0
\cr
LMCT-Laser  & I & 2011 &  3.1
\cr
LGS-Infrastructure  & I & 2012 &  5.5
\cr
NIR-Tip-Tilt  & I & 2014 &  3.4
\enddata
 \tablecomments{
From left to right: The name of the AO system or sub-system followed
by the telescope number on which it is located, the year of
commissioning and the cost for the project.  }
 \label{tab:AOFacility}
\end{deluxetable}

\section{Adaptive Optics}
\label{sec:AdaptiveOptics}

The ability to exquisitely align the 36 segments limited only by
the roughness of the segment surfaces (40\,nm to 80\,nm) allows the
Keck telescopes to take full advantage of the superb seeing of Mauna
Kea \citep{ctd+98,cot+00}. Provided the seeing cooperates the Keck
telescope can produce images with 0.4 arc-second full width at half
maximum in the visible \citep{w+94a}.  This exquisite performance
when combined with the large diameter, $D$,  of the Keck telescope
makes AO a natural strength\footnote{ The gains for AO grow as $D^n$
where $n=2$ to $6$, depending on what quantity is being measured.}
of the Observatory.  As a result, planning\footnote{see footnote
\ref{ft:COSTKECK}.} for AO began immediately after commissioning
of the first Keck telescope (Wizinowich et al. 1994b)\nocite{w+94b}.

In early 1999 an NGS AO system was commissioned on the Keck~II
telescope (being located at the left Nasmyth focus; \citealt{was+00}).
Routine observations began in the Fall of 1999.  The system was based
on a 349-actuator Xinetics deformable mirror and a 64$\times$64 pixel 
fast-readout CCD.  Following the commissioning of the AO system
``KCAM" (built primarily for engineering purposes and so lacked the
usual accoutrements of a science camera) served as the science
camera.  Starting 2001 NIRSPEC (and soon thereafter NIRC) was used
as the science instrument behind the AO system.   Two years later
an identical NGS AO system for Keck~I, located also at the left
Nasmyth station, was commissioned (see \citealt{was+03}).

The Observatory's AO roadmap called for a LGS assisted AO.  The
laser guide star can be used to infer most of the wave front
distortion but not the phase gradients (which lead to tip-tilt
errors). A natural guide star is still needed for this purpose but
it can be much fainter (approaching $V$ of 19) as compared to a purely
NGS AO system ($V\lesssim 13$).

A 13-watt Sodium dye laser supplied by the Lawrence Livermore
Laboratory was installed at the Keck~II telescope and LGS observations
began in 2004 \citep{vbl+06,wlb+06}.  In 2007 a major improvement
was undertaken for both the Keck~I and Keck~II AO systems.  The
wave-front sensor and wave-front electronics were upgraded.  As a
result the quality of correction (Strehl ratio for bright stars)
increased from 0.58 to 0.71 and the limiting magnitude for NGS AO
also improved ($V\lesssim 14$); see \citet{jvs+08}.

When first commissioned, the Keck~II  laser was launched using a
telescope mounted to the side of the Keck~II telescope. As a result,
there was a perspective elongation of the Laser Guide Star as seen
by the AO wave-front sensor, due to the thickness of the sodium
layer. The elongation naturally reduces the quality of corrections.
This elongation  can be reduced by having the launch telescope behind
the secondary mirror and thus aligned to the axis of the telescope.
A center-launch system is now in routine use
since mid 2015.

A program to replace the aging dye laser with a modern Raman
fiber-amplified  laser (made by Toptica Photonics; \citealt{fkw+12})
was completed recently.  The Toptica laser has been in routine use
since April 2016. The return signal is 19 times higher than that
of the dye laser owing to a combination of higher input laser power
and (expected) better coupling efficiency to the sodium layer (P.\ Wizinowich, 
pers. comm.).

%* Request: get official reference from Peter Wizinowich

The Keck~I AO system began shared risk observations in the summer
of 2012 \citep{cwc+12}. The National Science Foundation (NSF) funded
Lockheed Martin Coherent Technology (LCMT) to build 
lasers for the Gemini Observatory and WMKO.  The LMCT laser is a 20\,W
solid-state CW laser \citep{slj+10}.

The next improvement was to implement tip-tilt corrections based
on measurements undertaken in the NIR \citep{wsb+14}. The primary
advantage of using NIR tip-tilting sensitivity is both increased
Strehl ratio and sky coverage.  To this end an NIR tip-tilting
system based on a Hawaii-2RG detector (listed as NIR tip-tilt in
Table~\ref{tab:AOFacility}) was designed. 
In detail, dichroics are
used to send either the  Ks-band or H-band light, over a 100\,arc second
square field, to the NIR detector. Tip-tilt measurements are
undertaken on the AO-corrected core of the NGS image of the natural
guide star.  When using Ks-band light the sky fraction over which
the 1-D rms tip-tilt error is less than 20\,mas increases from the
older value of  45\% to 75\%.  This sub-system
was commissioned in 2014 and became routinely usable in 2015. 
The reader is referred to
Table~\ref{tab:AOFacility} for a comprehensive summary as well as
the timetable of both the AO systems.

%PCS (704)
%LWS (363), MIRLIN (64)
%MAPS (19), STEPS (22), OSCIR (21)
%KCAM (53), SHARC (37)
%V2 (224), Nuller (244), ASTRA (40)

\begin{deluxetable}{rrrlrr}[b]
\tabletypesize{\scriptsize}
\tablecaption{Allocation of Nights (1994B--2015B)}
\tablewidth{0pt}
\tablehead{
\colhead{\#} & \colhead{Instr.} & \colhead{Tel} & 
\colhead{Period} & \colhead{Nights}  & \colhead{OSF}}
\startdata
2 & NIRC & I & 1994--2010 & 926 & 0.5 \\
3 & LRIS & I & 1994-- & 3209 & 0.7 \\
4 & HIRES & I & 1994-- & 2692 & 0.8 \\
5 & ESI & II & 2000-- & 654 & 0.8 \\
6 & NIRC2 & II & 2000-- & 1185 & 0.4 \\
7 & NIRSPEC & II & 2002-- & 1533 & 0.6 \\
8 & DEIMOS & II & 2001-- & 1262 & 0.7 \\
9 & OSIRIS & I & 2005-- & 549 & 0.6 \\
10 & MOSFIRE & I & 2013-- & 456 & 0.6
\enddata
\tablecomments{
The allocation of nights for period starting with semester 1994B and ending with 2015B
Number of nights on Keck-I: 8050.
Number of nights on Keck-II: 7596.
The fraction of nights used by above instruments is 80\%.
\# is an internal index. The years over which the instrument was (and 
continues to be) used is given by ``Period'. We make no distinction
between NIRSPEC or NIRSPAO, LRIS-R or LRIS-B or LRIS-ADC or LRIS,
NIRC or NIRCs and HIRES, HIRESr or HIRESb.The last column is the
``open shutter fraction" -- the fraction of time, say over a typical night, that the shutter is open.
The OSF values reported are from R.\ Goodrich who undertook the
analysis in 2013. The OSF for 
MOSFIRE was provided by M. Kassis (measured in
2016).  }
\label{tab:AllocationOfNights}
\end{deluxetable}

\section{Data, Methodology \&\ Metrics}
\label{sec:PrimaryDataMethodology}

\subsection{Primary Data}
\label{sec:PrimaryData}

The primary data for the analysis is the bibliography\footnote{
\url{http://www2.keck.hawaii.edu/library/keck\_papers.html}} of
refereed papers maintained by Peggi Kamisato, the official librarian
of the W. M. Keck Observatory.  For every paper, Kamisato lists the
following attributes: Authors (limited to first six authors), Title
of the paper, Journal name, Volume, First page, Year of publication,
Instrument(s) used and the \texttt{bibcode}.\footnote{\label{ft:ADS}A
unique identifier to each paper by the SAO-NASA Astrophysics Data
System (ADS). See
\url{http://adsabs.harvard.edu}} The assignment of the instruments
were made by Kamisato based on her scanning the literature and
reading of the papers.  For the analysis presented here, I have
considered all Keck papers from 1993 through the end of 2015.

At the time I began my analysis,  the data base was expected to be
complete going forward from 1996.  Kamisato and  I did a search of
the literature and added papers for 1994 and 1995.  Next, about 150
papers lacked instrument entry. For about half the papers Kamisato
did not have easy access (primarily commercial publications for
which the WMKO did not carry a subscription) and those for which
an instrument assignment was not clear (see below).  I read these
papers and made the instrument assignments.  For a fraction of the
cases the assignment was difficult to make because the authors do
not provide sufficient details other than thanking the W. M.  Keck
Observatory.  Through patient reading, in most cases, I could discern
the instrument used.

Curiously, the same problem -- papers thanking WMKO but not citing
the instrument used --  has  arisen for a number of papers
published in the last few years (2013--2015).  I wrote letters to
authors  that I knew and Kamisato received clarifications (in most
cases). There still remain a total of about 30 papers that are yet
to be classified.

\subsection{Usage of Nights}
\label{sec:Usage}

Starting from the commissioning\footnote{The first official science
run appears to have taken place on 1-October-1994.} of Keck~I
through semester 2015B\footnote{A  year, as is the tradition in
many observatories, is divided into two semesters.  The ``A''
semester starts 1 February and the ``B'' semester starts 1 August.},
using the ``Query"
tool\footnote{http://www2.keck.hawaii.edu/schedule/schQuery.php}
provided by WMKO, I found a total of 8050 nights were available on
Keck~I.  The Query tool shows that between commissioning\footnote{The
first official science run appears to have taken place on
1-October-1996.} of the Keck~II telescope through the end of 2015B
a total of 7056 nights were available on Keck~II.  This tool shows the
{\it nominal} instrument for each night. 
However, for the purpose
of this paper, I used a spreadsheet maintained by Gloria Martin of WMKO
which properly apportions the night between multiple allocations (e.g.\ 
half nights used for science with the other half for engineering etc).

Sometimes the scheduling logs list, for the same night,  NIRC and
LWS. Both these instruments were sited at the forward Cassegrain
focus of the Keck~I telescope. The designation ``NIRC-LWS" meant
that the primary instrument for the night was LWS where the designation
``NIRC/LWS'' meant that the two instruments shared the night (R.
Campbell, WMKO, pers.  comm.). These nights were attributed equally
to NIRC and LWS (so that no night is double counted). 

The allocations
of nights by instrument is summarized in
Table~\ref{tab:AllocationOfNights}. From this table we can see that 
the
workhorse instruments were allocated nearly 80\% of the available nights.
The engineering
(telescope, commissioning, repairs, AO) 
represented 10\% of the total available nights. The remaining 10\% was
used as follows. The interferometer project which lasted from 01A through
12A  used 275 nights (sum of Keck~I and Keck~II nights) for
observing in either V2 or Nuller mode (and paltry nights for ``Ohana")
and 346 nights for associated engineering. LWS used 246 nights and
the remaining 159 nights went to guest and PI instruments.

\begin{figure*}[htbp]
   \centering
   \includegraphics[width=5.3in]{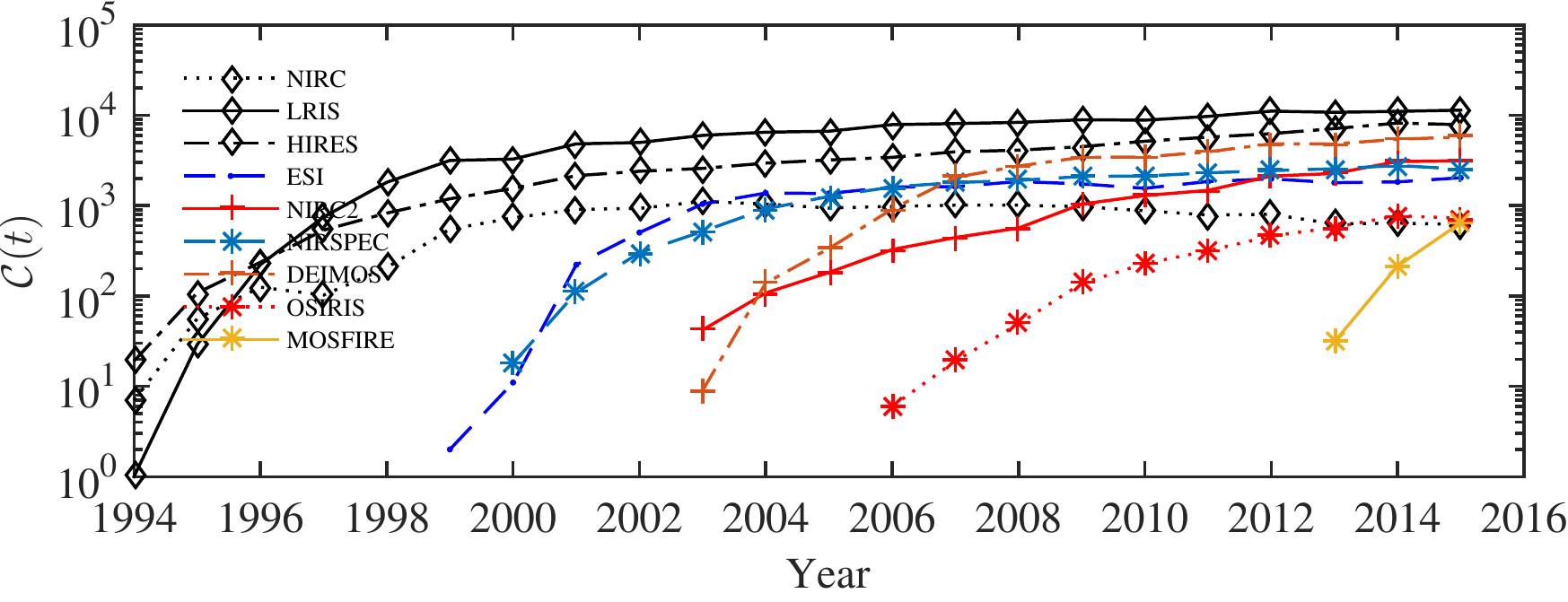}
   \caption{\small 
	The citation flux curve for every facility
   instrument (marked) of the W.~M.~Keck Observatory.  
    }
\label{fig:AllInstrumentsCitations} 
\end{figure*}

\subsection{Methodology}
\label{sec:Methodology}

I wrote a series of \texttt{MATLAB}  programs to analyze Kamisato's
database.  Each Keck paper is assigned a \texttt{structure}.  The
attributes of each paper in Kamisato's database are assigned to the
structure.  For each \texttt{bibcode} I wrote a program that queried
the ADS (see footnote~\ref{ft:ADS}) database and obtained information
of papers citing a given Keck paper.  The data thus obtained were
filtered to obtain $c_k(t_k,t)$, the number of citations in year
$t$ to Keck paper with index $k$ (whose year of publication is
$t_k$).  This list was added as an element to the structure.

The rest of the analyses worked off the structures.  All the analyses
programs use these structures as the inputs, filter them on instruments
and directly produce all the Tables (in \LaTeX\ format), the Appendix
(also in \LaTeX\ format) and all the Figures displayed in this
paper.

\begin{deluxetable}{rrrrr}
\tabletypesize{\scriptsize}
\tablecaption{Productivity \& Impact of Instruments}
\tablewidth{0pt}
\tablehead{
\colhead{Inst.} & \colhead{$N_P$}&
\colhead{$N_C$}&\colhead{$n_P^{-1}$}&\colhead{$n_C$}}
\startdata
 NIRC & 247 & 15564 &  3.7 & 16.8 \\ 
 LRIS & 1497 & 144585 &  2.1 & 45.1 \\ 
 HIRES & 1202 & 80455 &  2.2 & 29.9 \\ 
 ESI & 293 & 24220 &  2.2 & 37.0 \\ 
 NIRC2 & 461 & 19573 &  2.6 & 16.5 \\ 
 NIRSPEC & 484 & 27040 &  3.2 & 17.6 \\ 
 DEIMOS & 654 & 42936 &  1.9 & 34.0 \\ 
 OSIRIS & 104 & 3842 &  5.3 &  7.0 \\ 
 MOSFIRE & 54 & 1521 &  8.4 &  3.3   
\enddata
\tablecomments{Columns (from left to right):
$N_P$ is the total number of papers
$N_C$ is the sum of citations. However, rather than display fractional
numbers I display the inverse, 
$n_P^{-1}$  (or  the number of papers per night.
$n_C$ is the number of citations per night.}
\label{tab:Productivity}
\end{deluxetable}

\begin{deluxetable}{rrrr}
\tabletypesize{\scriptsize}
\tablecaption{Other Measures of Impact}
\tablewidth{0pt}
\tablehead{
\colhead{Inst.} & \colhead{$H$}&
\colhead{$M$}&\colhead{$\mu$}}
\startdata
 NIRC & 65 & 36 & 63 \\
 LRIS & 162 & 42 & 97 \\
 HIRES & 141 & 40 & 67 \\
 ESI & 81 & 41 & 83 \\
 NIRC2 & 67 & 23 & 42 \\
 NIRSPEC & 79 & 29 & 56 \\
 DEIMOS & 103 & 34 & 66 \\
 OSIRIS & 36 & 25 & 37 \\
 MOSFIRE & 24 & 22 & 28  
\enddata
\tablecomments{$H$ is the h-index, $M$ is the median 
 and $\mu=N_C/N_P$ is the mean number of citations per paper.}
\label{tab:ImpactOfInstruments1}
\end{deluxetable}

\subsection{Aggregate Metrics}

I define the {\it productivity} of an instrument  as the number of
nights taken to produce a paper (Table~\ref{tab:Productivity}). The
productivity is computed by taking the ratio of the total number
of papers ascribed to that instrument  to the  number of nights
{\it allocated}\footnote{Thus nights lost due to inclement weather
or instrument failure will adversely affect the productivity.}
to
the same instrument. The latter number can be found in
Table~\ref{tab:AllocationOfNights}.  The {\it impact} of the
instrument is measured by a number of attributes. One is the number
of citations per night of observing (Table~\ref{tab:Productivity}).
Other measures of impact are the H-index \citep{Hirsch05}, the mean
and median of the number of citations (Table~\ref{tab:ImpactOfInstruments1})
and the collection of the most cited papers (Appendix~\ref{sec:TopPapers}).

\subsection{Flux Curves}
\label{sec:FluxCurves}

Here I discuss functions of metrics  which capture the temporal
evolution of the productivity and impact of the Observatory.
 \begin{enumerate}
  \item The annual flux of {\it refereed} publications, $\mathcal{P}(t)$.
This curve is obtained by binning the list by the year of publication.
This is a widely used metric.
 \item
The sum of citations from publication to the present year ($t$) of the
$k$th paper is
  \begin{eqnarray}
    C_k(t)&=&\sum_{t\ge t_k}c_k(t_k,t).
   \label{eq:Ck}
  \end{eqnarray}
Colloquially,   $C_k(t)$ is referred
to as the ``number of citations'' and colloquially further simplified to ``citations''
for that paper. However, $C_k(t)$ changes with time (for young papers
$C_k$ usually increasing with $t$; for older papers it remains
constant with $t$; when a subject is revived, citations to an old
and dormant paper flourish again). As a result $C_k(t)$ does not
lend itself to a clean interpretation.  However, it does have some
limited use (see \S\ref{sec:ObservatoryLightCurve}).
 \item
The citation flux curve, $\mathcal{C}(t)$ measures the number of
new citations generated by a given list of Keck papers in a given
year ($t$).  The easiest way to understand this curve is to view
$c_k(t_k,t)$ as a response function of the $k$th paper, launched
at $t=t_k$.   In order to compute the citation flux curve in year
$t$ one needs to sum the response function of all the  relevant
Keck papers prior  to that year.  Mathematically,  the citation
flux curve is given by
  \begin{equation}
	\mathcal{C}(t) = \sum_{{k}, { t\le t_k}} c_k(t_k,t).  \label{eq:C}
  \end{equation}
 \end{enumerate}  
In \S\ref{sec:Analysis} I present the
paper and citation flux curves for the principal instruments of the
Keck Observatory.

I make some observations about the 
the flux time series curves\footnote{All the way up to the
pre-submission version I  used the term ``light" curves on the basis
that astronomers would both appreciate and understand the curves.
However, several colleagues found this term to be confusing and
so I have switched the pedantically correct term, flux time series
curve or flux curve for short.}: $\mathcal{P}(t)$
and $\mathcal{C}(t)$.  On general grounds we expect $\mathcal{P}(t)$
to rise slowly and then reach a plateau as users become familiar
with the instrument and data reduction tools mature. Once the ``low
hanging fruit'' projects are finished $\mathcal{P}(t)$ will likely decline
(unless a major discovery opens up new avenues of investigation).
Additionally, the decline will be precipitated by the arrival of
similar but more powerful instruments, usually, at other observatories.
In such a case, most  users of the Observatory will find themselves to be not competitive
and switch their attention to other projects.

In order to interpret the citation flux curve it is worth noting
that there is a lag between the publication of a paper and the
accrual of citations. Therefore, one generally expects  a typical
$\mathcal{C}(t)$ flux curve to rise  quite slowly, relative to
$\mathcal{P}(t)$,  enjoy a plateau and then gradually decline.
Next, an important paper is also durable which means that it keeps
getting cited for many years.  As a result, we can make three general
observations.
 \begin{enumerate}
  \item[I.] The higher the value of the peak flux (the value of the
  plateau flux) the higher  the impact of the instrument.  
\item[II.]
  The larger the duration of the plateau, as measured by the width
  of $\mathcal{C}(t)$, the higher  the productivity of that instrument.
 \item[III.]A decreasing $\mathcal{C}(t)$ almost always signifies
 that the instrument should be retired.  \end{enumerate}

%---------------------------------------------------------------------------------------
\section{Analysis: Flux Curves \&\ Performance Metrics}
\label{sec:Analysis}
%---------------------------------------------------------------------------------------

The  productivity and impact of the instruments of the Keck Observatory
(as defined in \S\ref{sec:Methodology}) are summarized in
Table~\ref{tab:Productivity} and Table~\ref{tab:ImpactOfInstruments1}.
The flux curves of all the instruments are summarized
in Figure~\ref{fig:AllInstrumentsCitations}. The flux curve of
each instrument can be found in \S\ref{sec:NIRCLC}--\S\ref{sec:AOLC}.

\begin{figure}[htbp]
%  \centering \includegraphics[width=3in]{NIRC.eps}
   \centering \includegraphics[width=3in]{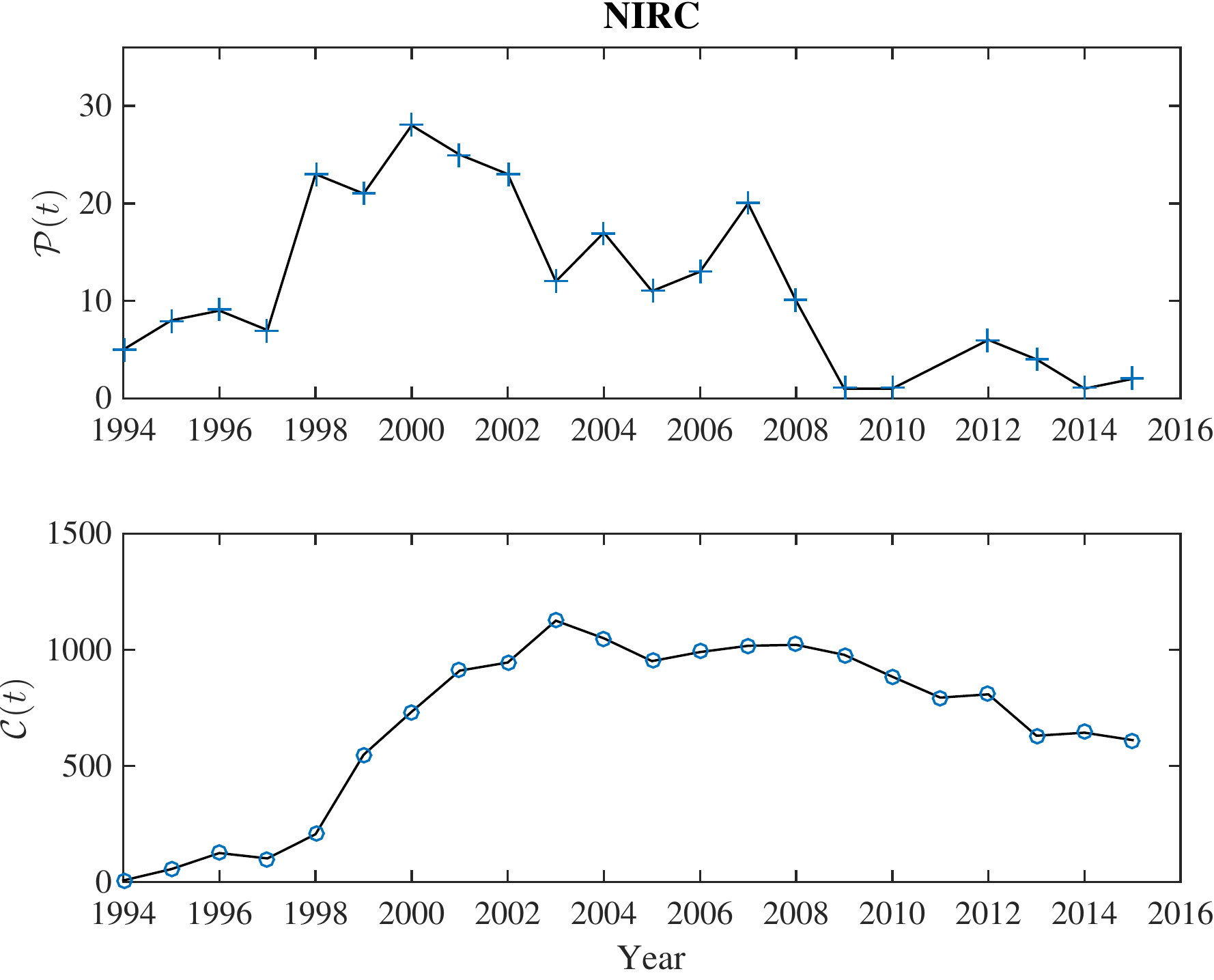}
   \caption{\small
    The annual paper flux, $\mathcal{P}(t)$  (top) and  the citation flux curve,
    $\mathcal{C}(t)$ (bottom) for NIRC.   See
    \S\ref{sec:PrimaryDataMethodology} for definition of these two
    quantities.
 }
  \label{fig:NIRC}
\end{figure}

\begin{figure}[htbp] 
   \centering
   \includegraphics[width=3in]{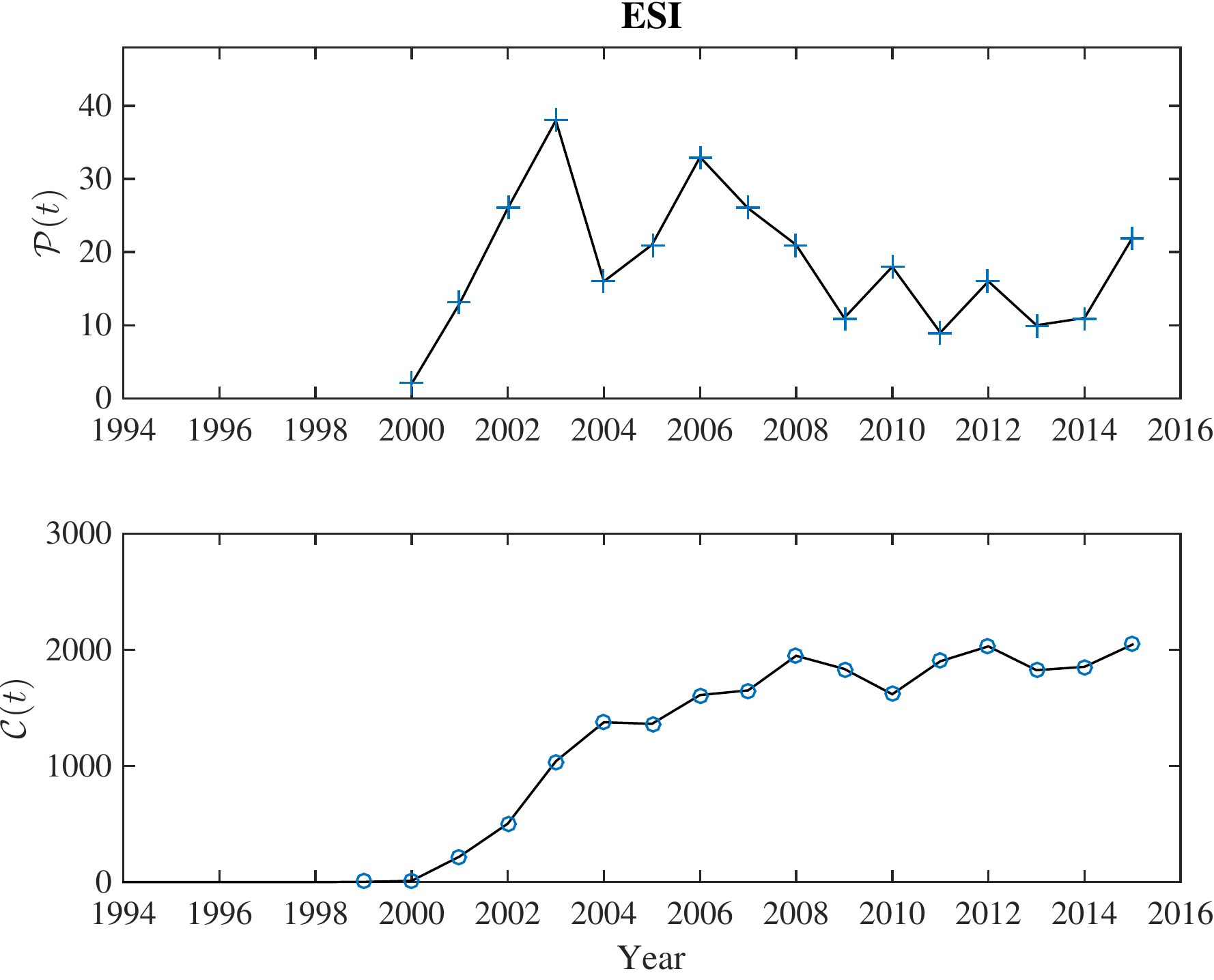}
   \caption{\small
   The paper flux curve (top) and the citation flux curve (bottom) for ESI.
   }
   \label{fig:ESI}
\end{figure}

\subsection{NIRC}
\label{sec:NIRCLC}

The flux curves of NIRC (Figure~\ref{fig:NIRC}) are worthy of
further study because NIRC did not undergo an upgrade whereas there
has been a steady increase in both the  format and performance  of
NIR detectors (and chronicled in Appendix~\ref{sec:NIR}). As a consequence,
NIRC has been subject to strong external forces.  Thus in some ways
NIRC provides an ideal ``test'' instrument for the purpose of this
paper.

The NIRC paper production reached a peak six years after commissioning
and this was followed by a linear decline. In contrast, the citation
flux curve reached a plateau nearly ten years after commissioning
and is now slowly declining. The lag between paper production and
garnering of citations is not unexpected. For future discussion I
note that the width of plateau of $\mathcal{C}(t)$ is in excess of
a decade.

\begin{figure}[htbp] 
   \centering
   \includegraphics[width=3in]{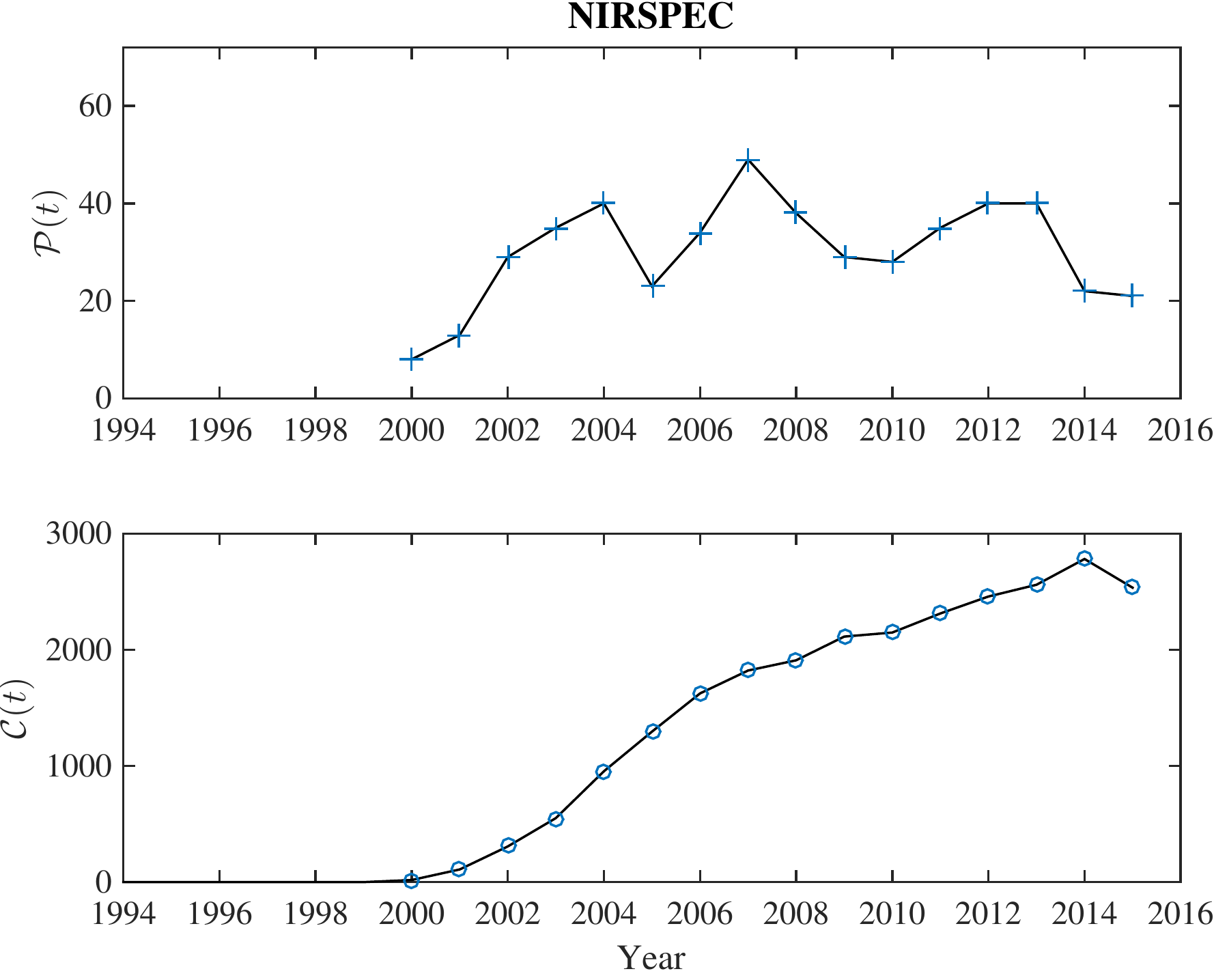}
   \caption{\small
   The paper flux curve (top) and the citation flux curve (bottom) for NIRSPEC.
   }
   \label{fig:NIRSPEC}
\end{figure}

\begin{figure}[htbp] 
   \centering
   \includegraphics[width=3in]{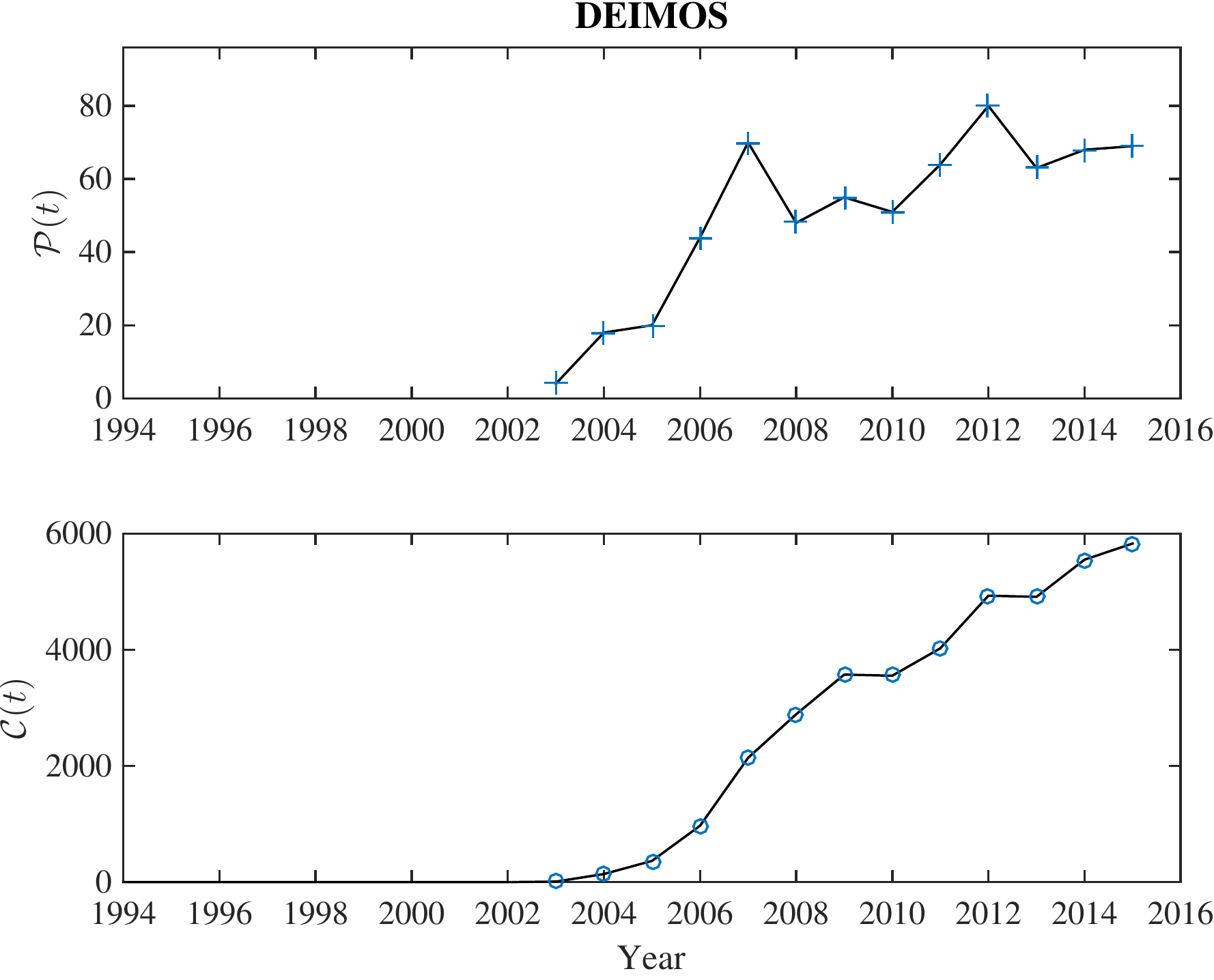}
   \caption{\small
   The paper flux curve (top) and the citation flux curve (bottom) for DEIMOS.
   }
   \label{fig:DEIMOS}
\end{figure}

\subsection{ESI, NIRSPEC, DEIMOS}

These three instruments are unified by the fact that they have not
undergone (significant) upgrades.  The paper curve of ESI mimics
that of NIRC (except shifted in time). The impact of ESI remains
quite high though (see Table~\ref{tab:ImpactOfInstruments1}).

The peak in paper flux of NIRSPEC appears to have been reached in
2007 (with a value of 48 papers per year). The paper flux averaged
over the last five years is 31 papers per year. So we conclude that
NIRSPEC peaked in paper production between seven to ten years post
commissioning.  However, unlike, NIRC, the citation flux  did not
plateau at the 10-year mark.  The flux rose, albeit slowly.

Within Poisson errors, DEIMOS has a steady rate of paper production
starting about five years after commissioning. The citation flux
has grown year after year. Arguably the citation flux is now peaking.

\begin{figure}[htbp] 
   \centering
   \includegraphics[width=3in]{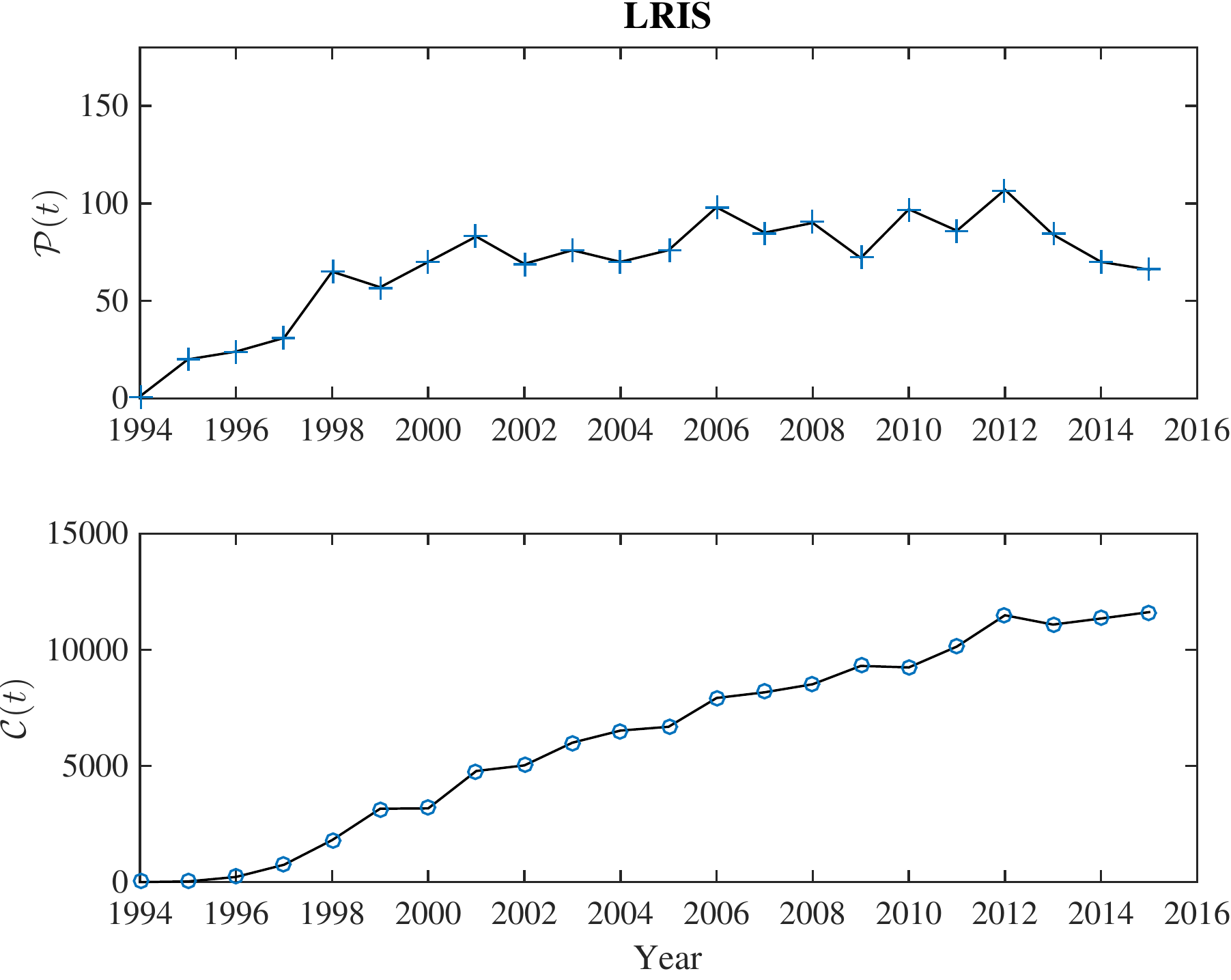}
   \caption{\small
   The paper flux curve (top) and the citation flux curve (bottom) for LRIS.
   }
   \label{fig:LRIS}
\end{figure}

\begin{figure}[htbp] 
   \centering
   \includegraphics[width=3in]{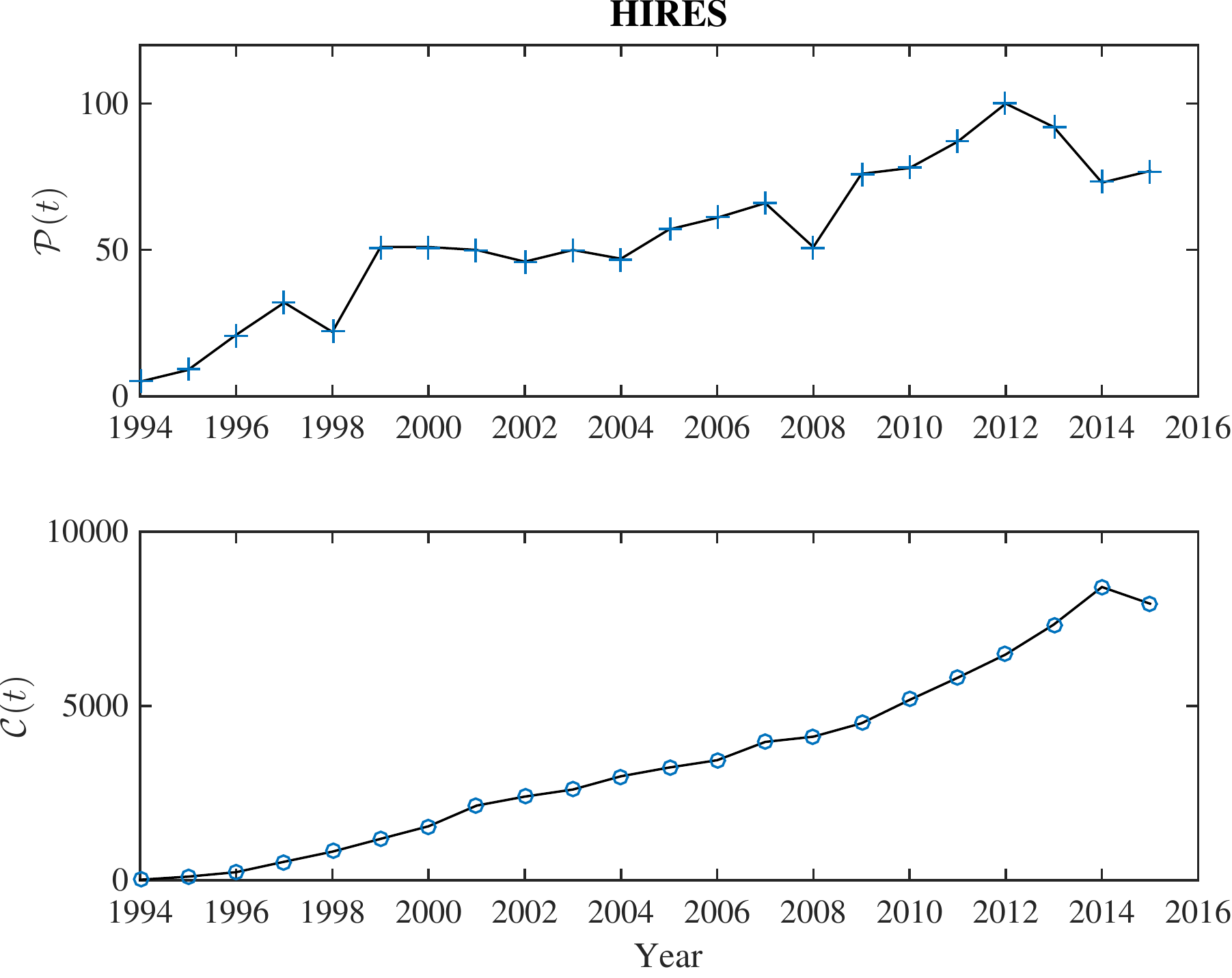}
   \caption{\small
   The paper flux curve (top) and the citation flux curve (bottom) for HIRES.
   }
   \label{fig:HIRES}
\end{figure}

\subsection{LRIS and HIRES}

LRIS and HIRES are remarkable instruments. These two first light
instruments show no fatigue in productivity. Perhaps this continued
fecundity is due to upgrades. After all, LRIS received upgrades in
2000, 2007 and 2010 (see \S\ref{sec:LRIS}) and HIRES was upgraded
in 2004 (see \S\ref{sec:HIRES}).

\begin{figure}[h] 
   \centering
   \includegraphics[width=3in]{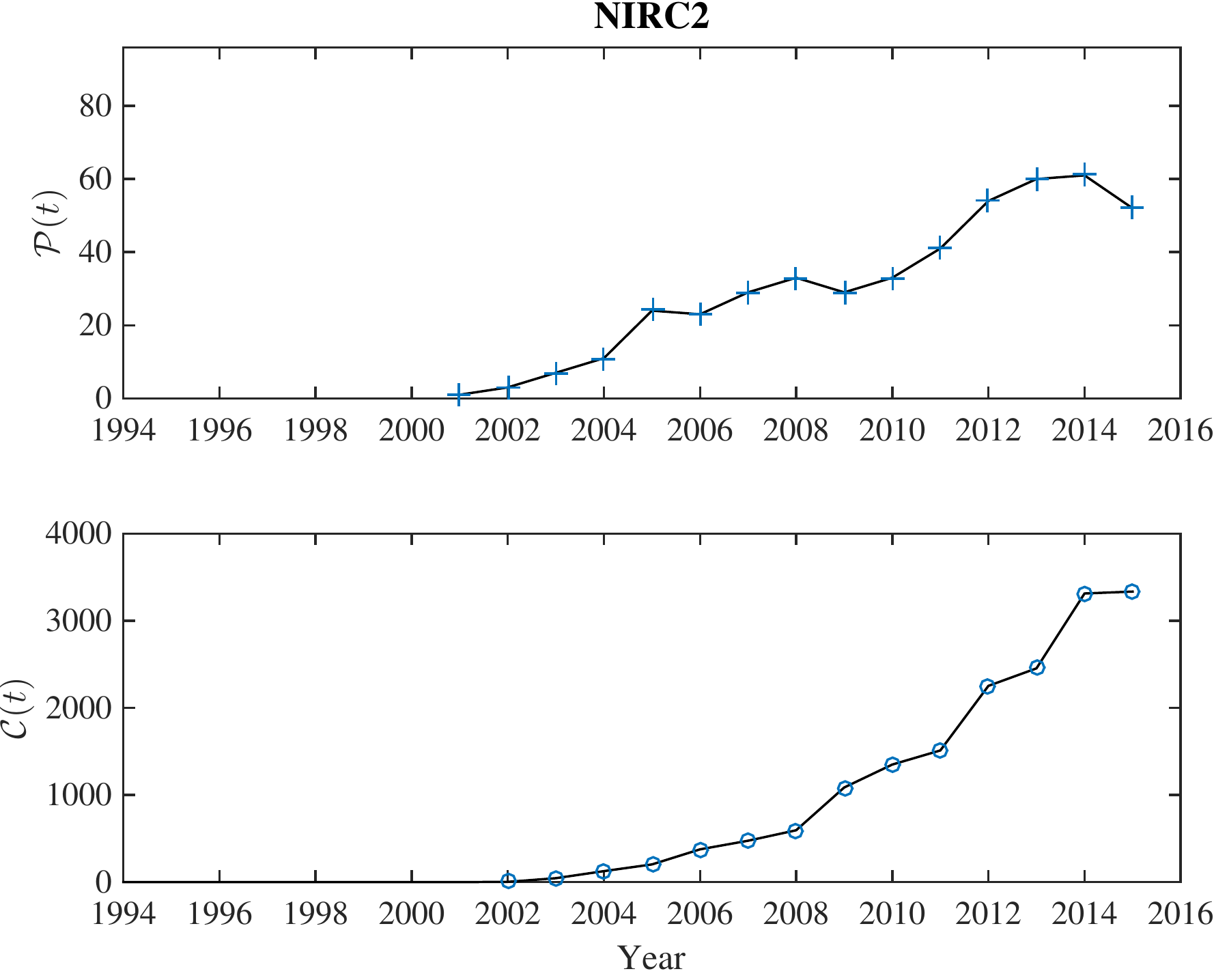}
   \caption{\small
   The paper flux curve (top) and the citation flux curve (bottom) for NIRC2.
   }
   \label{fig:NIRC2}
\end{figure}

\subsection{NIRC2, OSIRIS \& MOSFIRE}

The paper production of NIRC2, even ten years after commissioning,
is still rising as is the citation flux curve (Figure~\ref{fig:NIRC2}).
Since NIRC2 is only used behind the AO system  the fate of NIRC2
is firmly tied to improvements in the AO system. From
Table~\ref{tab:AOFacility} we note there has been significant
investment in improving AO (on both Keck~I and Keck~II) for the
past decade.  The continued rise of $\mathcal{P}(t)$ and $\mathcal{C}(t)$
is thus reasonable.  The modest flux of papers for OSIRIS has been
noted by several colleagues (see \S\ref{sec:Archives} for further
discussion).  MOSFIRE is too young an instrument to warrant a
detailed discussion.

\begin{figure}[h] 
   \centering
   \includegraphics[width=3in]{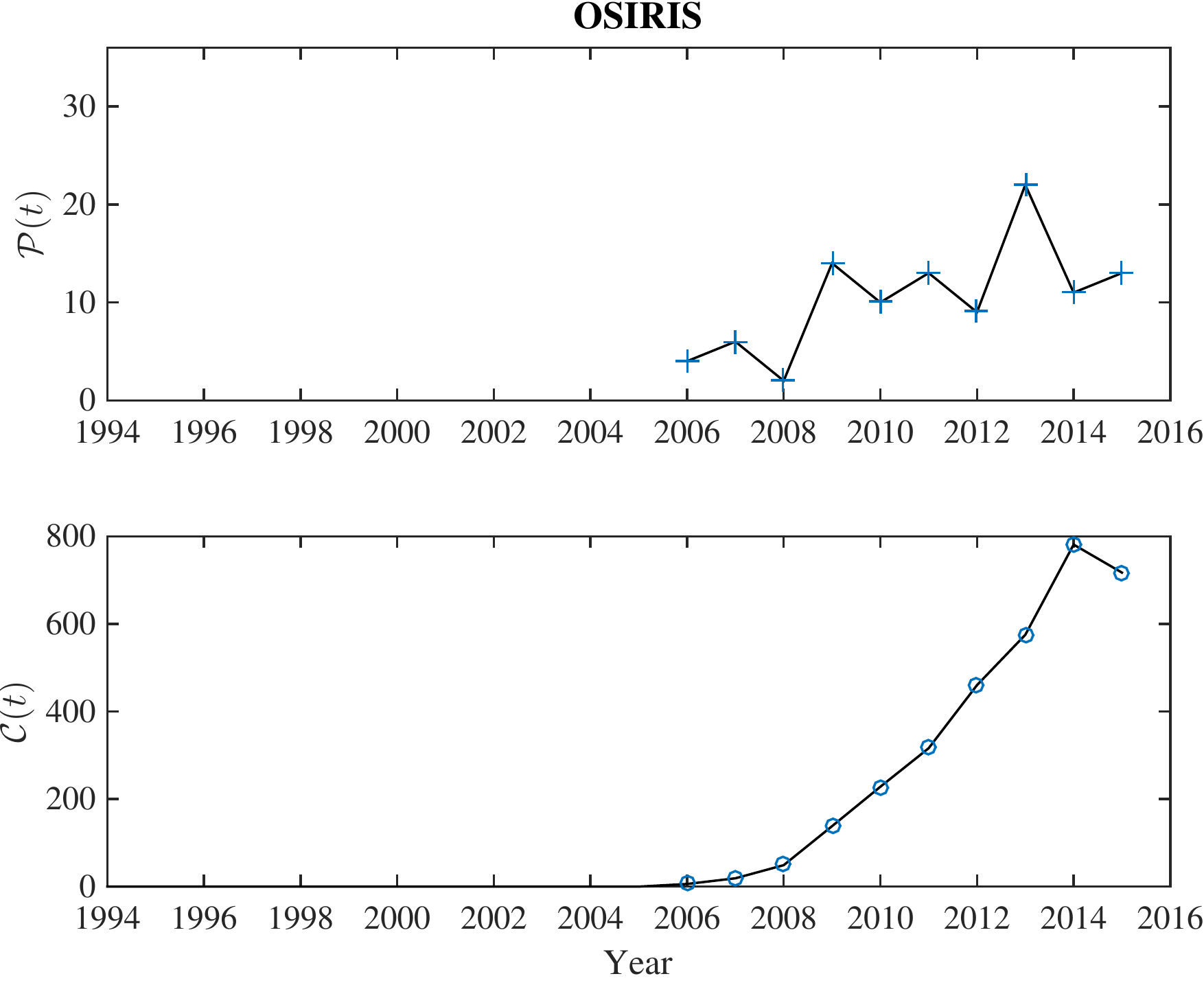}
   \caption{\small
   The paper flux curve (top) and the citation flux curve (bottom) for OSIRIS.
   }
   \label{fig:OSIRIS}
\end{figure}

\begin{figure}[h]
\centering
 \includegraphics[width=3in]{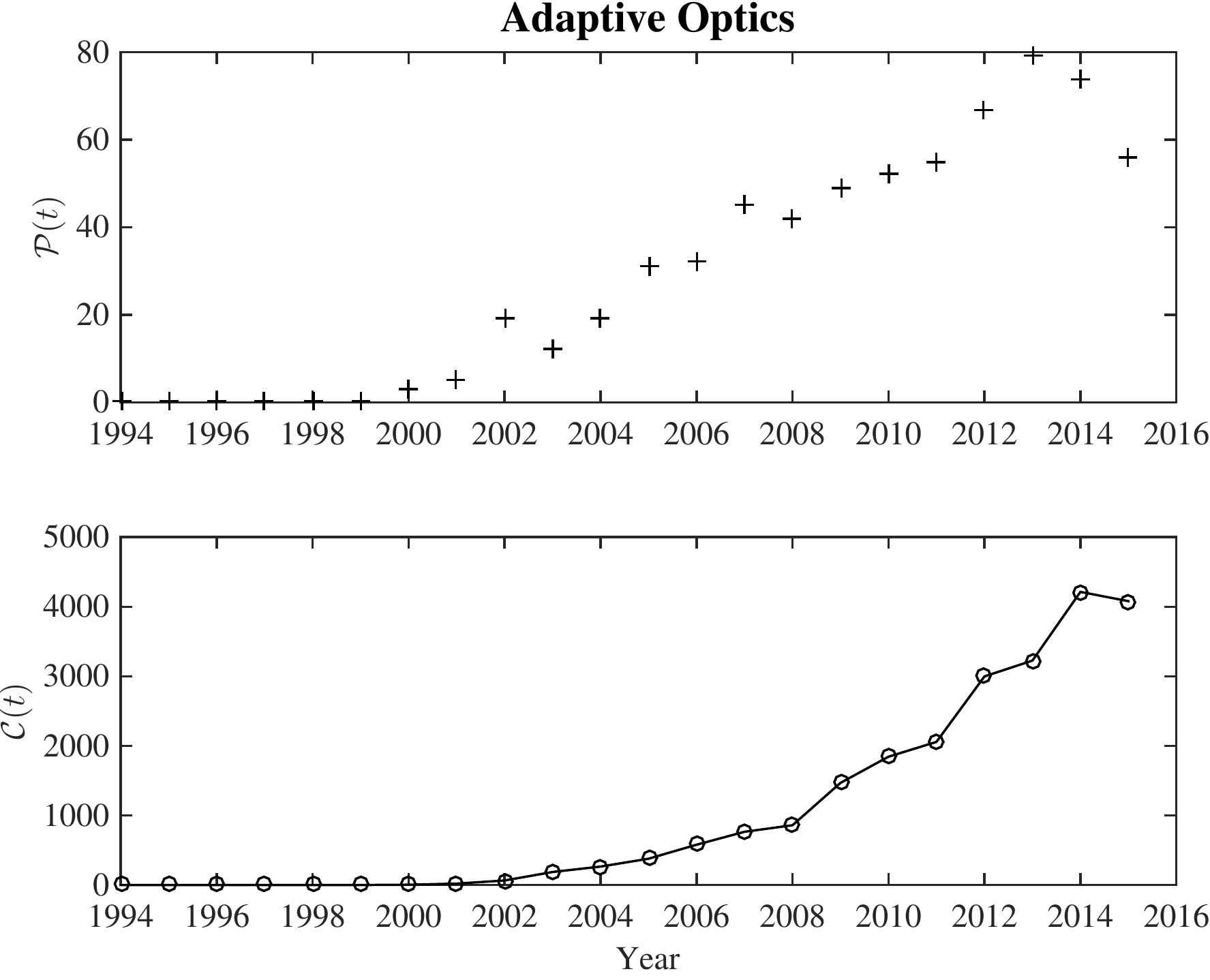}
\caption[]{\small
The flux curves of papers arising from AO methodology.
}
\label{fig:AOCurve}
\end{figure}

\subsection{Adaptive Optics}
\label{sec:AOLC}

The number of AO papers (which means both NGS and LGS) is 640 and
the total number of citations currently stands at 25,987. As can
be seen from Table~\ref{tab:Productivity} most of these contributions come
from NIRC2, OSIRIS and NGSPAO. The difference of about a hundred
papers are due to  Keck interferometry and KCAM.   The citation flux curve is
shown in Figure~\ref{fig:AOCurve}.  
The H-index of AO publications is 75 and the median of 
the number of citations is 23. About 8\% of the citations arise from the methodology
of AO.

%*Convert above paragraph to MATLAB

%--------------------------------------------------------------------------------------------
\section{Inferences}
\label{sec:Inferences}
%--------------------------------------------------------------------------------------------

\subsection{The Observatory Flux Curves}
\label{sec:ObservatoryLightCurve}

\begin{figure*}[htbp]
   \centering
   \includegraphics[width=5.3in]{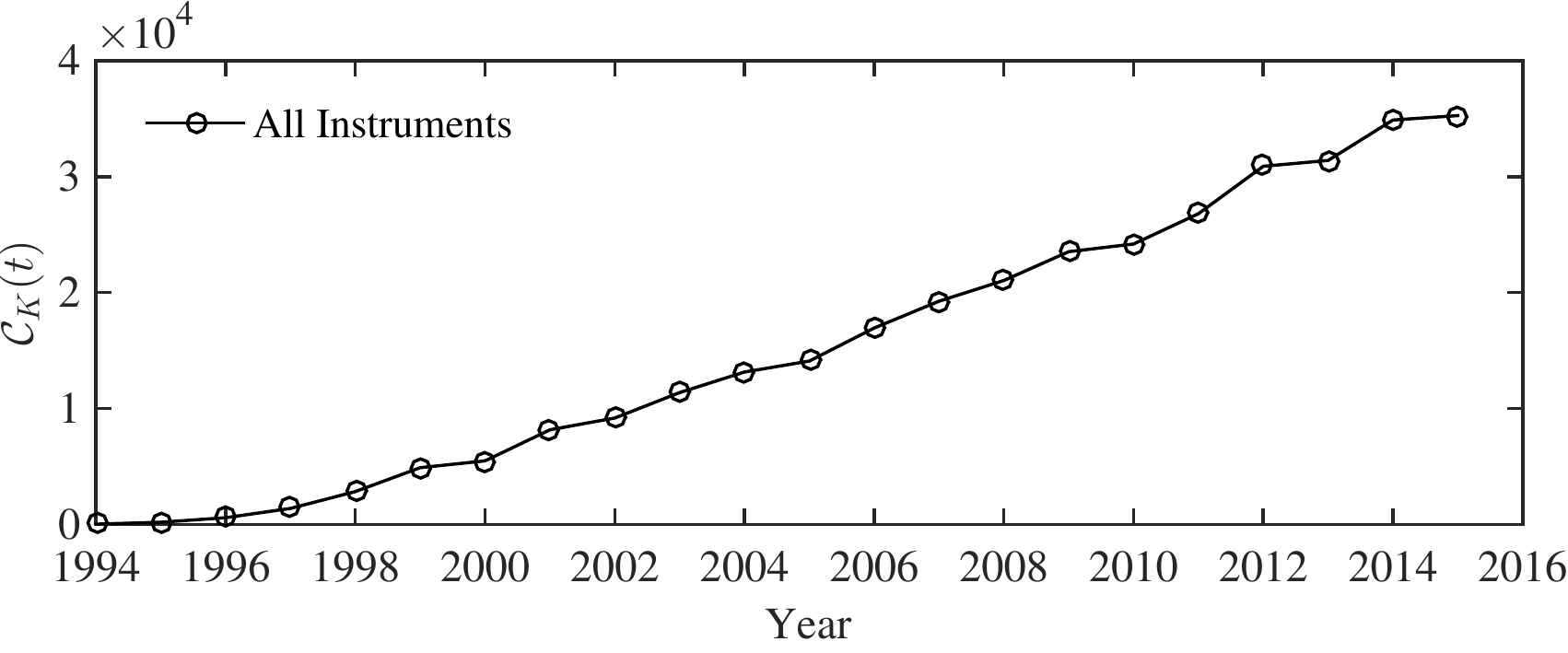}
   \caption{\small The citation flux curve of all the instruments,
   taken together, of the W.~M.~Keck Observatory.  }
 \label{fig:CitationSummaryCurve}
\end{figure*}

The annual paper flux, $\mathcal{P}(t)$ and $C(t)$, the total
citations nominally accrued in a given year (Equation~\ref{eq:Ck}),
are summarized in Table~\ref{tab:AllInstruments}; note that $C(t)$
is not the same as $\mathcal{C}(t)$ (see \S\ref{sec:FluxCurves}).
The citation flux curve for the Observatory as a whole (summing
over the instruments), $\mathcal{C}_K(t)$, is displayed in
Figure~\ref{fig:CitationSummaryCurve}. The annual flux in 2015 is
about 30,000 citations per year.  It is quite impressive to see a
linear growth lasting nearly two decades.  [I do note that the flux
curves for all instruments as well as the total number of papers
show either a reduction or no change between 2013 and 2014--2015.]

\begin{figure}[htbp]
% \centering \includegraphics[width=2.5in]{CitesPerPaper.eps}
  \centering \includegraphics[width=2.5in]{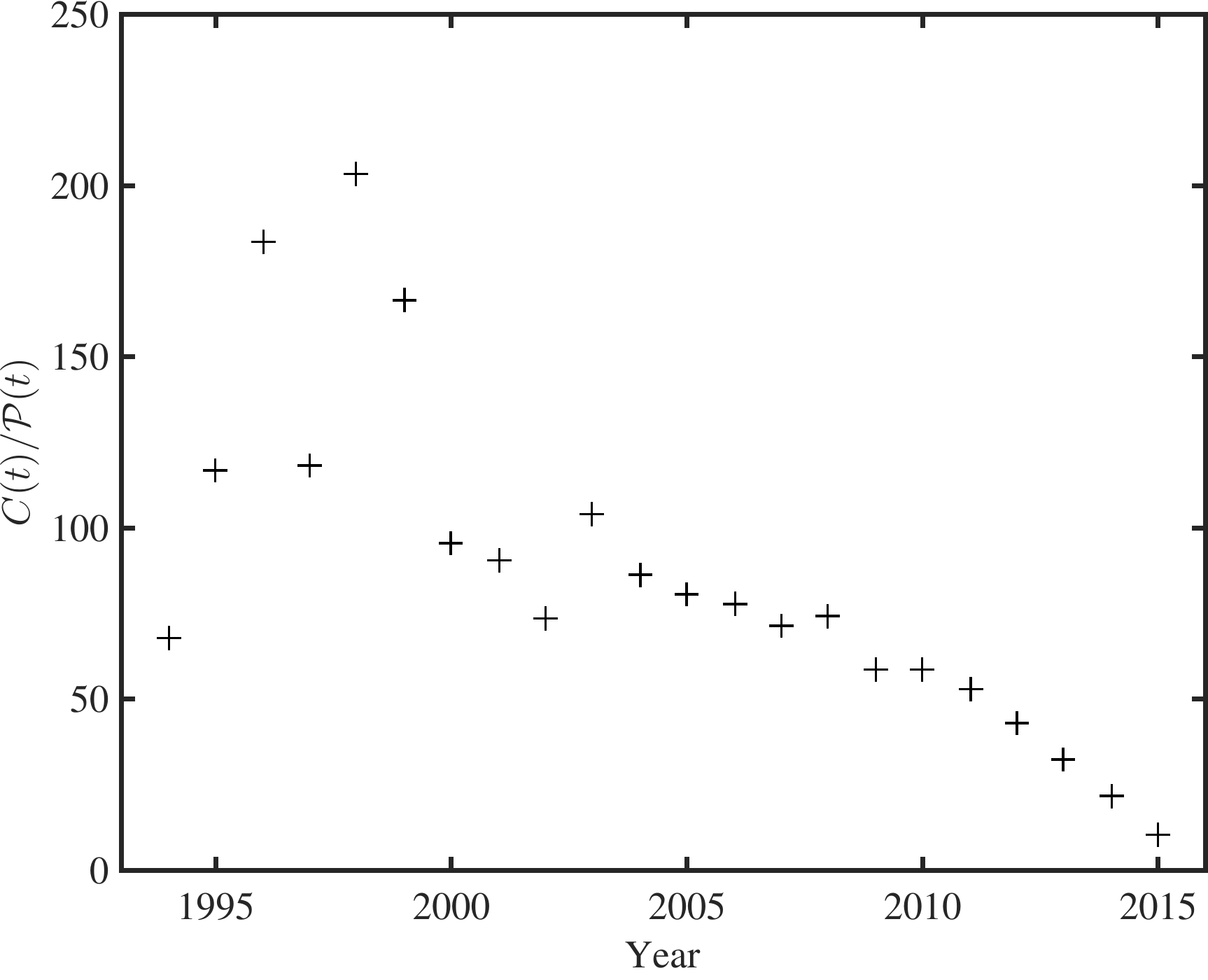}
   \caption{\small The abscissa is the ratio of the number of
   citations accrued in a given year, $C(t)$, to
   $\mathcal{P}(t)$, the number of refereed papers published in the
   same year.    }
 \label{fig:CitesPerPaper}
\end{figure}

In Figure~\ref{fig:CitesPerPaper} I plot $C(t)/\mathcal{P}(t)$. The
numerator is the sum of citations gathered by papers published in
year $t$ (see Equation~\ref{eq:Ck} and the discussion surrounding
it); it is {\it not} the citation flux curve, $\mathcal{C}(t)$.
The denominator is the number of papers published in the same year.
As can be seen from this
figure  papers published in the first six years of the Observatory's
beginnings (1994-2000)  had a distinctly higher impact relative to those 
published in later years. This plot is a dramatic
illustration of the great benefit enjoyed  by WMKO by being  ``first
on the block".

\begin{deluxetable*}{rr}
\tabletypesize{\scriptsize}
\tablecaption{Most Cited Papers}
\tablewidth{0pt}
\tablehead{
\colhead{Inst.} & \colhead{Tops}}
\startdata
NIRC & 633 540 538 507 455 440 420 382 350  \\
LRIS & 9000 9000 2983 1869 1736 1607 1438 1400 1286  \\
HIRES & 808 804 628 624 580 566 556 471 410  \\
ESI & 870 743 689 675 643 632 538 471 424  \\
NIRC2 & 828 719 713 535 422 399 360 240 223  \\
NIRSPEC & 2983 675 661 632 422 350 321 318 272  \\
DEIMOS & 1869 1192 767 743 646 509 481 471 422  \\
OSIRIS & 214 152 147 132 113 103 100 100 97  \\
MOSFIRE & 133 106 87 73 72 63 59 55 47   
\enddata
\tablecomments{The number of citations of the top 9 papers arising from each Keck facility instrument
}
\label{tab:ImpactOfInstruments2}
\end{deluxetable*}

In \S\ref{sec:MeasuringProgress} we noted that  the singular or
exceptional impact of an instrument (or an author, for that matter)
is measured by the highest cited papers. Initially I thought listing
the top five papers (for each instrument) would be adequate. However,
I realized that a few papers claimed the top spots for several
instruments. The most heavily cited papers from LRIS, DEIMOS and
ESI are all related to the same topic -- the use of supernovae for
cosmography. Progress in cosmography is important but like many
great  successes  in life there are numerous claimants. In particular
other observatories also assert their mighty contributions to
supernova Ia cosmography.  Thus in order to assess the unique
contribution of Keck, I expanded the list to the top nine papers
(Table~\ref{tab:ImpactOfInstruments2}).  The titles of these papers
can be found in the Appendix (\S\ref{sec:TopPapers}).  The reader
is urged to look at this list of papers to appreciate the singular
(and distinct) returns from each of these instruments.

\subsection{The High Impact of Optical Instruments}

As can be gathered from Tables~\ref{tab:Productivity} \&\
\ref{tab:ImpactOfInstruments1} and Figure~\ref{fig:AllInstrumentsCitations}
optical instruments are both productive and also have a larger
impact relative to NIR instruments as well as AO-assisted observations.
Along this line, I note that both ESI and NIRC did not receive any
upgrades since commissioning.  Yet ESI had a higher return relative
to NIRC.

There are two strengths that optical instruments enjoy relative to
NIR: (i) natural background that is orders of magnitude smaller in
the optical relative to NIR and (ii) detectors that are nearly
perfect in their response (with virtually no dark current).  NIR
instruments win only when the natural conditions favor them: objects
suffering from extinction (the poster child here is observations
of the stars in the center of our Galaxy) or when the diagnostics
are uniquely in the NIR band (e.g.\ cool objects such as brown
dwarfs, asteroid spectroscopy). While beyond the scope of this
paper it is worth noting that the IR/AO communities are smaller than
the optical community and this may introduce a bias \citep{pk12}.

\begin{deluxetable}{rrr}
\tabletypesize{\scriptsize}
\tablecaption{Papers \& Citations: All Instruments}
\tablewidth{0pt}
\tablehead{
\colhead{Year} & \colhead{Papers}& \colhead{$C(t)$}}
\startdata
1994 & 11 & 749 \\ 
1995 & 36 & 4202 \\ 
1996 & 53 & 9740 \\ 
1997 & 68 & 8040 \\ 
1998 & 109 & 22204 \\ 
1999 & 127 & 21159 \\ 
2000 & 169 & 16185 \\ 
2001 & 175 & 15884 \\ 
2002 & 193 & 14226 \\ 
2003 & 211 & 22004 \\ 
2004 & 214 & 18397 \\ 
2005 & 232 & 18659 \\ 
2006 & 277 & 21498 \\ 
2007 & 312 & 22373 \\ 
2008 & 262 & 19390 \\ 
2009 & 269 & 15671 \\ 
2010 & 289 & 16979 \\ 
2011 & 297 & 15811 \\ 
2012 & 337 & 14515 \\ 
2013 & 319 & 10375 \\ 
2014 & 291 & 6230 \\ 
2015 & 292 & 3075   
\enddata
\tablecomments{ columns from left to right: year, the total number
of papers published in the year and the number  of citations accrued
by {\it the papers published in that year.} As noted in
\S\ref{sec:FluxCurves} and Equation~\ref{eq:Ck} the value of $C(t)$
depends on the time at which the sum is evaluated. The exercise was
undertaken in May 2016.}
 \label{tab:AllInstruments}
\end{deluxetable}

\subsection{The Longevity of Instruments}
\label{sec:Longevity}

From an inspection of the paper generation curves I conclude that
instruments which have not undergone significant upgrades achieve
a peak between five to eight years after commissioning (e.g.\ NIRC,
ESI and NIRSPEC).  Some care should be exercised in interpreting
the flux curves of  NIRC2 and OSIRIS since the full power of these
instruments arises from the performance of the LGS AO system. As a
result the impact of NIRC2 and OSIRIS can be expected to track improvements in
the LGS AO system (which  is undergoing considerable improvements
since commissioning in 2004; see Table~\ref{tab:AOFacility}).

For the sake of argument we will accept the time for an instrument
without any upgrades to peak is six years (and perhaps as much as
ten years). Accepting this figure we ask the question: what sets
this timescale? Before I discuss possible explanations for this
duration I provide some background.

Progress in astronomy appears to  take place in three phases: ({\it i})
discovery, ({\it ii}) a search for patterns (made possible by many
measurements) and ({\it iii})  and the construction of a model to account for
the regularities (e.g.\ see \citealt{k12}).  The culmination is when
the model finds a natural explanation in known physics or leads
to new understanding of physics.
A famous example is (1) the recognition
of planets as a new phenomenon (namely they move, unlike stars),
(2) the gathering of exquisite  data by Tycho Brahe and others and
(3) a mathematical model by Johannes Kepler, culminating in a
physical explanation for the mathematical model by Isaac Newton.

A  modern and a far less dramatic example is the subject of brown
dwarfs. The first couple of years following the discovery of the
first brown dwarf  
constituted the period of ``low hanging fruits". Even a single
observation of a brown dwarf  resulted in a  nice paper. Following
this phase investigation shifted to  systematic study of large
samples. Naturally the paper production slows down during this
period.

With this background, we offer two reasons to explain the decrease
in $\mathcal{P}(t)$ with time. First,  following either a discovery
or the arrival of a powerful new instrument users exhaust ``low
hanging fruit" projects (in the sense as discussed above). Second,
it may well be that the instrument becomes unattractive because
other observatories start deploying instruments with larger reach
or higher sensitivity.  Users of the first telescope then do not
find it attractive to spend their precious allocation on a fading
asset.

I argue that the decline in productivity of NIRC is because of
increasing obsolescence.  The $256\times 256$ pixel InSb array detector of NIRC
was state-of-the-art in 1993.  However, the rapid growth in the
format and quality of NIR detectors (see Appendix~\ref{sec:NIR} for
a summary of the great progress in NIR detectors) hastened the
obsolescence of NIRC.

\subsection{Upgrades}
\label{sec:Upgrades}

The first light instruments are NIRC, LRIS and HIRES. NIRC shows
the expected classic behavior: peaking, as measured by paper
production, about six years  after first light and then gradually
declining.  In contrast, LRIS achieved a plateau six years later
and is maintaining the plateau. A simple explanation for this
continued productivity are the upgrades: Blue-channel (2000), ADC
(2007) and Red-channel (2010).  Likewise HIRES shows a rise to a
plateau in the year 2000 and then undergoes another rise starting
the year 2004. HIRES continues to show a sustained increase in both
productivity and impact. I  attribute this behavior in part to the 3-CCD
upgrade that was undertaken in 2004 (the other reason is the continued
blossoming of the extra-solar planet field).

\begin{deluxetable}{rrrr}[hbtp]
\tabletypesize{\scriptsize}
\tablecaption{Citations to Instrument Papers}
\tablewidth{0pt}
\tablehead{
\colhead{Instrument} &\colhead{Papers} & \colhead{Citations}& Q(\%)\\
& \colhead{$N_p$} & \colhead{$N_c$} & } 
\startdata
NIRC    & 247 & 223 & $-10$ \\ 
LRIS    & 1497 & 1699 & $13$ \\ 
HIRES    & 1202 & 891 & $-26$ \\ 
ESI    & 293 & 237 & $-19$ \\ 
NIRC2    & 461 & -- & -- \\ 
NIRSPEC    & 484 & 425 & $-12$ \\ 
DEIMOS    & 654 & 423 & $-35$ \\ 
OSIRIS    & 104 & 113 & $9$ \\ 
MOSFIRE    & 54 & 49 & $-9$ \\ 
AO      & 640 & 489 & $-24$ 
\enddata
\tablecomments{ Name of the instrument, number of refereed papers
($N_p$) arising from the instrument and the number of citations to
the fundamental paper(s) which describes the instrument ($N_c$).
$Q$ is defined by Equation~\ref{eq:Q}.  For each instrument, the
fundamental references are listed in various subsections of
\S\ref{sec:TheInstruments}.  In order these are NIRC (Matthews \&
Soifer 1994a-b)\nocite{ms94a}\nocite{ms94b}; LRIS
\citep{occ+95,mcb+98,ssp+04,rck+10}; HIRES \citep{vab+94}; ESI
\citep{smb+00,sbe+02}; NIRSPEC \citep{mbb+98}; DEIMOS \citep{fpk+03};
OSIRIS (Larkin et al. 2006a-b)\nocite{lbk+06a}\nocite{lbk+06b} and
the AO system (NGS \& LGS; \citealt{was+00,wlb+06,vbl+06}).  There
is no entry for $N_c$ for NIRC2 since the builders did not  publish
a paper describing the instrument. The quoted values were measured
at the time of the submission of this paper. }
 \label{tab:CitationsBuilders}
\end{deluxetable}

\subsection{Are Builders Well Recognized?}
\label{sec:BuildersRecognized}

Astronomy, particularly OIR astronomy, is perceived to have a culture
that does not reward astronomers with instrumentation skills.
Astronomers certainly appreciate the value of sophisticated
instruments.  However, whether this appreciation translates to
tangible rewards, especially those which are valuable (faculty
appointments) is unclear.  Some areas of astronomy --  radio astronomy
(particularly research related to Cosmic Background Radiation,
development of new facilities, pulsar research) -- have a long
tradition of rewarding astronomers with primary talent in
instrumentation. Perhaps the difference lies in the fact that in
the early history of optical astronomy (and extending through the
era of large telescopes in California) the instruments were relatively
simple and great value was (in effect) attributed to the astronomers
who were able to secure time and make discoveries.  However, over
the past several decades the complexity of OIR astronomy instrumentation
has dramatically increased and OIR now {\it needs} astronomers with
technical background.

In Table~\ref{tab:CitationsBuilders}, I present, for each Keck
facility instrument as well as the AO system (NGS, LGS)  the number
of published papers ($N_p$) that can be ascribed to that instrument.
As noted earlier (\S\ref{sec:TheInstruments}) some instruments have
multiple references to the performance of the instrument (usually
reporting a significant upgrade).  I have summed up the citations
from these papers (the papers are listed in the caption to
Table~\ref{tab:CitationsBuilders} and present the total number of
citations ($N_c$) for each instrument in
Table~\ref{tab:CitationsBuilders}).  Consider the quantity
 \begin{equation}
  Q \equiv \frac{N_c}{N_p} - 1.
   \label{eq:Q}
 \end{equation}

$Q=0$ means that every paper which used a particular instrument
acknowledged the builders of the said instrument.  $Q<0$ is the
fraction of astronomers who use a Keck instrument without acknowledging
the instrument team which made their observations possible.  The
users of NIRC, LRIS and OSIRIS and perhaps NIRSPEC can be argued
(within Poisson noise) to have been  grateful to the builders of the
instruments.  However, users of HIRES, ESI, DEIMOS and the AO
system(s) appear to be quite lax in acknowledging the instrumentation
teams that made their observations possible.

In case of LRIS we note $Q>0$.  The explanation for this curious
finding is that some of the observational papers refer to the
original LRIS paper \citep{occ+95} as well as one or more upgrades
\citep{mcb+98,ssp+04,rck+10}.  Finally, as illustrated by the
significant positive value of $Q$ for LRIS
(Table~\ref{tab:CitationsBuilders})  a major upgrade clearly benefits
by having its  own instrument paper.

While here I only address ``builders" in the usual sense of hardware
the fact remains that software engineering is increasingly a major
(and at times, even a dominant) aspect of modern instrumentation.
Clearly, any such future analysis  should also evaluate the returns
to those who, with ingenuity and hard work, build data acquisition,
data reduction pipelines and develop powerful software tools for
use by observers.

I end this section with an editorial remark. The research undertaken
for this project spread over many years and naturally over this
time I beavered away at many locations: airports, committee meetings
and visits to several institutions (domestic and otherwise). I came
to appreciate the value of society journals such as PASP and AJ in
terms of the ease of access from random sites.  Very few institutions
have paid subscription to commercial journals (especially the
unrefereed SPIE proceedings )  and access is an issue. I urge
instrument builders to bear this issue in mind and (1) publish their
key paper (the performance of their instrument) in journals that
are easily available at most institutions around the world and (2)
post a copy of their published papers on any archive server 
(such as arXiv).

\section{Archives \&\ Pipelines}
\label{sec:Archives}

It is now well demonstrated that a high quality archive\footnote{A
good archive is not merely a collection of FITS files but one with
an intelligent query interface and the ability to provide fully
calibrated data and higher level products.  In the absence of such
products, the archives are essentially write-only storage of data.}
enables additional exploitation of the data collected from the
observatories.  For instance, in 2011, the 4-telescope VLT facility
of ESO reported 550 refereed publications that were based on new
data. An additional 100 papers arose from archival data analysis.\footnote{
ESO Annual Report 2011, p. 30.  The report can be found at
\texttt{http://www.eso.org/public/products/annualreports/}.} Thus,
apparently, archival analysis can boost the productivity of a
ground-based facility by about 20\%.

The original operations model for WMKO did not include funding for
an archive. Fortunately, as noted in \S\ref{sec:HIRES}, starting
2004 (a decade after commissioning of the telescopes), NASA funded
a program -- the Keck Observatory Archive (KOA). This enterprise
is jointly operated by the NASA Exoplanet Science Institute (NexScI) and
WMKO.  KOA began with an ingestion of HIRES data. Within the annual budget
of KOA, the ingestion of data from other instruments could be
accommodated at a leisurely pace -- one instrument every other year (or so).
At the current time, KOA archives and serves public data\footnote{Public
data: data that no longer has any proprietary protection.  The
default proprietary period is 18 months though each partner can
request longer extensions.}  for all facility instruments \citep{bhm+15}.

\citet{pzs+08} is the first paper citing the use of data from KOA.
The reader should note the  four year lag between the year of the
publication of this paper and the launch of KOA. 
In 2015, forty three papers were
published or about 15\% of the total publications for that year.
It is anticipated that the archival papers for 2016 may reach a
fraction as large as 23\% (H. Tran, pers.  comm.). For comparison,
the Hubble Space Telescope (HST) archive,\footnote{\url{https://archive.stsci.edu/hst/bibliography/pubstat.html}}
widely reported to be the
most productive archive,
accounts
for about 54\% of HST papers. Returning to WMKO the late start of
KOA (nearly 10 years following routine astronomical usage of the
telescopes began) and the  slow ingestion means that KOA is a young
archive, relative to that of VLT and HST.  So likely KOA is on a
virtuous trajectory to boost the astronomical productivity of the
Observatory.

I bring up the importance and cost (both real and opportunity) of
DRPs.  The case study is OSIRIS. $\mathcal{P}(t)$ for  OSIRIS did
not show the expected strong early rise.  As noted in \S\ref{sec:OSIRIS},
the performance of OSIRIS at commissioning was lower due to grating
not manufactured to specifications. Thus at the very start
OSIRIS was at a disadvantage
relative to its competitor (ESO's SINFONI instrument which was
commissioned in late 2004).  The situation was further exacerbated
by the difficulty of extracting signal from IFU data.
Astronomers have come to appreciate that 
IFUs are inherently complex.  Developing the extraction 
algorithms requires 
requires a deep understanding of the instrument.
As a result, ordinary users {\it need} a quality DRP to reduce
the IFU data.
Unfortunately,  a robust DRP was not a part of the
OSIRIS commissioning.  As a result, in practice, OSIRIS data was
nearly un-reducible by the average user astronomer.  A GUI-based
DRP (with adequate documentation) that came several years later
helped alleviate the situation.  
The OSIRIS pipeline is
still a work in progress.  Astonishingly, a similar sad story
unfolded for an optical IFU that was built for the 60-inch telescope
(Spectral Energy Distribution Machine;
SEDM)\footnote{\url{https://nickkonidaris.com/sed-machine/}}.  In
contrast to OSIRIS, this is an ultra-low resolution seeing limited
spectrograph. The common problem was the lack of a quality DRP at
the time of commissioning followed by a lack of appreciation of the
scale of high quality manpower effort that is needed to
 extract signal from IFUs (whilst suppressing systematics).

\section{The Cost \&\ Value of a Night of Telescope Time}
\label{sec:CostAndValue}

Unlike radio astronomy (rather specifically, wavebands from decameter
through decimeter) very few, if any, optical observatories have a
truly ``open sky'' policy.  In optical astronomy it has long been
the tradition that access is primarily restricted to astronomers
from institutions which funded the construction of the telescope.\footnote{
Indeed, herein may lie the reason why the centroid of global OIR
astronomy shifted to the West Coast of California.  Access to the
Lick Observatory, the Mt.\ Wilson \&\ the Palomar Observatories, all
of which laid the astronomical foundation for the University of
California, Caltech and the Carnegie Observatories, was limited to
the investing institutions.} Once an observatory is successful it
is not unusual to find astronomers elsewhere  pushing their
institutions to obtain access to such telescopes.  Nor is it unusual
for observatories to seek new partners (either as a buy-in or a
limited time lease) to fund new initiatives or continue operations.

\subsection{TSIP}

Recognizing the above situation and also acknowledging inadequate
public investment in optical astronomy (at least relative to private
investment) in the United States, NSF commissioned a study. The
resulting ``McCray report" led to the establishment of the Telescope
System Instrumentation Program
(TISP).\footnote{\url{http://ast.noao.edu/system/tsip/}} This program
aimed to increase telescope access to the US community by funding
existing private observatories. The funding was either for building
new instruments or for compensating the operators a portion of
their running costs.  This
initiative directly raises the question of ``How should a night of
telescope access be valued?".

The TSIP framework was a landmark for US based OIR facilities. It
established a market place which may sound strange to astronomers
who tend to view their work as being outside the economic sphere.
The TSIP framework was constructed as follows. The cost for a night
of observations was derived from three contributions: the cost of
the telescope linearly amortized over twenty years, the cost of
instrumentation amortized over ten years and the current annual
operating cost. For the first two items ``then year'' dollars were
used whereas for the third item  current year dollars
are used.  For a telescope older than twenty years the recommendation
was to set the value of the telescope to the ``current estimated
cost to build a telescope of similar characteristics reduced by a
factor equal to inflation over the last ten years'' and then to
linearly amortize this estimate over the next twenty years.

\subsection{Re-examining  TSIP Framework}

The TSIP program was critical for WMKO. This program made it possible
for the Observatory to build OSIRIS, MOSFIRE, KCWI and underwrote
the considerable costs for the formulation of the ``Next Generation
Adaptive Optics" (NGAO) project.  The same program funded instrumentation
at other observatories as well. 

Returning back to the business at hand, overall, the 
the TSIP framework is reasonable. It is nonetheless useful
to review the three  assumptions. To start with,  the flux curve of
NIRC provides some justification for the TSIP 10-year amortization
rule. However, the flux curves of instruments which received
upgrades would favor a longer period for amortization.

Next, the primary function of a telescope is to collect light and
project it into a small image. This ability of the telescope need
not decay with age. I quote an example that I know very well -- the
Hale 5-m telescope (commissioned in 1949). Thanks to refurbishments
and a better ability to model the mechanical structure the primary
mirror of the Hale telescope is in better shape  today than it ever
was.  The pointing has been steadily improved and is now as good
as a modern telescope. The mirror coating is also up to modern
standards.  In my opinion and experience the primary danger to the
basic functioning of an older telescope is light pollution.  It is
possible to maintain aging facilities competitive, limited only by
the imagination of astronomers (for innovative projects) and the
ability of management to raise the necessary funding.

In defense of this assertion I give three examples of 
ground-based telescopes which continue to be of current value.
I start by noting the several reincarnations of
the Palomar 48-inch Oschin telescope (a Schmidt camera telescope;
commissioned in 1951)  -- photographic
all sky survey (POSS1, POSS2), robotic operation with CCD mosaic
(3-banger, PalomarQuest), Palomar Transient Factory (PTF) and soon
Zwicky Transient Factory (ZTF; FOV of 47 square degree, CCD mosaic
with 576\,Mpix, autofocus, improved pointing, rapid slewing
etc).\footnote{\url{http://www.ptf.caltech.edu/ztf}}   The Southern
counterpart, the  AAO 48-inch telescope (commissioned 1973), similarly underwent several
reincarnations: ESO/SERC Southern Photographic Survey, the pioneering
Fibre-Linked Array Image Reformatter
(FLAIR)\footnote{http://ftp.aao.gov.au/astro/flair.html} which
initiated the era of massively multiplexed spectroscopy,
6dF\footnote{http://ftp.aao.gov.au/ukst/6df.html} and
RAVE.\footnote{https://www.rave-survey.org/} 

Next, the Palomar 60-inch
telescope (commissioned 1970), originally built for student training, was robotized and
played a major role as a photometric (color) engine for the Palomar
Transient Factory and is now being reinvented for robotic spectral
classification of transients.  

The Palomar 200-inch has an  excellent
suite of workhorses and novel instruments (e.g.\ such as the Cosmic
Web Imager -- the fore-runner of the Keck Cosmic Web Imager; a
state-of-the-art coronagraph behind a 3,000-actuator AO system; an
upgrade of the current H2 detector to an H2RG along with a polarimetric
mode will result in a NIR imager very well suited to exoplanet
eclipses and weather on brown dwarfs). Indeed, the vibrancy of
the current partnership (Caltech, Jet Propulsion Laboratory, 
Yale University and National Astronomical Observatory of China)
shows the telescope offers {\it current} value.

Perhaps the most dramatic case for the proposition laid at the
start of the second  paragraph of this section is the Hubble Space
Telescope.  HST, when launched in 1990, carried the Wide Field \&\
Planetary Camera (WFPC; based on eight $800\times 800$ pixel array CCDs,
eighties vintage), Goddard High Resolution Spectrometer (GHRS; two
521-pixel Digimon light intensified detectors), High Speed Photometer
(HSP), Faint Object Camera (FOC; image intensifier technology) and
Faint Object Spectrograph (FOS; 512-pixel Digimon light intensified
detectors). Let us for a moment ignore the problem raising from the
flawed mirror (and discovered shortly after first light). Specifically let us
imagine a new world timeline in which HST was launched with a perfect
mirror but without the possibility of instrument upgrades.  In this world,
HST would have produced stunning results for the first five and perhaps
ten years. The march of technology, especially in improved QE (UV, optical, NIR), 
lower read noise (all bands) 
and larger format (all bands) and the development of Adaptive Optics
would have diminished HST's
standing relative to ground based astronomy. The only band where HST would have
had unique advantage would have been in the UV. Here, too, the gains
in QE (from image intensifiers with QE of tens of percent (at best)
to modern delta-doped CCDs
with near unity QE) has been dramatic. HST is a leader in faint object
wide-field astrometry (which will, for ever, remain a bastion of space
based projects, cf.\ Gaia).
It is the periodic updates of new
instruments (which take advantage of technological growth) that 
kept HST at the forefront of astronomy.

I would therefore suggest the following modification to the TSIP
framework: following an upgrade of an instrument the 20-year
amortization rule should be applied to the market value of the
upgraded instrument. A well maintained telescope should receive
similar consideration.

\subsection{Citations as basis for cost}

There is an entirely different approach to determine the value of
an observatory, namely the final output -- the scientific results
attributable to the observatory. In the spirit of this paper
(``astro-econometrics") I suggest that the citation flux,
$\mathcal{C}(t)$,  should form the basis of currency for optical
observatories. This market-based approach will  favor observatories
which build their telescopes at superior sites,  maintain their
telescopes to a high level of performance (so nights are not lost
due to telescope failures), undertake periodic infra-structure
improvements (so that the fraction of productive usage remains
high), build up a suite of powerful instruments  (optimized for
dark and bright time, for excellent and moderate seeing) and undertake
upgrades of instruments as detectors improve and so on and so forth.

The two fundamental quantities in a market are {\it cost} and {\it
value}.  The cost per citation, $C_1$,  is most simply computed as
the ratio of the citations accumulated up to a point of time to
that of the total money spent to that date (capital, operating
expenses, instruments and other investments; all inflated to the
end point).  

In contrast to cost, there is no simple basis to estimate value.
Fundamentally, value is intimately tied to the perception of the buyer
(``eyes of the beholder"). One simple approach 
is to accept the TSIP rate for a
night, $\mathcal{T}$ as a given. In this case the value per citation
is $V_1=\mathcal{N}(\mathcal{T}/\mathcal{C})$ where $\mathcal{C}$
is the annual flux of citations (Figure~\ref{fig:CitationSummaryCurve})
and $\mathcal{N}$ is the number of potentially usable nights (that
is after accounting for nights set aside for engineering and
commissioning).  It would be useful to carry out similar evaluations
for other recipients of TSIP grants. In a {\it rational} market (as
in a micro-economic sense) the values of $C_1$ and $V_1$ will be
consistent.

A high-level national study has noted that  there will be a high
demand for follow up facilities in the 
the Large Synoptic Survey Telescope (LSST) era  \citep{Elmegreen2015}.
If so, there will be demand for access to privately run facilities
by those who lack access (see \S\ref{sec:OptimizingTheSystem}).
Given this expectation, it would be most useful for NSF  to commission
a retrospective study of the influential TSIP program, particularly
addressing the ``business" side. Such a study would be extremely
helpful to the Chairs of astronomy departments as they build up a
strong case  for access to telescopes for their departments.  After
all trustees are usually practical people and appreciate sound
business arguments over any other type of argument.

The above formulation for $V_1$ is applicable for classical telescopes.
In particular, a night allocated to one party means that the same
night cannot be allocated to any other party. The above formulation
is not applicable to projects such as SDSS, PS-1 and PTF for which
the concept of a single night is not particularly meaningful.
Alternatively, $V_1$ for projects such as SDSS should be the value
computed above and divided by $\mathcal{M}$, the number of subscribers.

\section{The Future  Landscape}
\label{sec:FutureLandscape}

I had two objectives when I set out to undertake the investigations
leading to this paper: (1) quantify the productivity of observatories
by instruments and (2) explore a rational basis to determine the
value of a night of telescope time.  These two topics were addressed
in \S\ref{sec:Inferences}--\S\ref{sec:CostAndValue}.  In that sense
the previous section marks the formal end of the paper.

Here, I take the opportunity to use the conclusions drawn in this
paper to understand the future of large optical telescopes, both
in terms of opportunities and challenges. However, optical telescopes
(large or small) are only a part of the entire astronomical landscape.
It is, therefore, important to understand the larger landscape
before one can discuss the future of large optical telescopes. The
two main developments (of relevance to large optical telescopes)
are: the explosive growth of deep/wide  imaging/photometric/astrometric
surveys and the rise of of massively-multiplexed spectrographs.
These are discussed below, respectively, in \S\ref{sec:SynopticSurveys}
and \S\ref{sec:MassiveMultiplexing}.

\subsection{Imaging -- Synoptic Surveys}
\label{sec:SynopticSurveys}

Historically, all-sky (or large FOV) surveys have had a great impact.
For instance,  the plates or films (and later digitized versions)
of Palomar Observatory Sky Survey (POSS) were a fixture in any
respectable astronomy department;  see \cite{tc08} for more recent
examples. As additional support of the value of all-sky surveys I
draw the reader's attention to Appendix~\ref{sec:VLA} where I measure
the rate of return for two wide-field surveys undertaken with the
Very Large Array (VLA), an Observatory with which I have more than
a passing familiarity. I find the return rate of the two surveys
to be superior to those returned by PI-led projects.

In this context, the Sloan Digital Sky Survey (SDSS) deserves a
special mention.  Starting circa 2000, this project, based on a
2.5-m telescope, undertook the first large-area  (Northern Galactic
cap) digital survey in five optical bands.  SDSS inspired other
surveys (e.g.\ VLA/FIRST). SDSS is widely regarded as a  great
success story of modern optical astronomy \citep{mm09}. In the
North, the 1.8-m PanSTARSS \citep{kab+02} with its 1.4\,Gigapixel
(Gpix) mosaic  has concluded a 5-band Northern sky survey (with a
public release that is imminent) and the 1.35-m SkyMapper \citep{ksb+07}
with a 268\,Megapix (Mpix) imager is midway on a similar mission
for the Southern Sky.

Of all the bands, the optical band is the most mature in terms of
sky surveys.  The world is awash in
large FOV optical imagers -- thanks 
to the decreasing cost of sensors, data acquisition circuitry and
computing (when evaluated on a per unit basis; Moore's law). Here
is an incomplete listing of large FOV imagers: CHFT/CFH12K (96\,Mpix),
CFHT/MegaCam (324.5\,Mpix; \citealt{bca+03}), Subaru/Suprime-Cam
(80\,Mpix; \citealt{mks+02}), Blanco/Dark Energy Camera (520\,Mpix;
\citealt{fdh+15}) and  Subaru/HSC (870\,Mpix; \citealt{mkn+12}).

The sophistication and maturity of optical synoptic imaging can be
measured by the increasing number of {\it specialized} surveys.
Catalina Sky Survey and PS-1 are entirely devoted to the study of
Near-Earth Asteroids. We have ground- and space- missions {\it
dedicated} for exoplanets (e.g.\ WASP, HARPS, Kepler, TESS). The on-going
PTF  and the imminent ZTF have a singular goal of studying
optical transients. The CFH Supernova Legacy Survey (SNLS) was tuned
for Ia cosmology.  The Dark Energy Camera and the HyperSuprimeCam
were motivated by a single goal: probing Dark Energy through several
approaches.

Thanks to investments by NASA we now have full sky surveys in other
bands (e.g.\  2MASS/NIR,  GALEX/UV,  WISE/MIR).  European Space
Agency's (ESA) {\it Gaia} mission, with its precision photometry,
spectrophotometry and unparalleled astrometry of nearly $10^9$
objects, is poised to revolutionize stellar and Galactic astronomy.
The Russian-German Spektr-RG mission (expected to launch in 2017)
will present cadenced deep views of the entire sky in X-rays. The
bonanza of large FOV surveys will continue into the near and distant
future: TESS, CHEOPS and PLATO are wide field  precision synoptic
photometric surveys.  Euclid and WFIRST will undertake space-based
large FOV surveys in the optical and the NIR bands. Finally, the
LSST is expected to start routine
operations in 2022.

Entirely separately and truly exciting is that 2016 marks the opening
of the field of Gravitational Wave (GW) astronomy \citep{LIGO2016}.
The GW detectors, being essentially one-baseline interferometers,
have very large FOV (the primary beam) but poor localization (owing
to baselines of moderate length, relative to the wavelength).  Identification
of the electromagnetic counterpart of GW sources (involving neutron
stars) will benefit from archival data, require large FOV imagers
and rapid access to large optical telescopes for the  much sought
after spectroscopy of the GW events.

\subsection{Advances in Spectroscopy}
\label{sec:MassiveMultiplexing}

A traditional slit spectrograph does not make full use of the
available focal plane. The primary return is a single object spectrum
(since, nature rarely produces nebulae neatly lined up with the
slit). Multi-slits or use of fibers allow for spectra of large
numbers of objects to be obtained in one shot.  The  pioneering 2dF
spectrograph on the 3.9-m Australian Astronomical Observatory (AAO;
\citealt{cdm+01}) demonstrated how an existing telescope at a
mediocre site can undertake leading science projects.  The spectrograph
could obtain low resolution spectra of 400 objects over a 2-degree
field of view.  The 2dF Galaxy Redshift Survey (2dFGRS) measured
redshifts of 250,000 galaxies or stars over 2,000 square degrees
with a median redshift of 0.1.  The success of 2dFGRS has made it
now almost mandatory that all large optical telescopes be equipped
with multiplexed spectrographs (e.g.\ DEIMOS, MOSFIRE on Keck; IMACS
on Magellan; VMOS and KMOS on VLT and so on).  A recent development
is ``integral field unit" spectroscopy -- obtaining spectra of a
rectangular region (e.g.\ OSIRIS).  We are on the verge of the IFU
revolution -- soon astronomers will routinely have access to multiple
``deployable" IFUs on large telescopes.

The spectacular success of SDSS \citep{mm09} was in my opinion
entirely due to the resonance between imaging and massively multiplexed
spectroscopy (a pair of 320 fibers, upgraded to a pair of 500 fibers
in 2009 feeding a pair of two-armed spectrographs).  Indeed, without
the strong support of  highly multiplexed spectrographs the gains
of the synoptic surveys will go largely unrealized.  In this respect,
I admire the vision and courage of the
Subaru management for funding not just the HSC but also
PFS. The HSC/PFS combination is not only potent but durably so.

\subsection{NRC Study: Optimizing the System}
\label{sec:OptimizingTheSystem}

The focus of  this section -- namely the landscape of OIR astronomy --
has been discussed extensively and expansively by an National
Research Council (NRC) panel chaired by D.\ M.\ Elmegreen of Vassar
College. The panel goes further ahead and makes suggestions to
optimize the US-based OIR system, particularly in the LSST era
\citep{Elmegreen2015}.  The panel made seven recommendations and
here I bring up those relevant to this paper.

The panel recognizes the need for extensive follow up in the LSST
era.  It should not surprise the reader that the panel suggests
development of a wide-field, highly multiplexed spectroscopic
facility in the Southern hemisphere. Realistically, a full decade
will be needed to realize such a facility (and that is five years
after LSST has been in operation). Any such facility will be working
in an landscape of a range of highly multiplexed spectrographs (and
discussed in the next section). Clearly opportunities abound but
strategic analysis of the landscape is essential.

Another recommendation of the panel is to strengthen the US OIR
``system".  This recommendation follows directly from the value of
follow up of targets resulting from LSST.  Following up requires
access to telescopes and as noted by the panel the US community has
seen a decrease in the number of public telescopes. A simple way
to meet the panel's recommendation is for NSF to renew the ``TSIP"
program, in which case the discussion in \S\ref{sec:CostAndValue}
could be of some use.  I find the panel's recommendation of
``bartering" as not practical. Privately run observatories need
funds to run and improve their facilities. Separately, any great
opportunity for bartering will, in most cases, be recognized and acted
upon by the Directors of the observatories. Finally, the scale of
funding for a telescope access program (``TAP") that would make a
difference to the astronomical community and at the same time  have
the ability to influence the existing marketplace is about \$10M
to \$15M per year. This is a much larger sum than that discussed in the
report.

\section{Large Optical Telescopes: A Bright Future but  also Challenges}
\label{sec:LargeOpticalTelescopes}

As noted earlier (\S\ref{sec:SynopticSurveys}) the  astronomical
world is awash with sky surveys across the electromagnetic spectrum.
There is no doubt that considerable astronomical progress will
likely take place using the data obtained from each imaging (or
photometric or astrometric) survey and by cross-survey comparisons.
As an example, I note that the amazing progress in the field of
astero-seismology is primarily rooted in the precision photometric
data provided by the Kepler mission. In contrast,  the great progress
in exoplanet studies most certainly required extensive followup,
namely, precision radial velocity (RV) studies of stars which were
identified as {\it candidates} by the same Kepler mission.  In the
same spirit, time domain surveys such as ZTF (and eventually LSST)
are good at identifying variable stars and transient sources but
in many cases follow up is key to making progress beyond flux
curves.  Therefore, it stands to reason that ground-based optical/NIR
telescopes  will, at least for some areas of astronomy, become
increasingly sought after for followup studies.

Next, there now exists a class of instruments which I call as ``mega"
instruments.  Such instruments are expensive (\$30M and up) and are
usually built for a specific science goal (for which the instrument
is tuned to have an impressive reach).  A summary of the mega
instruments can be found in \S\ref{sec:MegaInstruments}.  Briefly,
these mega instruments come in three flavors: those with large
spectroscopic target throughput (e.g.\ SDSS, Prime Focus Spectrograph
on the Subaru telescope), those with large FOV imagers (Hyper Suprime
Camera, Dark Energy Camera)  and those associated with AO (e.g.\
SPHERE, GPI; both designed to address imaging of exoplanets).  Mega
instruments allow astronomers to undertake certain unique projects.
The Subaru telescope is increasingly defined by its large FOV imagers
(e.g.\ the SuprimeCam and the Hyper Suprime Camera or HSC).  GPI
appears to have made its mark in high contrast imaging of stars.

\begin{figure}[htbp]
 \centering
  \includegraphics[width=2.5in]{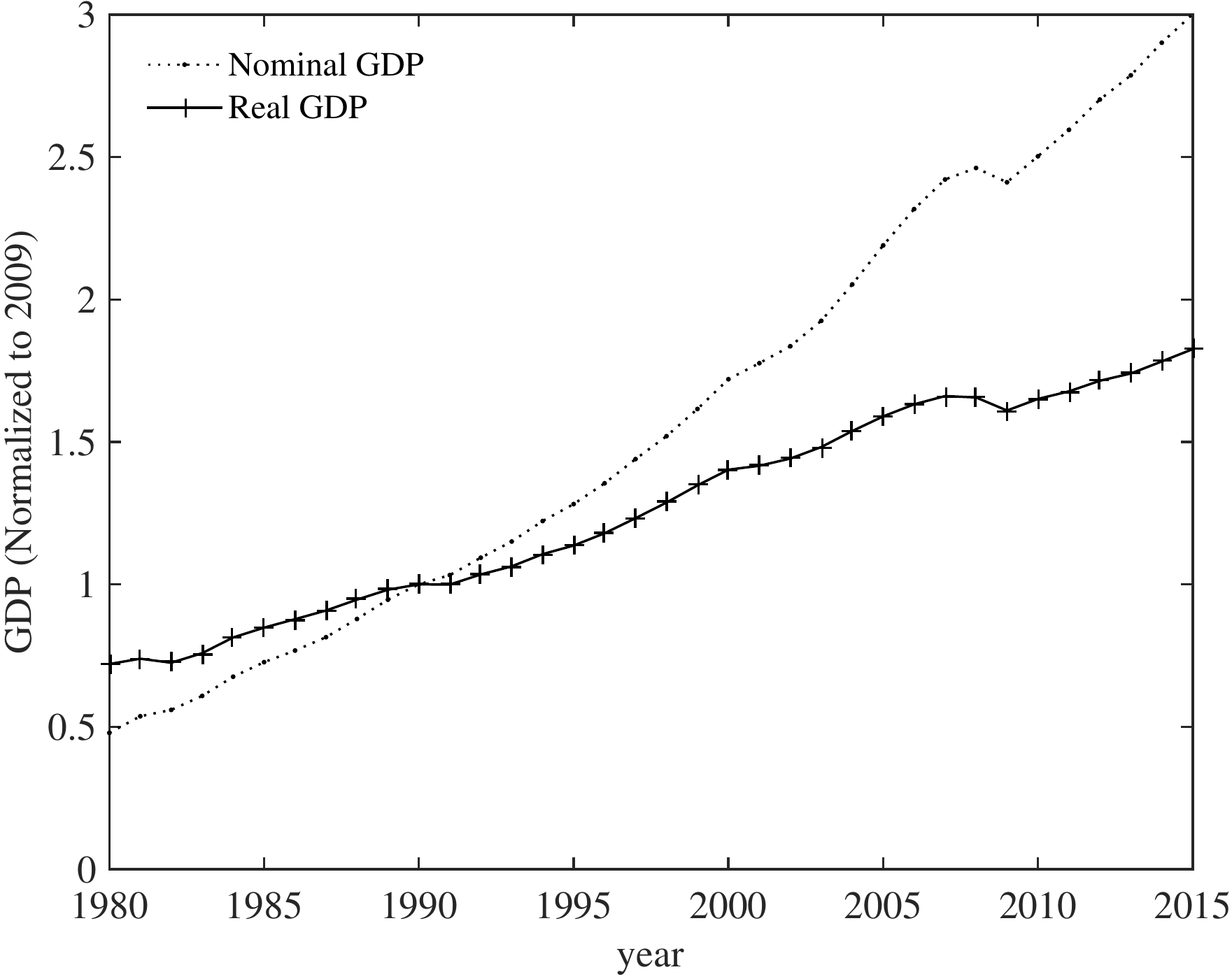} 
   \caption{\small Nominal and real GDP of the United States of
   America in the period 1980--2015.\footnote{Data from
\url{http://www.measuringworth.com}}
Nominal GDP is the value of production at current market prices.
Real GDP is the value of production using a given base year price;
the base year is set to 2009. I have normalized both measures by
dividing each measure by the corresponding value in 2009.  }
 \label{fig:GDP}
\end{figure}

\subsection{Challenge:  Cost of Instrumentation}
\label{sec:IncreasedCost}

It appears to be the case that every successive generation of
instruments, even in roughly the same category, are costing more
than those from the previous generation.  This assertion is justified
by a simple glance at Table~\ref{tab:TheInstruments}. For instance,
the cost of MOSFIRE is five times that of  NIRC. In contrast, as
can be seen from Figure~\ref{fig:GDP}, the nominal GDP increased
by a factor of 2.5 from 1990 to 2010 (during this period the real
GDP grew by only 1.5).

The increase in cost is easily explained: it arises from astronomer's
desire to get more out their fixed investment (the telescope) and
this desire is aided by rapid progress in technology.  Thanks to
Moore's law astronomers can now populate  increasingly larger
fraction of the focal plane with sensors. Thanks to improvements
in detectors, we can build useful wide-bandwidth instruments (e.g.\
X-shooter on VLT).  Technological developments have made new modes
possible as well as high quality measurements.  Examples include
integral field spectrographs and instruments with great stability
(e.g.\ flexure compensated spectrographs; precision radial velocity
spectrometers).

It is perhaps the case that large optical observatories  have
unwittingly entered into an arms race.  This situation has an uncanny
resemblance to the real arms race.\footnote{Successive generations
of fighter planes are vastly better than previous generation: greater
accuracy, lower mass,  more lethal power, smaller radar cross
section, and smaller risks to the pilot. However, the cost of fighter
planes has increased faster than the nominal GDP.  I recommend the
interested reader to carefully study a 2008 RAND report ``Why has
the cost of fixed-wing aircraft risen?" (RAND\_MG696).} I venture
to say that in both cases, the race is motivated by national pride,
the desire for global domination and is enabled by   rapid changes
in technology.

Thus, in the absence of planned upgrades, and bearing in mind  that it
takes about five years to design and develop even a typical instrument
for large telescopes, one should start working on the concept for
the next instrument a few years after the first generation instrument
is commissioned.  Should there be significant  technical innovations, of
relevance to the instrument, upgrades should be considered.
Failing this, a schedule for a replacement instrument must be launched
five years (the typical duration for construction of instruments
for the present generation of large telescopes)
before the anticipated obsolescence
of the current instrument (typically a decade after commissioning).

Observatory management would benefit having a bibliographic database
linked  to instruments. Ideally, the latter would include not merely
the name of the instrument (as has been and is currently the
situation with the WMKO bibliographic database) but also details
of the exposures undertaken during the run (integration times,
instrument mode, slewing time, seeing conditions, integration time).
Even if such a grand goal cannot achieved it is essential for
management to undertake retrospective analysis (of the sort undertaken
here) and use lessons learnt when making future choices. In this
regard, I draw the reader's attention, with some admiration, of
ESO's bibliometric portal\footnote{\url{http://telbib.eso.org/}}.
The portal is sufficiently sophisticated that the analysis I undertook
here can probably be done in less than a few weeks of time.

\subsection{Solution: Upgrades \&\ Common Development}

Upgrades can be  cost effective to maintain (if not increase) the
productivity of instruments (cf.\ LRIS-B, LRIS-R, HIRES; see
\S\ref{sec:Upgrades}).  Thus it would be useful to build into the
initial instrument the possibility for upgrades, especially
anticipating new and better detectors.  Next, instrumentation
projects encounter two types of cost challenges: the total cost and
the maximum burn rate. A phased approach to instrumentation would help
address the latter problem.  Indeed, in effect, this has been the
effective (if not planned) policy at WMKO (e.g.\ LRIS-R and then
LRIS-B; KCWI-B and then KCWI-R).

Finally, it would be useful to examine if reuse of either hardware
or software (especially) is possible. Reuse could consist of using
parts of instruments that are no longer competitive. For instance,
PTF uses the CHF12K detector (after extensive refurbishment; see
\citealt{rsv+08}).  A particularly innovative approach has been
undertaken by a collaboration between CHFT and Gemini-North
Observatory: ``Gemini Remote Access to CFHT ESPaDOnS"
(GRACES).\footnote{\url{http://www.gemini.edu/node/12131}} This
instrument combines the larger collecting area of Gemini with a
unique instrument (high resolving power, high efficiency, polarimetric
mode) at CHFT. GRACES is made possible by a 270-m length fiber which takes
starlight from Gemini and feeds to the spectrometer located in CFHT.

A real life example which avoids re-development and makes extensive
reuse is ``Collaboration for Astronomy Signal Processing and
Electronics" (CASPER)\footnote{\url{https://casper.berkeley.edu/}}.
The stated mission is ``to streamline and simplify the design flow
of radio astronomy instrumentation by promoting design reuse through
the development of platform-independent, open-source hardware and
software". CASPER is based on the idea of open source and community
development of hardware, gateware, gpuware, software (algorithms
and generic pipelines that can be easily adapted to various input
data formats).  The end goal is radio astronomy instrumentation for
pulsar search and timing, Fast Radio Burst (FRB) searches, aperture
synthesis, beam-forming, Search for Extraterrestrial Intelligence
(SETI) and Very Long Baseline Interferometer (VLBI).  CASPER
instrumentation is deployed world-wide: Arecibo (Puerto Rico), Green
Bank (West Virginia), Parkes (Australia), Effelsberg (Germany),
Giant Meter Wavelength Radio Telescope (GMRT, India), Submillimeter
Array (SMA; Hawaii), Long Wavelength Array \&\  Large Aperture
Experiment to Detect the Dark Ages (LEDA; both at Owens Valley Radio
Observatory, California), LWA/LEDA, PAPER (Precision Array for
Probing the Epoch of Reionization) \&\ HERA (Hydrogen Epoch of
Reionization Array), Very Long Baseline Array (NRAO), MeerKAT (South
Africa), Medicina Observatory (Sardinia, Italy), Allen Telescope
Array (Hat Creek Radio Observatory, California), Deep Space Network
(DSN; JPL/NASA), ALMA (Atacama Large Millimeter Array, Chile), Five
hundred meter Aperture Spherical Telescope (FAST, China), Shanghai
Observatory (China) and Infrared Spatial Interferometer (Mt.\ Wilson,
California).

I can personally attest to the impact of CASPER on radio astronomy.
As a student I either developed or was involved in several hardware
projects in radio astronomy:  correlators, hardware for pulsar
searching and timing, and long-baseline interferometry.  For of
each of these projects I spent a year just for the {\it development}
phase. During this summer I intend to  start a project for dipole-based
wide-angle FRB searches at OVRO and Palomar.  Thanks to CASPER I
expect that  the {\it implementation} phase for this project to be
less than 3 months.

In OIR astronomy, over the past fifteen years,  the AO community
undertook two ``roadmap" exercises were undertaken. Each exercise
led to collaborative developments.  This is not enough!  OIR astronomy
needs both a broader effort as well as a {\it sustained} effort,
similar to CASPER.  The success of CASPER would hopefully catalyze
similar common development programs.

\subsection{Mega Instruments: Swaps \& Vertical Integration}
\label{sec:Swaps}

As noted earlier, mega instruments, if chosen wisely and executed
well, can undertake spectacular science.  However, mega instruments
eponymously are expensive.  Say, for argument's sake, that a proposed
mega  instrument costs \$50M.  We will accept 10 years as a reasonable
peak lifetime.  Say, over this period, 1,000 nights are allocated
to the mega instrument.  Ignoring inflation, the instrument depreciates
at the rate of \$50K per night of usage, exceeding the ops cost for
a single night of a large optical telescope.  Thus naturally it only makes sense to
allocate all the time (subject only to lunations, if that is relevant)
to the mega instrument in question.  Swapping telescope time with
other facilities could then solve the problem of access to users
displaced by the arrival of the mega instrument.

A timely example is posed by the arrival of HSC on Subaru.  Since
HSC, until the commissioning of the LSST,
is unique it is the case that astronomers outside the Subaru
family  would be salivating at the prospect of using HSC. Thus, it
is desirable that in the era of mega-instruments significant time
swaps between major observatories be undertaken.\footnote{A joint
NAOJ-WMKO meeting  ``Keck-Subaru Strategy: Sendai 2015" to  explore
and discuss this topic was held in September 2015 at Sendai, Japan.}
The ultimate solution may well be to have several observatories
under one management (``vertical integration" in commercial parlance).
In the coming era of mega instrument, ESO, which is already a
vertically integrated observatory, may have an advantageous position
relative to stand alone observatories.

\section{A Future of the W.\ M.\ Keck Observatory}
\label{sec:FutureWMKO}

The great success of the  Keck Observatory can be traced to two advantages:
(1) an early start and (2) a suite of instruments consisting of
powerful workhorses and wisely chosen niche instruments.  Keck rode
the rising performance gains of adaptive optics (especially the
methodology of laser guide star adaptive optics).

Viewed in retrospect there were three clear weaknesses. First was
the lack of timely upgrades of  NIR instruments. After all, NIR
detectors  have been or continue to be on  a virtuous trajectory
(see Appendix~\ref{sec:NIR}).  Given this situation the lack of a
{\it timely} upgrade of NIRSPEC was particularly unfortunate.  Even
more so when there were magnificent follow up opportunities of
objects found in the 1-year cryogenic all-sky MIR survey of WISE
mission \citep{wem+10}.  The delay of NIRES (a one-shot NIR echelle
spectrometer; now scheduled for
first light in late 2016) only made matters worse.
I cannot help but wonder whether a timely
upgrade of NIRSPEC would have made this instrument as powerful as
the optical spectrographs.

Second was a lack of appreciation of the impact of quality DRPs on
the productivity of astronomical research. Over time DRPs were
developed (with HIRES and DEIMOS leading the way) but the lack of
high quality and {\it timely} DRPs appears to have hurt WMKO's
productivity (cf.\ see discussion of OSIRIS towards the end of
\S\ref{sec:Archives}).

Going forward, in my view, the  Keck Observatory should continue
its current course, namely, serving a wide swathe of astronomers
with interests that span from exoplanets to the early Universe.
Using a currently fashionable word, Keck has been and should continue
to be a {\it holistic} observatory. This approach leverages off
other investments (e.g.\ in the Hubble era, Keck undertook critical
spectroscopic observations of faint supernovae found by Hubble; see
\S\ref{sec:ObservatoryLightCurve}). Perhaps a future such ``resonance"
could be with the James Webb Space Telescope (which has an assured
launch in 2018).

Earlier in \S\ref{sec:SynopticSurveys}  I noted that we are solidly
in the era of synoptic surveys and squarely in the middle of time
domain astronomy. The very large  flux of candidates resulting from
these surveys offer a great many opportunities for the world's most
sensitive OIR telescope. For instance, it is expected that
a young ($<\,1$\,day) supernova will be found by ZTF every night.
This  assured flux of 
targets opens up new types of projects. For 
instance LRISp\footnote{Polarimetric module;  see \citet{gcp95}.}
would be provide powerful diagnostics for asymmetries in the
progenitor and explosions.  WMKO observers can not only reap
low-hanging fruits in transient object astronomy but get ready for
highly nuanced and sophisticated usages when LSST turns on (first
survey, 2022).

Next, I note  the insatiable demand for multiplexed spectroscopy.
A modern version of DEIMOS (using the entire field of view) would
be unrivaled (given the large aperture of the telescope; slits,
relative to fibers, allow for fainter targets).  WFIRST, in particular,
will need highly multiplexed spectroscopy at extremely faint levels.
Next, {\it Gaia} is poised to revolutionize stellar and Galactic
astronomy (or more fashionably, ``near-field cosmology"). A moderate
resolution single object spectrograph operating from 0.3\,$\mu$m
to J-band (and employing EMCCDs and modern NIR detectors) is ideally
suited to exploiting {\it Gaia} data. Either a rebuild of ESI or a
new spectrograph would be a great addition to the Observatory.  

As noted earlier the
world is awash in large FOV imagers. Nonetheless, given the strong
red bias of existing large FOV detectors, a Keck U-band imager based
on highly efficient delta-doped CCDs \citep{jho+15} would be unique
and enable a wide range of astronomy (from SN shock breakout to UV
bright galaxies).

WMKO should be extremely  cautious of mega projects. As in ordinary
life, big investments have two costs: the cost of the investment
and the opportunity cost of the investment. In my view, after having
analyzed the market place and understood the grave risks of opportunity
cost, I do not find a compelling mega instrument for WMKO (although the
cost of a new version of DEIMOS tailored to WFIRST may well cross the \$30M mark).

\begin{figure}[htbp]
 \centering
  \includegraphics[width=2.5in]{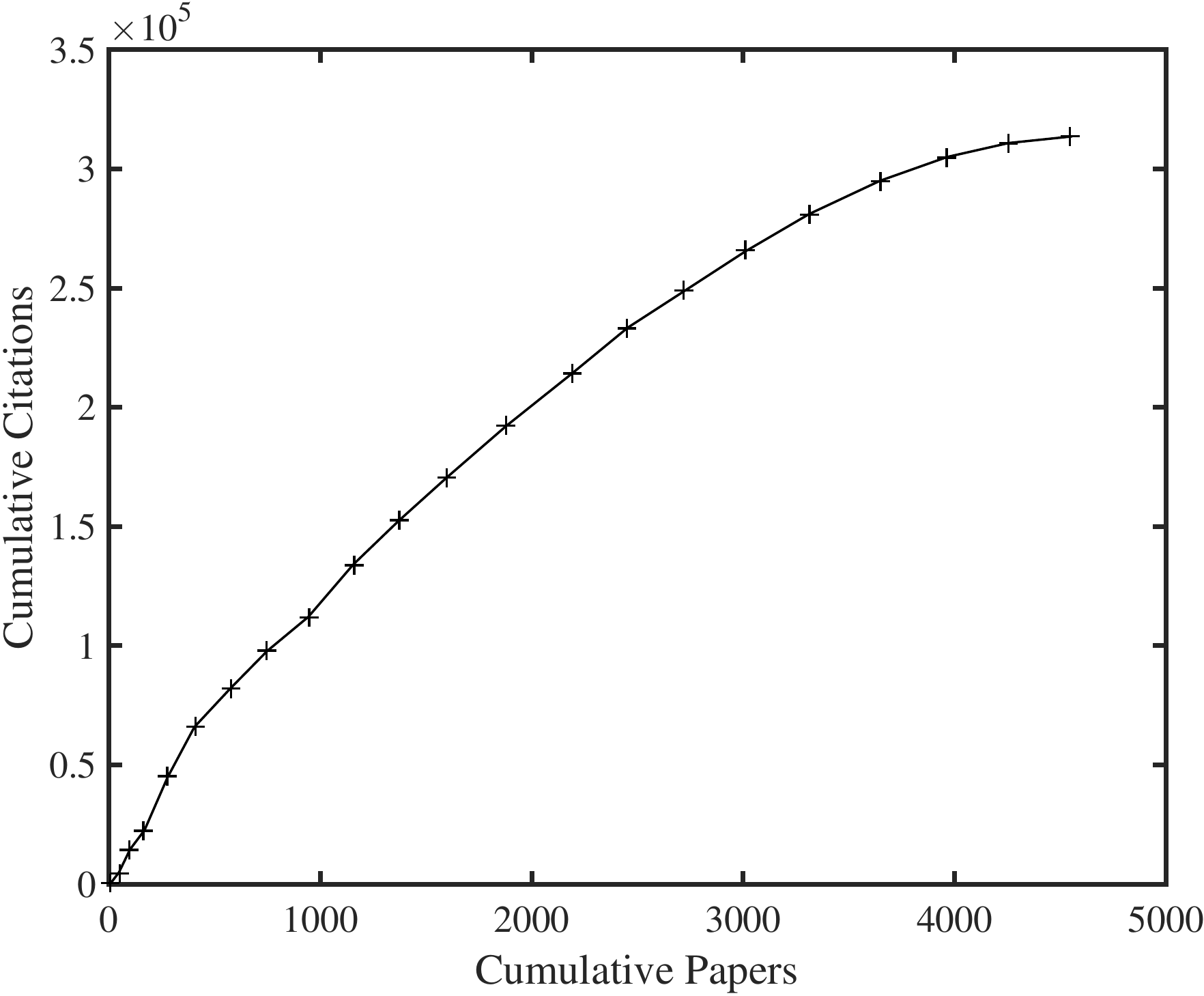}
   \caption{\small The cumulative of $\mathcal{C}_K(t)$ (abscissa)
   versus the cumulative of the
   annual flux of refereed published papers. The cumulative number of
   papers can serve as a proxy for time.  It appears that the 
   ratio has plateaued to a value of about 68 citations per paper.   }
 \label{fig:CumPaperCumCites}
\end{figure}

Even with all these suggested improvements it is important to
recognize that a continued growth in productivity will not be easy.
Indeed, as can be seen from  Figure~\ref{fig:CumPaperCumCites} there
is good evidence that the productivity of the Observatory has
plateaued.  (A flattening of the citation annual flux is also hinted
at in Figure~\ref{fig:CitationSummaryCurve}).

The growing Keck Observatory (\S\ref{sec:Archives}) archive can be
counted on to boost the productivity of the Observatory. Some help
may come from the soon-to-be-commissioned deployable tertiary on
Keck~I (the K1DM3 project).  This project allows for finer division
of nights, ``cadenced" observing and an increased number of TOOs
-- all of which, if properly leveraged, can contribute to increased
productivity.  New instruments -- Near-Infrared Echellete Spectrograph
(NIRES; summer 2016), Keck Cosmic Web Imager (KCWI; Fall 2016),
Keck Planet Finder  (2019) and the on-going
and planned upgrades (OSIRIS, NIRSPEC) -- have powerful capabilities.
These instruments combined with the enormous collecting area of
telescope  along with the superb site  means that astronomers who
are fortunate to have access to the Observatory cannot 
but continue to make great discoveries.  The astronomical
future of the W. M. Keck Observatory is bright, limited only by
financing, the ability of astronomers to innovate in observing
styles, and the competence of management.

\acknowledgements 

I am grateful to the Peggi Kamisato (WMKO Librarian) for providing
the Keck bibliography in a machine readable format, to Barbara
Schaefer (WMKO Scheduler) for patiently clarifying my queries, to
Hien Tran (WMKO Support Astronomer) for direct contribution to
\S\ref{sec:Archives}, to  Don Hall (IfA, UH) for providing the
material in Appendix~\ref{sec:NIR}, to Dan Werthimer (UCB) for
discussions of the CASPER vision and program and to Dr. Marten van
Kerkwijk (U. Toronto) for help with programming to query ADS. I
would like thank the scientists who led the projects for providing
estimates of cost and other details of their projects: K.\ Matthews
(NIRC, NIRC2), J.\ Cohen (LRIS, LRIS-B), R.\ Campbell (LWS), I.\
McLean (NIRSPEC), X.\ Prochaska (ESI), M.\ Bolte (ESI, ADC), S.\
Faber (DEIMOS), J.\ Larkin (OSIRIS), P.\ Wizinowich (AO), S.\ Adkins
(MOSFIRE), S.\ Vogt (HIRES), C.\ Rockosi (LRIS-R) and G.\ Chanan
(Phasing and quality of images).  I acknowledge useful discussions
with the following: A.~Barth, R.~Campbell, G.~Doppmann, D.~Elmegreen,  E.~Kirby,
A.~Kinney, H.~Lewis, J.~Lyke and V.~Trimble.  I am grateful to H.~A.~Abt,
D.~A.~Frail,  \&\  R.~Goodrich, A.~Ho, A.~Mahabal, C.~Max, H.~Vedantham,
M.~Strauss \&\ P.~Wizinowich for their careful reading. Their
suggested corrections and/or 
constructive criticisms greatly improved the paper.

I am grateful to Wendy Freedman of the Carnegie Observatories for
hosting my mini-sabbatical in the summer of 2013.  It was during
this time that I took the occasion to convert my preliminary informal
analysis into a paper. The serene surroundings combined with an
elegant library steeped in astronomical history minimized the pain
of tedious investigations, coding and collation of bibliographic
references.

As usual, I very much appreciate the excellent work undertaken by
librarians at various centers and Universities who maintain the ADS
data base.  The ADS is 
now a corner stone of astronomical research world wide. In particular,
without ADS this paper would not have been possible.

%\bibliographystyle{aasjournal}

%\bibliographystyle{apj1b}
%\bibliography{KeckInstruments}

%\input{KeckInstruments.bbl}

\appendix

\section{A Brief History of IR detectors}
\label{sec:NIR}

The industry notation is as follows: Short Wavelength Infrared
(SWIR) covering the range 1--2.5\,$\mu$m; Near Infrared (NIR)
covering the range 1--5\,$\mu$m; and Mid Infrared (MIR) covering
the range 5--25\,$\mu$m.  The eighties saw an explosion of IR
detector technologies. The eighties also marked the time for
technology transfer from the military to astronomers.

In optical astronomy, Silicon serves both as the ``sensor" (exciting
electrons to the conduction band) as well as the  ``reader" (converting
the electron count to digital values). However, the band-gap  of
Silicon is 1.05\,$\mu$m. Clearly, Silicon is not suitable for NIR
detectors. NIR detectors need  two distinct materials: a sensing
layer (with band-gaps appropriate for NIR photons) and a Silicon
layer for reading.  A mechanism has to be identified to connect
these two layers (``hybridizing").

Three families of detectors cover the full range of IR astronomy.
InSb detectors are used for SWIR and NIR bands.  HgCdTe works from
0.8\,$\mu$ to a long wavelength cutoff that can be engineered to
as low as H-band and to as high as 10\,$\mu$m.  Si:As IBC cover the
range 6\,$\mu$m to nearly 30\,$\mu$m and are the detectors of choice
for the MIR band.  As noted above, in all cases, the readout is
done by Silicon based circuitry.  Here, we review the development
of NIR detectors during the period 1990--2015.

In the early nineties the NIR detectors of choice were ``Astronomical
Large Area Detector Development on InSb" (ALADDIN) InSb detectors
manufactured by Hughes Santa Barbara Research Corporation (SBRC).
See \citealt{fgv+94} for the history of this line of detectors since
introduction of the first detector in 1986. It was only natural
that NIRC (commissioned in 1993) was based on a $256\times 256$ pixel
ALADDIN array detector.  Fowler et al. ({\it ibid})  talk of the development
of $1024\times 1024$ pixel array detector (and indeed was deployed  in NIRSPEC).

The HAWAII-1 (HgCdTe Astronomical Wide Area Infrared Imager) was
developed by a partnership between U.\ Hawaii (D.~Hall and K.~Hodapp)
and Rockwell (now Teledyne).  The Rockwell HAWAII (1K$\times$1K
pixel array) and the HAWAII-2 (2K$\times$2K pixel) arrays were based on the
$2.5\,\mu$m cutoff LPE HgCdTe on Sapphire substrate technology.
The HAWAII-1 was first used in the
UH Quick Infra-Red Camera (QUIRC) in July, 1994 to observe the Comet
Shoemaker-Levy impact with Jupiter \citep{hhh+96}.

The PICNIC detector used the sensing layer of InSb but a readout
circuit based on the HAWAII array technology.  Thus PICNIC detectors
(unlike NICMOS) can only reset a line of pixels but not a single
pixel. However, PICNIC gained the lower read noise advantage  of
HAWAII and the ability to turn off the circuitry during an exposure
to reduce amplifier glow. Rockwell also developed the 
$256^2$ pixel array for Near Infrared Camera and Multi-Object
Spectrometer (NICMOS). NICMOS was commissioned on HST in 1997.
NICMOS uses HgCdTe sensing layer bonded to sapphire substrate and the read out was
based on HAWAII technology.

The ability to butt became possible with a Hawaii-2 (2\,K$\times$2\,K
pixel)
which was first produced in 1998. These detectors saw saw limited
use, notably in the UKIRT mosaic camera.  Towards the end of the
nineties Rockwell declassified the read and guide mode technology
which made possible rapid guiding (RG) mode \citep{hhg+00}.  HAWAII-2RG
became available in 2001.  The Sidecar ASIC control chip was
introduced in 2003. H1R (reference pixels) flew on Deep-Impact
(launched in 2005). H1RG (reference pixels plus guide sub-array)
was employed by the Orbiting Carbon Observatory (OCO; launched in
early 2009) and the Wide Infrared Survey Explorer mission (WISE;
launched in late 2009).  Currently, U. Hawaii  is being funded by
NSF to develop HAWAII-4RG detectors for use by large ground based
telescopes.  As we go to press, the first H4RG will be field tested
at a telescope (D.\ Hall, pers. comm.).

In summary, apart from the fantastic increase in pixel count (256$\times$
256 pixel) from the NIRC InSb ALADDIN array to the H4RG-15, there have
been major improvements in dark current (achieved at significantly
higher operating temperature), in lower read noise, in higher QE,
in reduction of persistent image effects and in reduction in radiation
effects through removal of the CZT substrate.

\section{Mega Instruments}
\label{sec:MegaInstruments}

SDSS  marks perhaps  the first instance wherein the capital cost
of the instrument was comparable to the cost of the telescope itself
(and dominated the project cost if software expenses were included).
Also relative to the size of the telescope the annual operation
cost was very high. I use the following criteria to classify an
instrument as a mega instrument: a capital cost  approaching that
of the telescope or a cost, say, of \$30M or more.

Mega instruments come in two flavors: those with large reach and
those which are designed to answer specific but important questions.
Related to the first category are facilities built around highly
multiplexed spectrographs:   the Sloan Digital Sky Survey (SDSS)
and the Large Area Multi-Object Fibre Spectrograph
(LAMOST)\footnote{\url{http://www.lamost.org/public/?locale=en}}.
 The backend for LAMOST is an impressive  4000-channel
dual beam spectrograph and this is fed by a wide FOV 4-m Schmidt
camera (which also happens to be the large Schmidt telescope in the
world).

HERMES  has 390-fibers feeding {\it four} spectrographic arms on
the 3.9-m AAT telescope\footnote{\url{http://www.aao.gov.au/HERMES/}}.
HETDEX consists of 150 (!) IFU spectrographs mounted at the prime
focus of the HET\footnote{\url{http://hetdex.org/hetdex/}}.  WEAVE
is  a 1000-channel spectrograph on the 4.2-m William Herschel
telescope\footnote{\url{http://www.ing.iac.es/weave/about.html}}.
The planned MS-DESI is a 5000-channel spectrograph on the 4-m Mayall
telescope\footnote{\url{http://desi.lbl.gov/}}. The planned Prime
Focus Spectrograph
(PFS/SuMIRE)\footnote{\url{http://sumire.ipmu.jp/en/2652}} on the
Subaru 8.2-m telescope has 2400 fibers feeding a 3-arm spectrograph.
Then we have large FOV cameras: the Dark Energy
Camera (3\,square degrees) on the Blanco 4-m telescope and the
Hyper-Suprime Camera\footnote{\url{http://www.naoj.org/Projects/HSC/}}
(HSC; FOV of 1.7\,square degrees) on the Subaru 8.2-m telescope.
These are major undertakings with  costs\footnote{including the
cost of modifying the primary support system, the camera optics,
the data taking system and pipelines}  that place them squarely in
the mega-instrument category.

In the second category, the instruments appeared to be rooted in
Adaptive Optics.  The Gemini Planet Imager
(GPI)\footnote{\url{http://planetimager.org/}} on the Gemini South
8-m telescope is a high-contrast AO imager for high dynamic range
(extra-solar planets) studies.
SPHERE\footnote{\url{https://www.eso.org/sci/facilities/develop/instruments/sphere.html}}
is also a ``planet finder'' but on the VLT 8-m telescope.  The
run-out cost of GPI and Sphere are estimated to be \$26M and \$50M,
respectively.  The Gemini Multi-Conjugate Adaptive Optics System
(GeMS)\footnote{\url{http://www.gemini.edu/sciops/instruments/gems/}} aims
to deliver a very well corrected beam over one arc-minute field-of-view.

%From Bruce Macintosh (June 22, 2016)
%One minor comment/question - you quote $30M for a total cost for
%GPI. The number I would use is $25M (from the start of construction
%contracts to first light); the conceptual design study in 2004 would
%add $0.2M and the verification and performance characterization
%phase about $0.5M, so perhaps $26M with rounding.

\section{Top Papers}
\label{sec:TopPapers}

\subsection{NIRC}
\begin{enumerate}
\item
{\it The Rest-Frame Optical Spectra of Lyman Break Galaxies: Star Formation, Extinction, Abundances, and Kinematics}  (2001), {\it ApJ } {\bf 554}, 981
\item
{\it Submillimetre-wavelength detection of dusty star-forming galaxies at high redshift}  (1998), {\it Nature} {\bf 394}, 248
\item
{\it A Survey of $z>$5.7 Quasars in the Sloan Digital Sky Survey. II. Discovery of Three Additional Quasars at $z>6$}  (2003), {\it AJ } {\bf 125}, 1649
\item
{\it Stellar Orbits around the Galactic Center Black Hole}  (2005), {\it ApJ } {\bf 620}, 744
\item
{\it The Rest-Frame Optical Properties of $z\sim 3$ Galaxies}  (2001), {\it ApJ } {\bf 562}, 95
\item
{\it High Proper-Motion Stars in the Vicinity of Sagittarius A*: Evidence for a Supermassive Black Hole at the Center of Our Galaxy}  (1998), {\it ApJ } {\bf 509}, 678
\item
{\it The unusual afterglow of the $\gamma$-ray burst of 26 March 1998 as evidence for a supernova connection}  (1999), {\it Nature } {\bf 401}, 453
\item
{\it Multiwavelength Observations of Dusty Star Formation at Low and High Redshift}  (2000), {\it ApJ } {\bf 544}, 218
\item
{\it The Stellar, Gas, and Dynamical Masses of Star-forming Galaxies at $z \sim 2$}  (2006), {\it ApJ } {\bf 646}, 107
\end{enumerate}

\subsection{LRIS}
\begin{enumerate}
\item
{\it Observational Evidence from Supernovae for an Accelerating Universe and a Cosmological Constant}  (1998), {\it AJ } {\bf 116}, 1009
\item
{\it Measurements of {$\Omega$} and {$\Lambda$} from 42 High-Redshift Supernovae}  (1999), {\it ApJ } {\bf 517}, 565
\item
{\it Type Ia Supernova Discoveries at $z>1$ from the Hubble Space Telescope: Evidence for Past Deceleration and Constraints on Dark Energy Evolution}  (2004), {\it ApJ } {\bf 607}, 665
\item
{\it The Supernova Legacy Survey: measurement of $\Omega_{M}, \Omega_{\Lambda}$ and $w$ from the first year data set}  (2006), {\it A\&A } {\bf 447}, 31
\item
{\it High-redshift galaxies in the Hubble Deep Field: colour selection and star formation history to $z\sim 4$}  (1996), {\it MNRAS } {\bf 283}, 1388
\item
{\it Discovery of a supernova explosion at half the age of the universe}  (1998), {\it Nature } {\bf 391}, 51
\item
{\it Cosmological Results from High-z Supernovae}  (2003), {\it ApJ } {\bf 594}, 1
\item
{\it Lyman-Break Galaxies at $z\gtrsim 4$ and the Evolution of the Ultraviolet Luminosity Density at High Redshift}  (1999), {\it ApJ } {\bf 519}, 1
\item
{\it The Keck Low-Resolution Imaging Spectrometer}  (1995), {\it PASP } {\bf 107}, 375
\end{enumerate}

\subsection{HIRES}
\begin{enumerate}
\item
{\it The Planet-Metallicity Correlation}  (2005), {\it ApJ } {\bf 622}, 1102
\item
{\it Spectroscopic Properties of Cool Stars (SPOCS). I. 1040 F, G, and K Dwarfs from Keck, Lick, and AAT Planet Search Programs}  (2005), {\it ApJS } {\bf 159}, 141
\item
{\it Attaining Doppler Precision of 3\,m\,s$^{-1}$}  (1996), {\it PASP } {\bf 108}, 500
\item
{\it A Transiting ``51 Peg-like'' Planet}  (2000), {\it ApJ } {\bf 529}, L41
\item
{\it Catalog of Nearby Exoplanets}  (2006), {\it ApJ } {\bf 646}, 505
\item
{\it Planetary Candidates Observed by Kepler. III. Analysis of the First 16 Months of Data}  (2013), {\it ApJS } {\bf 204}, 24
\item
{\it Star-Formation Histories, Abundances, and Kinematics of Dwarf Galaxies in the Local Group}  (2009), {\it ARA\&A } {\bf 47}, 371
\item
{\it The Deuterium Abundance toward Q1937-1009}  (1998), {\it ApJ } {\bf 499}, 699
\item
{\it Planet Occurrence within 0.25 AU of Solar-type Stars from Kepler}  (2012), {\it ApJS } {\bf 201}, 15
\end{enumerate}

\subsection{ESI}
\begin{enumerate}
\item
{\it Spectra and Hubble Space Telescope Light Curves of Six Type Ia Supernovae at 
$0.511< z < 1.12$ and the Union2 Compilation}  (2010), {\it ApJ } {\bf 716}, 712
\item
{\it Observational Constraints on the Nature of Dark Energy: First Cosmological Results from the ESSENCE Supernova Survey}  (2007), {\it ApJ } {\bf 666}, 694
\item
{\it Evidence for Reionization at $z\sim 6$: Detection of a Gunn-Peterson Trough in a 
$z=6.28$ Quasar}  (2001), {\it AJ } {\bf 122}, 2850
\item
{\it The Farthest Known Supernova: Support for an Accelerating Universe and a Glimpse of the Epoch of Deceleration}  (2001), {\it ApJ } {\bf 560}, 49
\item
{\it A Survey of $z>5.8$ Quasars in the Sloan Digital Sky Survey. I. Discovery of Three New Quasars and the Spatial Density of Luminous Quasars at $z\sim 6$}  (2001), {\it AJ } {\bf 122}, 2833
\item
{\it Constraining the Evolution of the Ionizing Background and the Epoch of Reionization with $z\sim 6$ Quasars. II. A Sample of 19 Quasars}  (2006), {\it AJ } {\bf 132}, 117
\item
{\it A Survey of $z>5.7$ Quasars in the Sloan Digital Sky Survey. II. Discovery of Three Additional Quasars at $z>6$}  (2003), {\it AJ } {\bf 125}, 1649
\item
{\it The Observed Offset Distribution of Gamma-Ray Bursts from Their Host Galaxies: A Robust Clue to the Nature of the Progenitors}  (2002), {\it AJ } {\bf 123}, 1111
\item
{\it Damped Ly {$\alpha$} Systems}  (2005), {\it ARA\&A } {\bf 43}, 861
\end{enumerate}

\subsection{NIRC2}
\begin{enumerate}
\item
{\it Direct Imaging of Multiple Planets Orbiting the Star HR 8799}  (2008), {\it Science } {\bf 322}, 1348
\item
{\it Measuring Distance and Properties of the Milky Way's Central Supermassive Black Hole with Stellar Orbits}  (2008), {\it ApJ } {\bf 689}, 1044-1062
\item
{\it Optical Images of an Exosolar Planet 25 Light-Years from Earth}  (2008), {\it Sci } {\bf 322}, 1345
\item
{\it Confirmation of the Remarkable Compactness of Massive Quiescent Galaxies at $z \sim 2.3$: Early-Type Galaxies Did not Form in a Simple Monolithic Collapse}  (2008), {\it ApJ } {\bf 677}, L5
\item
{\it The First Measurement of Spectral Lines in a Short-Period Star Bound to the Galaxy's Central Black Hole: A Paradox of Youth}  (2003), {\it ApJ } {\bf 586}, L127
\item
{\it Images of a fourth planet orbiting HR 8799}  (2010), {\it Nature } {\bf 468}, 1080
\item
{\it A Close Look at Star Formation around Active Galactic Nuclei}  (2007), {\it ApJ } {\bf 671}, 1388
\item
{\it The W. M. Keck Observatory Laser Guide Star Adaptive Optics System: Overview}  (2006), {\it PASP } {\bf 118}, 297
\item
{\it The Hawaii Infrared Parallax Program. I. Ultracool Binaries and the L/T Transition}  (2012), {\it ApJS } {\bf 201}, 19
\end{enumerate}

\subsection{NIRSPEC}
\begin{enumerate}
\item
{\it Type Ia Supernova Discoveries at $z>1$ from the Hubble Space Telescope: Evidence for Past Deceleration and Constraints on Dark Energy Evolution}  (2004), {\it ApJ } {\bf 607}, 665
\item
{\it The Farthest Known Supernova: Support for an Accelerating Universe and a Glimpse of the Epoch of Deceleration}  (2001), {\it ApJ } {\bf 560}, 49
\item
{\it The Mass-Metallicity Relation at $z>2$}  (2006), {\it ApJ } {\bf 644}, 813
\item
{\it A Survey of $z>5.8$ Quasars in the Sloan Digital Sky Survey. I. Discovery of Three New Quasars and the Spatial Density of Luminous Quasars at $z\sim 6$}  (2001), {\it AJ } {\bf 122}, 2833
\item
{\it The Structure and Kinematics of the Circum-galactic Medium from Far-ultraviolet Spectra of 
$z \sim$ 2--3 Galaxies} (2010), {\it ApJ} {\bf 717}, 289
\item
{\it The Stellar, Gas, and Dynamical Masses of Star-forming Galaxies at $z \sim 2$}  (2006), {\it ApJ } {\bf 646}, 107
\item
{\it Toward Spectral Classification of L and T Dwarfs: Infrared and Optical Spectroscopy and Analysis}  (2002), {\it ApJ } {\bf 564}, 466
\item
{\it Multiwavelength Constraints on the Cosmic Star Formation History from Spectroscopy: The Rest-Frame Ultraviolet, H{$\alpha$}, and Infrared Luminosity Functions at Redshifts 
$1.9 \lesssim z \lesssim 3.4$}  (2008), {\it ApJS } {\bf 175}, 48
\item
{\it H{$\alpha$} Observations of a Large Sample of Galaxies at $z \sim 2$: Implications for Star Formation in High-Redshift Galaxies}  (2006), {\it ApJ } {\bf 647}, 128
\end{enumerate}

\subsection{DEIMOS}
\begin{enumerate}
\item
{\it The Supernova Legacy Survey: measurement of $\Omega_M, \Omega_{\Lambda}$ and $w$ from the first year data set}  (2006), {\it A\&A } {\bf 447}, 31
\item
{\it New Hubble Space Telescope Discoveries of Type Ia Supernovae at $z\ge 1$: Narrowing Constraints on the Early Behavior of Dark Energy}  (2007), {\it ApJ } {\bf 659}, 98
\item
{\it Star Formation in AEGIS Field Galaxies since $z=1.1$: The Dominance of Gradually Declining Star Formation, and the Main Sequence of Star-forming Galaxies}  (2007), {\it ApJ } {\bf 660}, L43
\item
{\it Observational Constraints on the Nature of Dark Energy: First Cosmological Results from the ESSENCE Supernova Survey}  (2007), {\it ApJ } {\bf 666}, 694
\item
{\it Cosmos Photometric Redshifts with 30-Bands for 2-deg$^{2}$}  (2009), {\it ApJ } {\bf 690}, 1236
\item
{\it The Kinematics of the Ultra-faint Milky Way Satellites: Solving the Missing Satellite Problem}  (2007), {\it ApJ } {\bf 670}, 313
\item
{\it The Mass Assembly History of Field Galaxies: Detection of an Evolving Mass Limit for Star-Forming Galaxies}  (2006), {\it ApJ } {\bf 651}, 120
\item
{\it Galaxy Stellar Mass Assembly Between $0.2 <z < 2$ from the S-COSMOS Survey}  (2010), {\it ApJ } {\bf 709}, 644
\item
{\it Metallicities of $0.3<z<1.0$ Galaxies in the GOODS-North Field}  (2004), {\it ApJ } {\bf 617}, 240
\end{enumerate}

\subsection{OSIRIS}
\begin{enumerate}
\item
{\it The Kiloparsec-scale Kinematics of High-redshift Star-forming Galaxies}  (2009), {\it ApJ } {\bf 697}, 2057
\item
{\it Two ten-billion-solar-mass black holes at the centres of giant elliptical galaxies}  (2011), {\it Nature} {\bf 480}, 215
\item
{\it Clouds and Chemistry in the Atmosphere of Extrasolar Planet HR8799b}  (2011), {\it ApJ } {\bf 733}, 65
\item
{\it The Lick AGN Monitoring Project: The $M_{BH}$-$\sigma_{*}$ Relation for Reverberation-mapped Active Galaxies}  (2010), {\it ApJ } {\bf 716}, 269
\item
{\it Near-infrared Spectroscopy of the Extrasolar Planet HR 8799 b}  (2010), {\it ApJ } {\bf 723}, 850
\item
{\it Dynamics of Galactic Disks and Mergers at $z \sim 1.6$: Spatially Resolved Spectroscopy with Keck Laser Guide Star Adaptive Optics}  (2009), {\it ApJ } {\bf 699}, 421
\item
{\it Integral Field Spectroscopy of High-Redshift Star-forming Galaxies with Laser-guided Adaptive Optics: Evidence for Dispersion-dominated Kinematics}  (2007), {\it ApJ } {\bf 669}, 929
\item
{\it The formation and assembly of a typical star-forming galaxy at redshift $z\sim 3$}  (2008), {\it Nature} {\bf 455}, 775
\item
{\it High Angular Resolution Integral-Field Spectroscopy of the Galaxy's Nuclear Cluster: A Missing Stellar Cusp?}  (2009), {\it ApJ } {\bf 703}, 1323
\end{enumerate}

\section{Very Large Array}
\label{sec:VLA}

The National Radio Astronomy Observatory (NRAO) librarian(s) maintain
a  data base\footnote{\url{https://find.nrao.edu/papers/}} of papers
published using NRAO facilities.  I am informed that librarians
pore through papers in journals and use a uniform criterion for
including papers in the NRAO data base.  The classification is quite
detailed (key projects, archival research, papers arising from
surveys etc).  Librarian(s) maintain a data base of papers published
on data obtained from NRAO facilities.  Two major surveys were
undertaken with the VLA:  Northern VLA Sky Survey (NVSS; eponymously
the entire Northern Sky) and FIRST (conforming to the SDSS footprint
of about 10,000 square degree of the Northern Galactic cap).

\begin{figure*}[htbp] 
   \centering
   \includegraphics[width=3.7in]{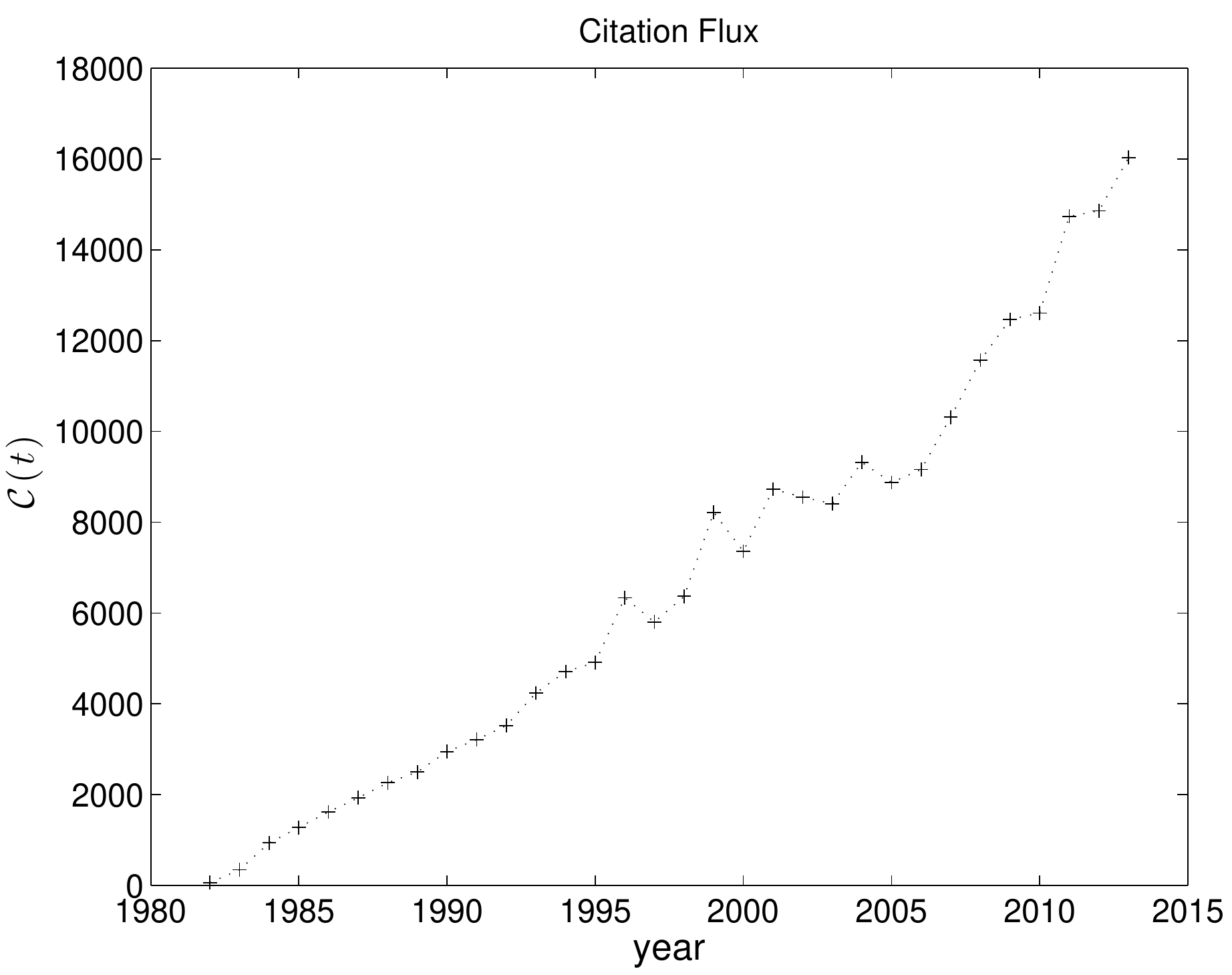}
   \caption{\small The citation flux arising from VLA refereed
   papers belonging to the following collections:
   ``VLA'', ``eVLA'', ``ArchVLA", ``FIRST" and  ``NVSS".
   }
\label{fig:VLALC} \end{figure*}

\begin{figure*}[htbp] 
   \centering
   \includegraphics[width=2.7in]{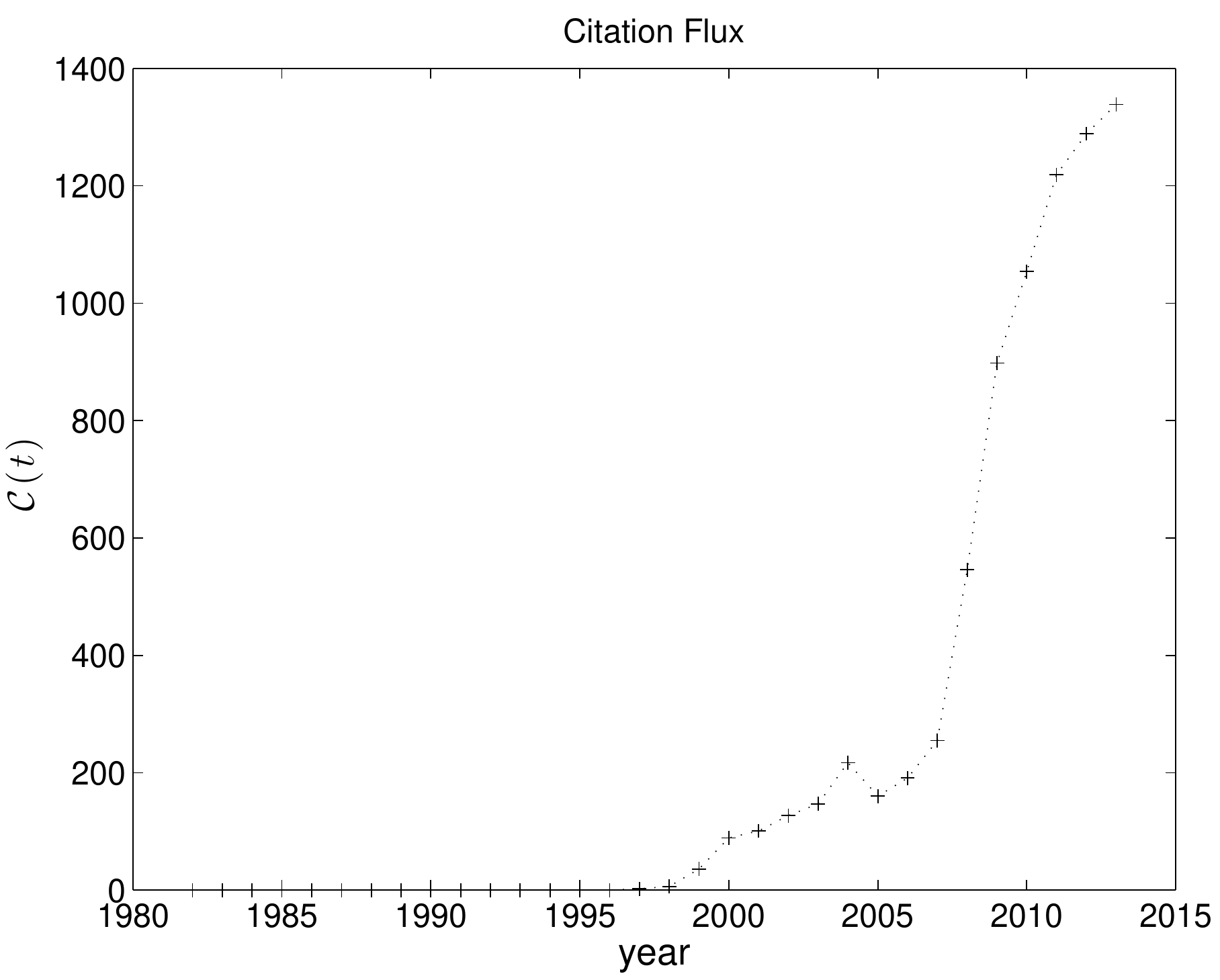}
   \hbox{\qquad}
   \includegraphics[width=2.7in]{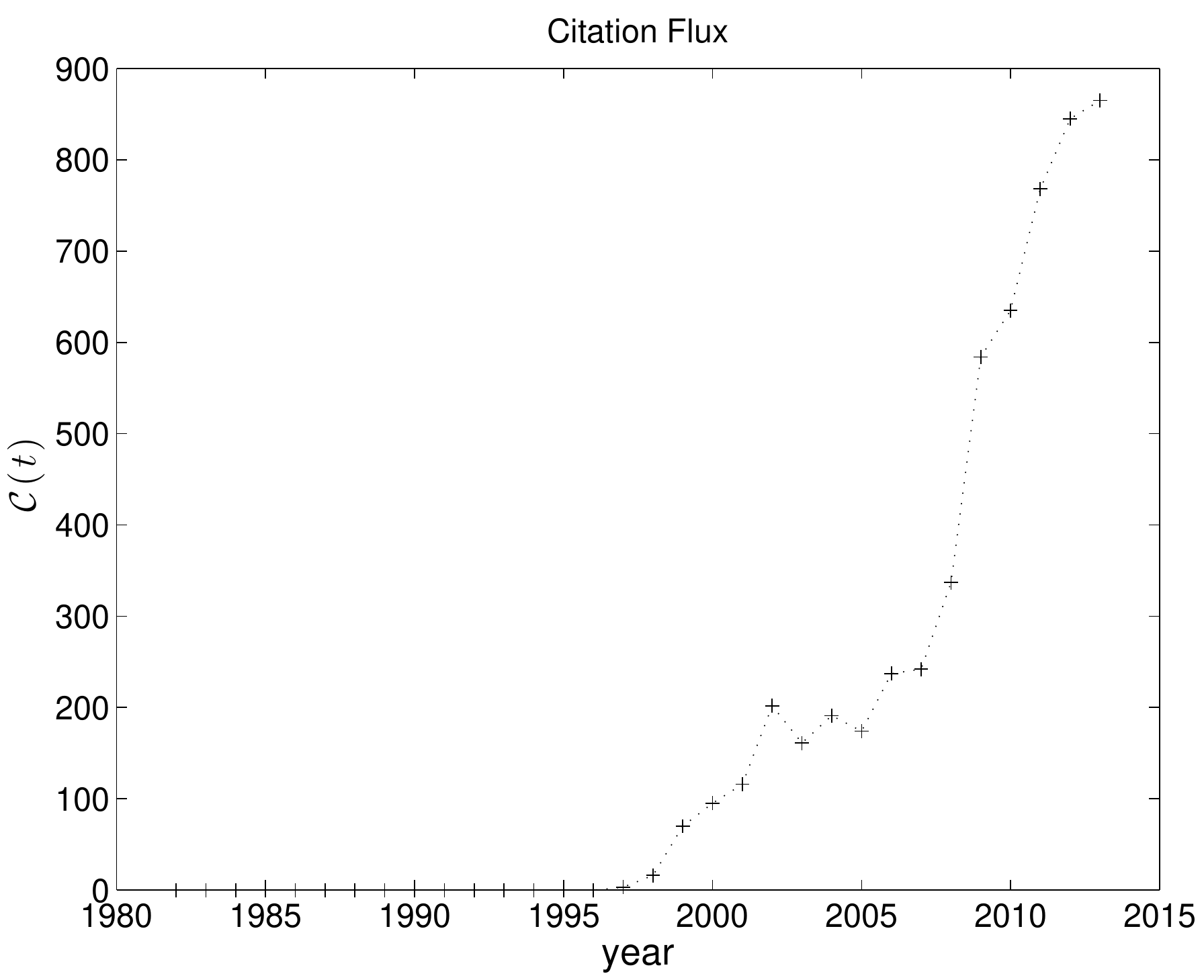}
   \caption{\small
   The citation flux arising from papers attributed as ``NVSS"
   (left) and ``FIRST'' (right).  The total number of citations of
   papers which use NVSS is 8249 and that for  FIRST is 5867.  
   }
 \label{fig:VLASurveysLC} 
\end{figure*}

Circa mid April 2014 I downloaded from the NRAO database (mentioned
above) the  output of the following collections (relevant to the
VLA): ``VLA" (5665 papers), ``eVLA" (268), ``FIRST" (266),  ``NVSS"
(442) and ``Archival VLA" (608). The total number of papers of these data sets
is 7249\footnote{I verified that there are no overlapping
papers between the data sets.}.  I applied the machinery developed
for this paper to the NRAO papers.  The citation flux curve for the five
data sets is given in Figure~\ref{fig:VLALC} while the citation flux curves
of the NVSS and FIRST can be found in Figure~\ref{fig:VLASurveysLC}.

To start with, the two papers describing the surveys are the most
cited papers in the approximately 30-year history of the VLA: NVSS
is explicitly cited by 2664 papers whereas FIRST is cited by 1300
papers.  NVSS was granted 2700 hours (Condon et al. 1998). The
number of papers which made use  of NVSS is, as of mid April 2014,
8249.  The total VLA citations at the same epoch  stand at 228,949.
The VLA was officially commissioned in 1982. Assuming an efficiency
factor of 0.,7 I find the mean citation production for the VLA is
1.16 per hour whereas that for NVSS (even after excluding citations
to the NVSS paper itself) is 3.06 per hour.

\end{document}